\documentclass[chap,11pt]{usthesis} 
\usepackage{epsfig,amssymb,multirow,amsmath} 

\newcommand{\afg}[2]{\frac{d#1}{d#2}} 
\newcommand{\comm}[2]{\left[#1,#2\right]} 
\newcommand{\acomm}[2]{\left\{#1,#2\right\}} 
\newcommand{\inp}[3]{\left\langle #1\left|#2\right|#3\right\rangle} 
\newcommand{\inpns}[3]{\langle #1|#2|#3\rangle} 
\newcommand{\ket}[1]{\left|#1\right\rangle}  
\newcommand{\bra}[1]{\left\langle#1\right|} 
\newcommand{\braket}[2]{\left\langle#1|#2\right\rangle} 
\newcommand{\ave}[1]{\left<#1\right>}
\newcommand{\pafg}[2]{\frac{\partial #1}{\partial #2}}
\newcommand{\bs}[1]{\boldsymbol{#1}}
\newcommand{\hbs}[1]{\hat{\boldsymbol{#1}}}
\newcommand{\eref}[1]{(\ref{#1})}
\newenvironment{narrow}[2]{%
\begin{list}{}{%
\setlength{\topsep}{0pt}%
\setlength{\leftmargin}{#1}%
\setlength{\rightmargin}{#2}%
\setlength{\listparindent}{\parindent}%
\setlength{\itemindent}{\parindent}%
\setlength{\parsep}{\parskip}}%
\item[]}{\end{list}}

\begin{document}

\thesistitle{Non-perturbative flow equations from continuous unitary transformations}
\author{Johannes Nicolaas Kriel}         
\degree{Master~of~Science} 
\supervisor{Professor F. G. Scholtz} 
\cosupervisor{Professor H. B. Geyer}
\submitdate{December 2005}  

\titlepage 
\declaration

\specialhead{ABSTRACT}
The goal of this thesis is the development and implementation of a non-perturbative solution method for Wegner's flow equations. We show that a parameterization of the flowing Hamiltonian in terms of a scalar function allows the flow equation to be rewritten as a nonlinear partial differential equation. The implementation is non-perturbative in that the derivation of the PDE is based on an expansion controlled by the size of the system rather than the coupling constant. We apply this method to the Lipkin model and obtain very accurate results for the spectrum, expectation values and eigenstates for all values of the coupling and in the thermodynamic limit. New aspects of the phase structure, made apparent by this non-perturbative treatment, are also investigated.  The Dicke model is treated using a two-step diagonalization procedure which illustrates how an effective Hamiltonian may be constructed and subsequently solved within this framework.
\vspace{3cm}
\begin{center}
{\large {\bfseries OPSOMMING}}
\end{center}
Die doel van hierdie tesis is die ontwikkeling en implementering van nie-steuringsteoretiese metodes vir die oplos van Wegner se vloeivergelykings. Ons toon dat 'n parameterisasie van die vloeiende Hamilton operator in terme van 'n skalare funksie daartoe lei dat die vloeivergelyking herskryf kan word as 'n nie-lini\^{e}re parsi\"{e}le differensiaal vergelyking. Hierdie implementering is nie op steuringsteorie gebaseer nie, en maak slegs gebruik van 'n reeks uitbreiding wat deur die grootte van die sisteem, eerder as die koppelingsterkte, beheer word. Ons pas die metode toe op die Lipkin model en verkry akkurate resultate vir die spektrum, verwagtingswaardes en eietoestande vir alle waardes van die koppelingsterkte en in die termodinamiese limiet. Nuwe aspekte van die fase struktuur wat deur hierdie benadering na vore gebring is, word ook bespreek. Die Dicke model word ondersoek deur 'n twee-stap diagonalisasie prosedure wat dien as voorbeeld van hoe effektiewe operatore binne hierdie raamwerk gekonstrueer en opgelos kan word.
\tableofcontents         
\listoffigures
\specialhead{Introduction}
The renormalization group and its associated flow equations \cite{Zinn} have become an indispensable tool in the study of modern physics. Its applications range from the construction of effective theories to the study of phase transitions and critical phenomenon. It also constitutes one of the few potentially non-perturbative techniques available for the treatment of interacting quantum systems. Our interest lies with a recent addition to this framework proposed by Wegner \cite{Wegner} and separately by Glazek and Wilson \cite{Glazek}, namely that of flow equations obtained from continuous unitary transformations. The flow equation in question describes the evolution of an operator, typically a Hamiltonian, under the application of a sequence of successive infinitesimal unitary transformations. These transformations are constructed so as to steer the evolution, or flow, of the operator towards a simpler, possibly diagonal, form. The major advantage of this approach is that no prior knowledge of these transformations is needed as they are generated dynamically at each point during the flow. These attractive properties have led to applications to several diverse quantum mechanical problems, including that of electron-phonon coupling \cite{Wegner}, boson and spin-boson models \cite{Kehrein,Vidal4,Vidal5}, the Hubbard model \cite{Stein1}, the Sine-Gordon model \cite{Kehrein1} and the Foldy-Wouthuysen transformation \cite{Bylev}.  The Lipkin model has also been particularly prominent among applications \cite{PirnerFriman,Stein2, Mielke,Vidal1,Vidal2}. More recently the flow equations have been used to construct effective Hamiltonians which conserve the number of quasi-particles or elementary excitations in a system \cite{Knetter3}. Examples of quasi-particles treated in this manner include triplet bonds on a dimerized spin chain \cite{Knetter,Knetter2,Schmidt,Schmidt2} and particle-hole excitations \cite{Heidbrink,Heidbrink2} in Fermi systems. Studies of these equations in a purely mathematical context, where they are known as double bracket flows, have also been conducted \cite{Brockett,Chu}.

The versatility of this approach should be clear from the references above. Unfortunately the practical implementation is hampered by the fact that the Hamiltonian typically does not preserve its form under the flow, and that additional operators, not present in the original Hamiltonian are generated \cite{Zinn,Wegner}. In general we are confronted with an infinite set of coupled nonlinear differential equations, the truncation of which is a highly non-trivial task. Perturbative approximations do allow one to make progress, although the validity of the results is usually limited to a single phase. 

It is the aim of this thesis to develop methods for the non-perturbative treatment of the flow equations. We will do so via two routes, both of which involve the representation of the flowing Hamiltonian as a scalar function and a systematic expansion in $1/N$, where $N$ represents the system size. This allows the original operator equation to be rewritten as a regular partial differential equation amenable to a numeric or, in some instances, analytic treatment. The bulk of this work comprises of the detailed application of these methods to two simple but non-trivial models. These calculations are found to reproduce known exact results to a very high accuracy.

The material is organized as follows. Chapter 1 provides a brief overview of the flow equation formalism with particular emphasis on the role of the generator. In Chapter 2 we present the two solution methods. The first is based on an expansion in fluctuations controlled by the system size, while the second makes use of non-commutative coordinates to rewrite the flow equation in terms of the Moyal bracket \cite{Moyal}. The application of these methods to the Lipkin model constitutes the third chapter. We are able to calculate both eigenvalues and expectation values non-perturbatively and in the thermodynamic limit. New aspects of the phase structure, made apparent by the non-perturbative treatment, are also investigated. These results, published in \cite{Kriel}, have subsequently led to further studies of the phase structure in \cite{Heiss}. In Chapter 4 the Dicke model is treated using a novel two-step diagonalization procedure. Although we derive the flow equation non-perturbatively its complexity necessitates a partially perturbative solution. Despite this our approach serves as a valuable example of how an effective Hamiltonian may be constructed and subsequently solved within this framework.

\chapter{Flow equations}
\label{S1}
\section{Overview}
\label{SS1.1}
The central notion in Wegner's flow equations \cite{Wegner} is the transformation of a Hamiltonian $H$ through the application of a sequence of consecutive infinitesimal unitary transformations. It is the continuous evolution of $H$ under these transformations that we refer to as the flow of the Hamiltonian. These transformations are constructed to bring about decoupling in $H$, leading to a final Hamiltonian with a  diagonal, or block diagonal, form. The major advantage of this approach is that the relevant transformations are determined dynamically during the flow, and no \textit{a priori} knowledge about them is required. 

We are led to consider a family of unitary transformations $U(\ell)$ which is continuously parametrized by the flow parameter $\ell\in[0,\infty)$. $U(\ell)$ constitutes the net effect of all the infinitesimal transformations applied up to the point in the flow labelled by $\ell$. At the beginning of the flow $U(0)$ equals the identity operator. The evolution of $U(\ell)$ is governed by 
\begin{equation}
\label{flowEqForU}
	\afg{U(\ell)}{\ell}=-U(\ell)\eta(\ell)
\end{equation}
where $\eta(\ell)$ is the anti-hermitian generator of the transformation. Applying $U(\ell)$ to $H$ produces the transformed Hamiltonian $H(\ell)=U^\dag(\ell)H U(\ell)$ for which the flow equation reads
\begin{equation}
  \label{flowEqForH}
	\afg{H(\ell)}{\ell}=\comm{\eta(\ell)}{H(\ell)}.
\end{equation}
To be consistent in the calculation of expectation values the same transformation needs to be applied to the relevant observables. An eigenstate $\ket{\phi}$ of $H$ transforms according to
\begin{equation}
	\ket{\phi,\ell}=U^\dag(\ell)\ket{\phi},
\end{equation}
while the flow of a general observable $O(\ell)=U^\dag(\ell)O U(\ell)$ is governed by
\begin{equation}
  \label{flowEqForO}
	\afg{O(\ell)}{\ell}=\comm{\eta(\ell)}{O(\ell)}.
\end{equation}
Expectation values in the original and transformed basis are related in the usual way:
\begin{equation}
\inp{\phi}{O}{\phi}=\inpns{\phi}{U(\ell)U^\dag(\ell)OU(\ell)U^\dag(\ell)}{\phi}=\inp{\phi,\ell}{O(\ell)}{\phi,\ell}.
\end{equation}
\section{The role of $\eta(\ell)$}
Much of the versatility of the flow equation method stems from the freedom that exists in choosing the generator $\eta(\ell)$. Several different forms have been employed in the literature, and we will explore the consequences of some of these next. For the moment we restrict ourselves to the finite dimensional case.

\subsection{Wegner's choice}
\label{wegnerChoice}
In Wegner's original formulation \cite{Wegner} $\eta(\ell)$ was chosen as the commutator of the diagonal part of $H(\ell)$, in some basis, with $H(\ell)$ itself, i.e. 
\begin{equation}
\eta(\ell)=\comm{H_{diag}(\ell)}{H(\ell)}.
\end{equation}
It was shown that in the $\ell\rightarrow\infty$ limit $H(\ell)$ converges to a final Hamiltonian $H(\infty)$ for which 
\begin{equation}
	\comm{H_{diag}(\infty)}{H(\infty)}=0.
\end{equation}
We conclude that the effect of the flow is to decouple those states which correspond to differing diagonal matrix elements. In general this leads to a block-diagonal structure for $H(\infty)$. 

We will not use this formulation as other choices exist which offer greater control over both the type of decoupling present in $H(\infty)$ (i.e. the fixed point of the flow) and the form of $H(\ell)$ during flow (i.e. the path followed to the fixed point).

\subsection{$\eta(\ell)=\comm{G}{H(\ell)}$}
\label{gChoice}
An alternative to Wegner's formulation is 
\begin{equation}
	\eta(\ell)=\comm{G}{H(\ell)}
\end{equation}
where $G$ is a fixed ($\ell$-independent) hermitian operator of our choice. It is straightforward to show that $H(\ell)$ converges to a final Hamiltonian which commutes with $G$. The proof rests on the observation that
\begin{equation}
 \frac{d}{d\ell}{\rm tr}(H(\ell)-G)^2=-2\,{\rm tr}([G,H(\ell)]^\dagger[G,H(\ell)])<0,
 \label{flowIneq}
\end{equation}
where the positivity of the trace norm has been used. It follows that ${\rm tr}(H(\ell)-G)^2$ is a monotonically decreasing function of $\ell$ that is bounded from below by $0$, and so its derivative must vanish in the $\ell\rightarrow\infty$ limit. The right-hand side of \eref{flowIneq} is simply the trace norm of $\eta(\ell)=\comm{G}{H(\ell)}$, and so we conclude that $\eta(\infty)=0$. Choosing a diagonal $G$ clearly leads to a block-diagonal structure for $H(\infty)$ where only states corresponding to equal diagonal matrix elements of $G$ are connected. Put differently, $G$ assigns weights to different subspaces through its diagonal matrix elements. The flow generated by $\eta(\ell)$ then decouples subspaces with differing weights. In particular, a non-degenerate choice of $G$ will lead to a complete diagonalisation of $H$. Furthermore, it can be shown \cite{Brockett} that the eigenvalues of $H$, as they appear on the diagonal of $H(\infty)$, will have the same ordering as the eigenvalues (diagonal matrix elements) of $G$. We can summarize this by saying that the flow equation generates a transformation that maps the eigenstates of $H$ onto the eigenstates of $G$ in an order preserving fashion.

It is worth noting that this ordering can only take place within subspaces that are irreducible under $G$ and $H$. The reason for this is that the flow equation clearly cannot mix subspaces that are not connected by either $H$ or $G$. We will see several examples of this later on.\\

\subsection{The structure preserving generator}
\label{etaChoice}
Whereas the formulation above provides a good deal of control over the structure of $H(\infty)$, the form of $H(\ell)$ at finite $\ell$ is generally unknown. This is due to non-zero off-diagonal matrix elements appearing at finite $\ell$ that are not present in either the initial or final Hamiltonian. For example, a band-diagonal Hamiltonian may become dense\footnote{A dense matrix is one of which the majority of elements is non-zero.} during flow and still converge to a diagonal form. From a computational point of view it is clearly desirable that $H(\ell)$ assumes as simple a form as possible. Of particular interest in this regard are generators which preserve a band diagonal, or more generally band block diagonal structure present in the original Hamiltonian. These types of generators have been applied to a wide range of models \cite{Knetter,Knetter2,Schmidt,Schmidt2,Heidbrink,Heidbrink2}, and are particularly attractive in that they allow for a clear physical interpretation of the transformed Hamiltonian $H(\infty)$. 

\begin{figure}[ht]
\begin{center}
\begin{tabular}{ccc}
	\epsfig{file=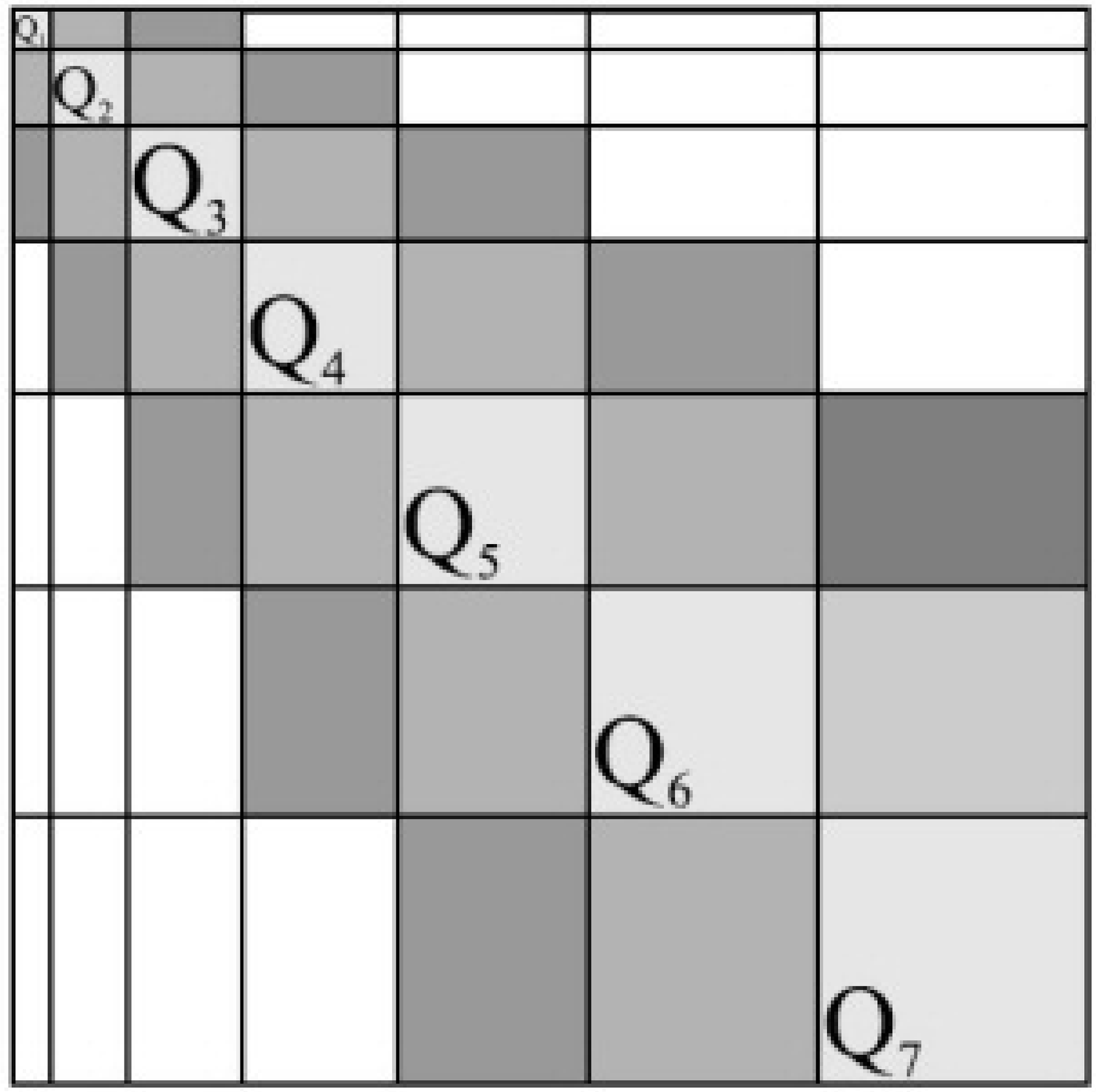,height=6cm,clip=,angle=0} & \makebox[1.2cm]{ }&
	\epsfig{file=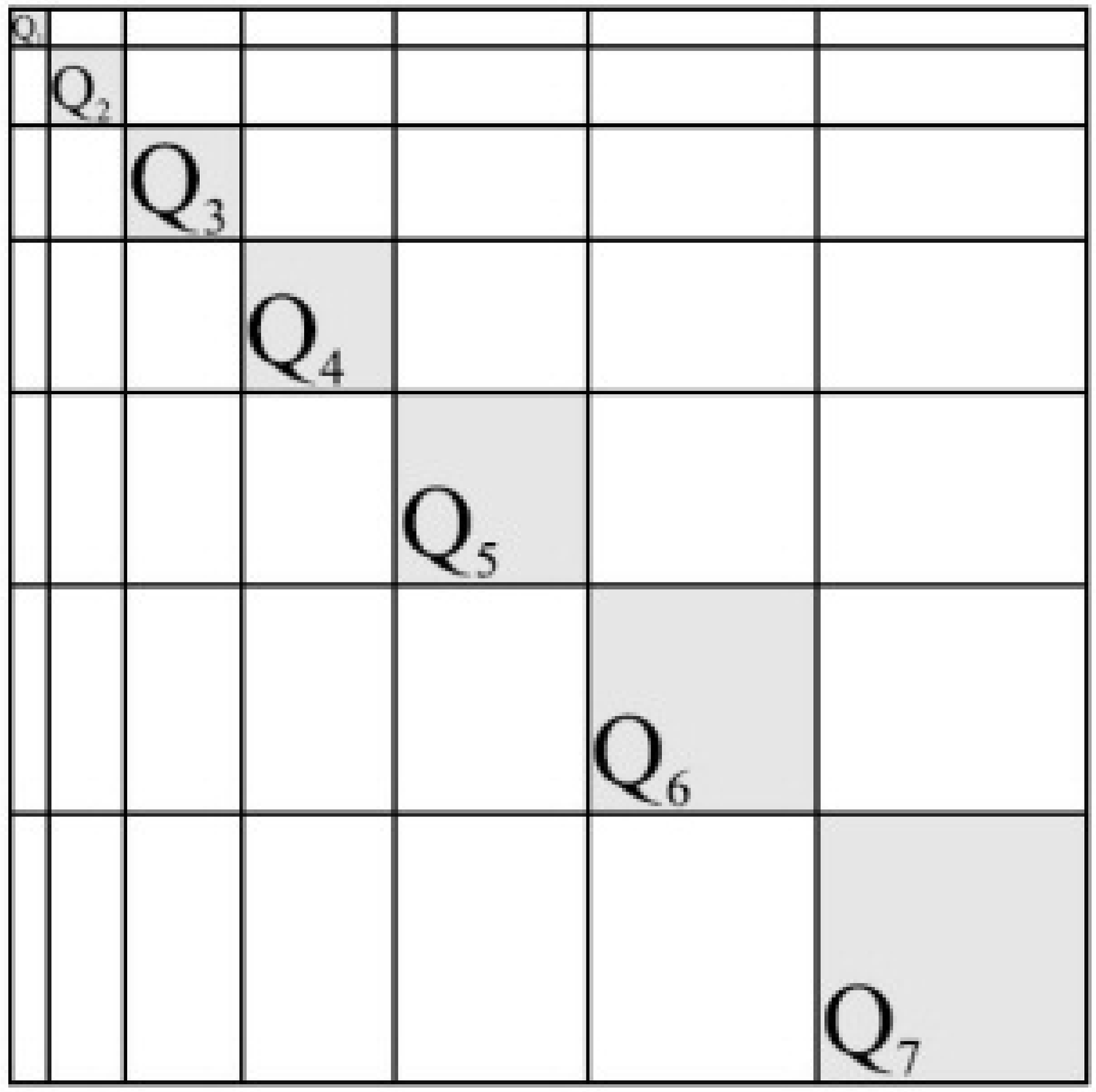,height=6cm,clip=,angle=0} \\
	(a) &   & (b)
	\end{tabular}
	\caption[The structure of band block diagonal matrices.]{The typical structure of (a) band block diagonal and (b) block diagonal matrices. Darker shaded blocks correspond to matrix elements which connect increasingly distant $Q$-sectors. Unshaded areas are filled with zero elements. For purposes of illustration we used $N=2$ in (a).}
	\label{bandblock}
\end{center}
\end{figure}

First we introduce an operator $Q$ with integer eigenvalues which will serve as a labelling device for different subspaces in the Hilbert space. Associated with each distinct eigenvalue $q_i$ of $Q$ is the corresponding subspace of eigenstates $Q_i$ where $i=1,2,3,\ldots$. We also assume, without loss of generality, that $q_i<q_{i+1}$ for all $i$. The Hamiltonian $H$ is said to possess a band block diagonal structure with respect to $Q$ if there exists an integer $N$ such that $\inp{i}{H}{j}=0$ for all $\ket{i}\in Q_i$ and $\ket{j}\in Q_j$ whenever $\left|q_i-q_j\right|>N$. This selection rule clearly places a bound on the amount by which $H$ can change $Q$. The matrix representation of such an operator in the $Q$ basis typically has a form similar to that shown in Figure \ref{bandblock} (a). For the cases we will consider, and for those treated in the literature, it is possible to 
group terms in the Hamiltonian together based on whether they increase, decrease or leave unchanged the value of $Q$. This leads to the form
\begin{equation}
	H=T_0+\underbrace{\sum_{n=1}^{n=N}T_n}_{T_+}+\underbrace{\sum_{n=-1}^{n=-N}T_n}_{T_-}
	\label{hform}
\end{equation}
where $T_n$ changes $Q$ by $n$, i.e. $\comm{Q}{T_n}=nT_n$. Clearly $T_0$ is hermitian, while $T_n$ and $T_{-n}$ are conjugates. $T_0$ is responsible for scattering within each $Q$-sector. 

Flow equations are used to bring this Hamiltonian into a form which conserves $Q$, i.e. for which $\comm{H(\infty)}{Q}=0$. This effective Hamiltonian will be block diagonal, similar to Figure \ref{bandblock} (b), with each block containing new interactions generated during the flow. We require that the band block diagonal structure of $H$ is retained at finite $\ell$, and so the flowing Hamiltonian should be of the form
\begin{equation}
H(\ell)=T_0(\ell)+\underbrace{\sum_{n=1}^{n=N}T_n(\ell)}_{T_+(\ell)}+\underbrace{\sum_{n=-1}^{n=-N}T_n(\ell)}_{T_-(\ell)}.
	\label{hformflow}
\end{equation}
The generator which achieves this is
\begin{equation}
	\eta(\ell)=T_+(\ell)-T_-(\ell).
\end{equation}
The corresponding flow equation reads
\begin{equation}
	\afg{H(\ell)}{\ell}=2\comm{T_+(\ell)}{T_-(\ell)}+\comm{T_+(\ell)-T_-(\ell)}{T_0(\ell)}.
	\label{flowtpm}
\end{equation}
The combinations of $T_+,T_-$ and $T_0$ appearing in \eref{flowtpm} clearly cannot generate scattering between $Q$-sectors differing by more than $N$, and so the band block diagonal structure of $H(0)$ is preserved. It can be shown \cite{Knetter} that this generator guarantees convergence to the desired form of $H(\infty)$ which conserves $Q$. 
\\

Before proceeding, let us point out why the choice of the previous section $\eta(\ell)=\comm{Q}{H(\ell)}$ does not lead to a flow of the form \eref{hformflow}, although it does produce the correct fixed point. Using the form of $H(0)$ we see that
\begin{equation}
	\eta(0)=\comm{Q}{H}=T'_+-T'_- 
\end{equation}
where $T'_\pm=\sum_{n=1}^{n=N}nT_{\pm n}$, and so
\begin{equation}
\left.\afg{H(\ell)}{\ell}\right|_{\ell=0}=\comm{T'_+-T'_-}{T_0+T_++T_-}.
\label{flowtpmG}
\end{equation}
Whereas commutators between terms which increase (decrease) $Q$ dropped out in \eref{flowtpm} this is not the case in \eref{flowtpmG}. In general $\comm{T'_\pm}{T_\pm}$ will contain terms which change $Q$ by up to $2N-1$, thus destroying the band block diagonal structure. This process will continue until, at finite $\ell$, $H(\ell)$ will connect all possible $Q$-sectors.
\\

Although our treatment has dealt with $Q$ largely in abstract terms, this is not generally the case in the literature. Previous applications of this method present a physical picture of $Q$ as a counting operator for the fundamental excitations describing the low energy physics of a system. Examples of these include triplet bonds on a dimerized spin chain \cite{Knetter,Knetter2,Schmidt,Schmidt2} and quasi-particles (particle-hole excitations)  \cite{Heidbrink,Heidbrink2} in Fermi systems. The flow equation is used to obtain an effective Hamiltonian that conserves the number of these excitations. These ``effective particle-conserving models" \cite{Knetter3} have been studied in considerable detail within a perturbative framework. 

Finally we point out that the approaches of Sections \ref{gChoice} and \ref{etaChoice} coincide when $H$ only connects sectors for which $Q$ differs by some fixed amount $N$, i.e. $\inp{i}{H}{j}=0$ when  $\left|q_i-q_j\right|\notin \nolinebreak \{0,N\}$.

\section{Flow equations in infinite dimensions}
Proofs concerning the convergence properties of the flow equations discussed thus far rely on the invariance of the trace under unitary transformations. In infinite dimensions the trace is not generally well defined, and the question of convergence depends on special properties of the Hamiltonian. The most important property in this regard is the boundedness of the spectrum. It has been shown that all the methods described above will converge provided that the Hamiltonian possesses a spectrum bounded from below \cite{Wegner,Knetter}. Aspects of the flow equations in infinite dimensions were investigated in \cite{Dusuel} and \cite{Ohira} using the bosonic Hamiltonian
\begin{equation}
	H=a^\dag a+\frac{\alpha}{2}\left(a^2+{a^\dag}^2\right),
\end{equation}
and we digress for a moment to discuss this case in more depth. It is well known that for $\alpha<1$ this model possesses a harmonic spectrum which can be found by applying the Bogoliubov transformation. When $\alpha>1$ the spectrum forms a continuum which is unbounded from both above and below. We will contrast the behaviour of the flow equations resulting from Wegner's choice of $\eta(\ell)=\comm{H_{diag}(\ell)}{H(\ell)}$ and the choice $\eta(\ell)=\comm{a^\dag a}{H(\ell)}$. In both these cases $H(\ell)$ has the simple form
\begin{equation}
	H(\ell)=\sigma(\ell)+\omega(\ell)a^\dag a+\frac{\alpha(\ell)}{2}\left(a^2+{a^\dag}^2\right),
\end{equation}
and the flow equation closes on a coupled set of three differential equations for $\sigma(\ell),\ \omega(\ell)$ and $\alpha(\ell)$. When $0\leq\alpha\leq1$ in the starting Hamiltonian we find that both choices of $\eta$ lead to the fixed point
\begin{equation}
	H(\infty)=\frac{1}{2}\left(\omega(\infty)-\alpha\right)+\omega(\infty)a^\dag a
\end{equation}
where $\omega(\infty)=\sqrt{1-\alpha^2}$; a result matching that of the Bogoliubov transformation.

Next, consider the $\alpha>1$ case. When we solve the flow equation for $\eta(\ell)=\comm{a^\dag a}{H(\ell)}$ we no longer find convergence. Surprisingly, Wegner's choice for the generator still produces a convergent flow, although not one leading to a diagonal form:
\begin{equation}
	H(\infty)=-\frac{1}{2}+\frac{1}{2}\sqrt{\alpha^2-1}\left(a^2+{a^\dag}^2\right).
\end{equation}
Note that this does not contradict the assertion made earlier that $\comm{H_{diag}(\infty)}{H(\infty)}=0$.

Admittedly this example does represent an extreme case. For physically sensible Hamiltonians the spectrum is always bounded from below, as one would expect in order to have a well-defined ground state. For the models that we will consider the flow equations may be safely applied without further modification.

\chapter{Solving the flow equation}
We have seen the theoretical capabilities of flow equations with regard to the diagonalization of Hamiltonians and the construction of effective operators. The practical implementation of this method is, however, hampered by the difficulty of solving the resulting operator differential equation. On the operator level this is due to the generation of additional operators during the flow that were not present in the original Hamiltonian. This leads to an extremely large set of coupled differential equations for the coupling constants of these terms. Generally some kind of approximation is required in order to continue. The usual approach consists of replacing $H(\ell)$ by a simpler parametrized form for which the flow equation closes on a set of equations of a tractable size. A particular parametrization is usually selected on the basis of a perturbative approximation, or by using some knowledge of the relevant degrees of freedom in the problem. In general this approach is only valid for a limited range of the coupling constant, and tends to break down when the system exhibits non-perturbative features, i.e. non-analytic behaviour in the coupling constant. 

We will introduce two new approaches to this problem which allow us to treat the flow equation in a non-perturbative way. The first involves a systematic expansion in fluctuations, which is controlled by the size of the system rather than the coupling constant. The second makes use of non-commutative variables to recast the flow equation as a regular partial differential equation. This approach also involves an expansion controlled by the system size. Although there are similarities between these two methods we will treat them separately and then show how they produce the same results in specific cases.

\section{Expansion in fluctuations}
\label{fluxExpand}
We consider the flow of a time reversal invariant hermitian operator $H(\ell)$ which acts on a Hilbert space ${\cal H}$. We assume ${\cal H}$ to be finite dimensional since the expansion \eref{genparamh} below is easily proved in this case (see Appendix \ref{appa}). This restriction is, however, by no means essential. If the Hilbert space is infinite dimensional the expansion \eref{genparamh} still applies to bounded operators \cite{Blank} and subsequently one may approach the infinite dimensional case by studying the flow of a bounded function of the Hamiltonian, instead of the Hamiltonian itself. Alternatively one can introduce a cut-off in some basis, e.g., a momentum cut-off and study the behaviour of the system as a function of the cut-off. 

The aim of this section is to develop a parameterization of the flowing Hamiltonian which allows for a systematic expansion controlled by the fluctuations, rather than the coupling constant.  This is in the same spirit as the semi-classical expansion of quantum mechanics, based on an expansion in orders of $\hbar$, and corresponds to resumming certain classes of diagrams in the perturbation series of the coupling constant.    

Let $H_0$ and $H_1$ be hermitian operators acting on ${\cal H}$ which together form an irreducible set.  What follows holds for any irreducible set, however, the case of two operators appears naturally in flow equations as the Hamiltonian can usually be written in the form $H=H_0+H_1$ where the spectrum and eigenstates of $H_0$ are known.  Note that if $H_0$ and $H_1$ are reducible on ${\cal H}$, but irreducible on a proper subspace of ${\cal H}$, the problem can be restricted to this smaller subspace making the irreducibility of $H_0$ and $H_1$ a very natural requirement. In Appendix \ref{appa} it is shown that any operator acting on ${\cal H}$ can be written as a polynomial in $H_0$ and $H_1$. In particular this holds for $H(\ell)$, and so we may write
\begin{equation}
	H(\ell)=\Gamma+\Gamma^iH_i+\Gamma^{ij}H_iH_j+\Gamma^{ijk}H_iH_jH_k+\ldots
	\label{genparamh}
\end{equation}
where each $\Gamma^{ij\ldots}$ coefficient is a function of $\ell$ and repeated indices indicate sums over $0$ and $1$. Since $H(\ell)$ is both hermitian and real it follows that the $\Gamma$'s are invariant under the reversal of indices:  
\begin{equation}
 	\Gamma^{n_1n_2\ldots n_k}=\Gamma^{n_kn_{k-1}\ldots n_1}\ \ \ \forall k>0,\ n_i\in\{0,1\}.
	\label{hermiteconstr}
\end{equation}
Now define $\Delta H_i=H_i-\ave{H_i}$ as the fluctuation of $H_i$ around the expectation value $\ave{H_i}=\inp{\phi}{H_i}{\phi}$, where $\ket{\phi}$ is some arbitrary state. By setting $H_i=\Delta H_i+\ave{H_i}$ in equation \eref{genparamh} we obtain $H(\ell)$ as an expansion in these fluctuations 
\begin{equation}
\label{expansion1}
	H(\ell)=\bar{\Gamma}+\bar{\Gamma}^i\Delta H_i+\bar{\Gamma}^{ij}\Delta H_i\Delta H_j+\bar{\Gamma}^{ijk}\Delta H_i\Delta H_j\Delta H_k+\ldots,
\end{equation}
where
\begin{eqnarray}
\bar{\Gamma}&=&\Gamma+\Gamma^i\ave{H_i}+\Gamma^{ij}\ave{H_i}\ave{H_j}+\Gamma^{ijk}\ave{H_i}\ave{H_j}\ave{H_k}+\ldots \label{gammabardef}\\
\bar{\Gamma}^i&=&\Gamma^i+\left(\Gamma^{ij}+\Gamma^{ji}\right)\ave{H_j}+\left(\Gamma^{ijk}+\Gamma^{jik}+\Gamma^{kji}\right)\ave{H_j}\ave{H_k}+\ldots\\
\bar{\Gamma}^{ij}&=&\Gamma^{ij}+\left(\Gamma^{kij}+\Gamma^{ikj}+\Gamma^{ijk}\right)\ave{H_k}+\ldots
\end{eqnarray}
and so forth. Note that $\bar{\Gamma}$, when viewed as a function of $\ell$, $\ave{H_0}$ and $\ave{H_1}$, encodes information about all the expansion coefficients $\Gamma^{ij\ldots}$ appearing in equation \eref{genparamh}. A natural strategy that presents itself is to set up an equation for $\bar{\Gamma}$ as a function of $\ell$, $\ave{H_0}$ and $\ave{H_1}$ in some domain ${\cal D}$ of the $\ave{H_0}$--$\ave{H_1}$ plane, determined by the properties of the operators $H_0$ and $H_1$.  In particular we require that this domain includes values of $\ave{H_0}$ ranging from the largest to the smallest eigenvalues of $H_0$.  For this purpose we require a family of states $\ket{\phi(\lambda)}$, parameterized by a continuous set of variables $\lambda\in \bar {\cal D}$, such that as $\lambda$ is varied over the domain $\bar {\cal D}$, $\ave{H_0}$ and $\ave{H_1}$ range continuously over the domain ${\cal D}$. A very general set of states that meets these requirements are coherent states \cite{Klauder}. In this representation we may consider $\ave{H_0}$ and $\ave{H_1}$ as continuous variables with each $\bar{\Gamma}$ a function of $\ave{H_0}$, $\ave{H_1}$ and the following relations hold generally: 
\begin{equation}
\label{2ndorder}
	\bar{\Gamma}^i=\frac{\partial\bar{\Gamma}}{\partial \ave{H_i}}\ \ {\rm and}\ \ \bar{\Gamma}^{ij}=\frac{1}{2}\frac{\partial^2 \bar{\Gamma}}{\partial\ave{H_i}\partial\ave{H_j}}.
\end{equation}
Coefficients with three or more indices cannot be written in this way, since, for example, $\bar{\Gamma}^{101}$ need not be equal to $\bar{\Gamma}^{011}$. The general relationship between the true coefficients and derivatives of $\bar{\Gamma}$ are
\begin{equation}
 \frac{\partial^{n+m} \bar{\Gamma}}{\partial\ave{H_0}^n\partial\ave{H_1}^m}=n!m!\sum_{p}\bar{\Gamma}^{p(n,m)}
\end{equation}
where the sum over $p$ is over all the distinct ways of ordering $n$ zeros and $m$ ones. For example
\begin{equation}
	\frac{1}{3!}\frac{\partial^{3} \bar{\Gamma}}{\partial\ave{H_0}^2\partial\ave{H_1}^1}=\frac{1}{3}\left(\bar{\Gamma}^{001}+\bar{\Gamma}^{010}+\bar{\Gamma}^{100}\right).
\end{equation}
Replacing $\bar{\Gamma}^{001}$ by the left-hand side of the equation above is equivalent to approximating $\bar{\Gamma}^{001}$ by the average of $\bar{\Gamma}^{001}$,$\bar{\Gamma}^{010}$ and $\bar{\Gamma}^{100}$. In general this amounts to approximating $\bar{\Gamma}^{n_1\ldots n_\ell}$ by the average of the coefficients corresponding to all the distinct reorderings of $n_1,\ldots,n_\ell$. 

To set up an equation for $\bar{\Gamma}(\ave{H_0},\ave{H_1},\ell)$ we insert the expansion \eref{expansion1} into the flow equation with $\eta(\ell)=[H_0,H(\ell)]$ and take the expectation value with respect to the state $\ket{\phi(\lambda)}$. The left-and right-hand sides become a systematic expansion in orders of the fluctuations $\ave{\Delta H_{n_1}\ldots \Delta H_{n_\ell}}$:

\begin{eqnarray}
\label{flow1}
\frac{\partial\bar{\Gamma}}{\partial\ell}+\frac{\partial\bar{\Gamma}^{ij}}{\partial\ell}\ave{\Delta H_i\Delta H_j}+\ldots&=&\bar{\Gamma}^{i}\bar{\Gamma}^{j}\ave{[[\Delta H_0,\Delta H_{i}],\Delta H_{j}]} \nonumber\\ 
&+&\bar{\Gamma}^{i}\bar{\Gamma}^{j_1 j_2}\ave{[[\Delta H_0,\Delta H_{i}],\Delta H_{j_1}\Delta H_{j_2}]}\nonumber\\
&+&\bar{\Gamma}^{i_1 i_2}\bar{\Gamma}^{j}\ave{[[\Delta H_0,\Delta H_{i_1}\Delta H_{i_2}],\Delta H_{j}]}\\ 
&+&\bar{\Gamma}^{i_1 i_2}\bar{\Gamma}^{j_1 j_2}\ave{[[\Delta H_0,\Delta H_{i_1}\Delta H_{i_2}],\Delta H_{j_1}\Delta H_{j_2}]}\nonumber\\ 
&+&\ldots\nonumber
\end{eqnarray}
Note that writing $H_0$ or $\Delta H_0$ in the first position of the double commutator on the right-hand side of equation \eref{flow1} is a matter of taste. The expectation values appearing above are naturally functions of $\lambda$ and may be written as functions of $\ave{H_0}$ and $\ave{H_{1}}$ by inverting the equations $\ave{H_i}=\inp{\phi(\lambda)}{H_i}{\phi(\lambda)}$ to obtain $\lambda(\ave{H_0},\ave{H_1})$. With an appropriately chosen state, such as a coherent state, the higher orders in the fluctuation can often be neglected, as the expansion is controlled by the inverse of the number of degrees of freedom (see Appendix \ref{appc} for an explicit example). A useful analogy is the minimal uncertainty states in quantum mechanics which minimizes the fluctuations in position and momentum. The choice of state above aims at the same goal for $H_0$ and $H_1$.  Clearly it is difficult to give a general algorithm for the construction of these states and this property has to be checked on a case by case basis. If this is found to be the case we note that replacing the ${\bar\Gamma^{ij\ldots}}$'s with more than two indices by a derivative will introduce corrections in \eref{flow1} of an order higher than the terms already listed, or, on the level of the operator expansion, corrections higher than second order in the fluctuations. Working to second order in the fluctuations we can therefore safely replace ${\bar\Gamma^{ij\ldots}}$ by the derivatives of a function $f\left(\ave{H_0},\ave{H_1},\ell\right)\equiv \bar{\Gamma}(\ave{H_0},\ave{H_1},\ell)$ and write for $H(\ell)$:
\begin{eqnarray}
	H(\ell)&=&f+f^{10}\Delta H_0+f^{01}\Delta H_1+\frac{1}{2}f^{11}\left\{\Delta H_0,\Delta H_1\right\}+\frac{1}{2}f^{20}\Delta H_0^2+\frac{1}{2}f^{02}\Delta H_1^2+\cdots \nonumber \\
	&=&\sum_{i,j=0}^\infty\frac{f^{ij}}{(i+j)!}\sum\left({\rm Distinct\ orderings\ of\ } i\ \Delta H_0{\rm 's}\ {\rm and}\ j\ \Delta H_1's \right)
	\label{fexpandflux}
\end{eqnarray}
where $\{\cdot,\cdot\}$ denotes the anti-commutator and
\begin{equation}
	 f^{ij}=\frac{\partial^{\,i+j} f}{\partial\ave{H_0}^i\partial\ave{H_1}^j}.
\end{equation}
This turns the flow equation \eref{flow1} into a nonlinear partial differential equation for $f\left(\ave{H_0},\ave{H_1},\ell\right)$, correct up to the order shown in \eref{flow1}.  The choice of coherent state, the corresponding calculation of the fluctuations appearing in equation \eref{flow1} and the identification of the parameter controlling the expansion are problem specific. There are, however, a number of general statements that can be made about the flow equation and the behaviour of $f\left(\ave{H_0},\ave{H_1},\ell\right)$.  The first property to be noted is that since $H(\infty)$ is diagonal in the eigenbasis of $H_0$ it should become a function of only $H_0$, provided that the spectrum of $H_0$ is non-degenerate. This is reflected in the behaviour of $f$ by the fact that $f\left(\ave{H_0},\ave{H_1},\ell=\infty\right)$ should be a function of $\ave{H_0}$ only.  This is indeed borne out to high accuracy in our later numerical investigations. This function in turn provides us with the functional dependence of $H(\infty)$ on $H_0$. Keeping in mind the unitary connection between $H(\infty)$ and $H$ this enables us to compute the eigenvalues of $H$ straightforwardly by inserting the supposedly known eigenvalues of $H_0$ into the function $f\left(\ave{H_0},\infty\right)$.    

A second point to note is that the considerations above apply to the flow of an arbitrary hermitian operator with time reversal symmetry.  It is easily verified that the transformed operator $O(\ell)= U^\dagger(\ell)O(0)U(\ell)$ satisfies the flow equation:
\begin{equation}
\label{flow2}
\frac{d O(\ell)}{d\ell}=[\eta(\ell),O(\ell)]\,
\end{equation} 
where the same choice of $\eta(\ell)$ as for the Hamiltonian has to be made, i.e., $\eta(\ell)=[H_0,H(\ell)]$.  The expansion \eref{fexpandflux} can be made for both operators $H(\ell)$ and $O(\ell)$.  Denoting the corresponding function for $O(\ell)$ by $g\left(\ave{H_0},\ave{H_1},\ell\right)$ the flow equation \eref{flow2} turns into a linear partial differential equation for $g\left(\ave{H_0},\ave{H_1},\ell\right)$ containing the function $f\left(\ave{H_0},\ave{H_1},\ell\right)$, which is determined by \eref{flow1}. The expectation value of $O=O(0)$ with respect to an eigenstate $\ket{E_n}$ of $H$ can be expressed as
\begin{equation}
	\inp{E_n}{O}{E_n}=\inp{E_n}{U(\ell)U^\dag(\ell) OU(\ell)U^\dag(\ell)}{E_n}=\inp{E_n,\ell}{O(\ell)}{E_n,\ell}
\end{equation}
where $\ket{E_n,\ell}=U^\dag(\ell)\ket{E_n}$. In the limit $\ell\rightarrow\infty$ the states  $\ket{E_n,\infty}$ are simply the eigenstates of $H_0$, which are supposedly known.  In this way the computation of the expectation value $\inp{E_n}{O}{E_n}$ can be translated into the calculation of expectation values of the operator $O(\infty)$, obtained by solving the flow equation \eref{flow2}, in the known eigenstates of $H_0$.   

\section{Moyal bracket approach}
\label{methodMoyal}
Next we present an approach based on the Moyal bracket formalism \cite{Moyal}. Let ${\cal H}$ denote the $D$-dimensional Hilbert space of the Hamiltonian under consideration. We define two unitary operators $h$ and $g$ that act irreducibly on ${\cal H}$ and satisfy the exchange relation
\begin{equation}
	hg=e^{-i\theta}gh.
\end{equation}
Since $g$ is unitary its eigenvalues are simply phases. Let $e^{i\sigma}$ be one such eigenvalue, and consider the action of $gh$ on the corresponding eigenstate $\ket{\sigma}$:
\begin{equation}
	gh\ket{\sigma}=e^{i\theta}hg\ket{\sigma}=e^{i(\theta+\sigma)}h\ket{\sigma}.
\end{equation}
We see that $h\ket{\sigma}$ is again an eigenstate of $g$ with eigenvalue $e^{i(\theta+\sigma)}$. Since $g$ and $h$ act irreducibly on ${\cal H}$ all the eigenstates of $g$ can be obtained by the repeated application of $h$ to $\ket{\sigma}$. Furthermore, we may scale $g$ so that it has an eigenvalue equal to one.

It follows that the eigenvalues and eigenstates of $g$ take the form
\begin{equation}
	g\ket{n}=e^{i\theta n}\ket{n}\ \ \ {\rm where}\ \ \ n=0,1,2,\ldots,D-1
	\label{gEV}
\end{equation}
while $h$ acts as a ladder operator between these states:
\begin{equation}
	gh\ket{n}=e^{i\theta}hg\ket{n}=e^{i\theta(n+1)}h\ket{n}\ \ \Longrightarrow\ \  h\ket{n}\propto\ket{n+1}.
		\label{gLadder}
\end{equation}
The allowed values of $\theta$ are found by taking the trace on both sides of $h^{-1}gh=e^{i\theta}g$, which leads to the requirement 
\begin{equation}
	{\rm tr}(g)=\sum_{n=0}^{D-1}e^{in\theta}=0. 
\end{equation}
This fixes $\theta$ at an integer multiple of $2\pi/D$. We choose $\theta=2\pi/D$ as this ensures that $g$ is non-degenerate, which is crucial for the construction that follows.

In similar fashion to the previous section, we wish to represent flowing operators in terms of $g$ and $h$. The  main result in this regard is that the set
\begin{equation}
	{\cal P}=\left\{D^{-\frac{1}{2}}g^nh^m\;:\;n,m=0,1,2,\ldots,D-1\right\}
	\label{setP}
\end{equation}
forms an orthogonal basis for the space of linear operators acting on ${\cal H}$. The orthogonality of ${\cal P}$ follows from applying the trace inner product in the $g$-basis to two members of ${\cal P}$:
\begin{equation}
	{\rm tr}\left[\left(g^nh^m\right)^\dag g^{n'}h^{m'}\right]=\delta_{m,m'}{\rm tr}\left[g^{(n'-n)}\right]=\delta_{m,m'}\sum_{q=0}^{D-1}e^{iq(n'-n)\theta}=\delta_{m,m'}\delta_{n,n'}D,
\end{equation}
where $\delta_{m,n}$ equals one if $m=n\ {\rm mod}\ D$ and is zero otherwise. This, together with the observation that the dimension of the linear operator space equals $|{\cal P}|=D^2$, proves the claim.

Consider two arbitrary operators $U$ and $V$ expressed in the ${\cal P}$ basis as
\begin{equation}
	U=\sum_{n,m}C_{n,m}g^nh^m\ \ {\rm and}\ \ V=\sum_{n',m'}C'_{n',m'}g^{n'}h^{m'}
\end{equation}
where $C_{n,m}$ and $C'_{n',m'}$ are scalar coefficients. We use the convention of always writing the $h$'s to the right of the $g$'s. The product of $U$ and $V$ then gives
\begin{equation}
	UV=\sum_{n,m,n',m'}C_{n,m}C'_{n',m'}g^{m+m'}h^{n+n'}e^{-inm'\theta}.
	\label{genmul}
\end{equation}
Note the similarity in form between this product and the product of functions of regular commuting variables. Only the phase factor, the result of imposing our ordering convention on the product, distinguishes the two. In fact, we may treat $g$ and $h$ as regular scalar variables provided that we modify the product rule to incorporate this phase. Convenient variables for this procedure are $\alpha$ and $\beta$, which are related to $g$ and $h$ (now treated as scalars) through $g=e^{i\alpha}$ and $h=e^{i\beta}$. Having replaced operators by functions of $\alpha$ and $\beta$ the modified product rule reads
\begin{equation}
	U(\alpha,\beta)\ast V(\alpha,\beta)\equiv U(\alpha,\beta)e^{i\theta\stackrel{\leftarrow}{\partial_\beta}\:\stackrel{\rightarrow}{\partial_\alpha}}V(\alpha,\beta),
\end{equation}
where the $\alpha$ and $\beta$ derivatives act to the right and left respectively. This is seen to be of the required form by using the fact that both $g$ and $h$ are eigenfunctions of $\partial_\alpha$ and $\partial_\beta$:
\begin{eqnarray}
U(\alpha,\beta)e^{i\theta\stackrel{\leftarrow}{\partial_\beta}\:\stackrel{\rightarrow}{\partial_\alpha}}V(\alpha,\beta)&=&\sum_{n,m,n',m'}C_{n,m}\:C'_{n',m'}\:e^{im\alpha}\: e^{in\beta}\:\underbrace{e^{i\theta\stackrel{\leftarrow}{\partial_\beta}\:\stackrel{\rightarrow}{\partial_\alpha}}}_{e^{-i\theta m'n}}\:e^{im'\alpha}\:e^{in'\beta},
\end{eqnarray}
which agrees with \eref{genmul}.

The $\ast$-operation is known as the Moyal product \cite{Moyal}, while the corresponding commutator $\comm{U}{V}_\ast=U\ast V-V\ast U$ is the Moyal bracket. When $H(\ell)$ and $\eta(\ell)$ are represented in this manner the flow equation becomes a partial differential equation in $\alpha$, $\beta$ and $\ell$:
\begin{equation}
	\afg{H(\alpha,\beta,\ell)}{\ell}=\comm{\eta(\alpha,\beta,\ell)}{H(\alpha,\beta,\ell)}_\ast.
\end{equation}
In its exact form this formulation is not of much practical value, since the operator exponent involved in the Moyal product is very difficult to implement numerically. A significant simplification is achieved by expanding the operator exponent to first order in $\theta$, which is known to scale like one over the dimension $D$ of the Hilbert space. We expect this to be a very good approximation provided that the derivatives do not bring about factors of the order of $D$. This translates into a smoothness condition: we require that the derivatives of the relevant functions remain bounded in the thermodynamic limit as $D$ goes to infinity. 

Using this approximation the Moyal product becomes, to leading order,
\begin{equation}
	U\ast V=U V+i\theta U_\beta V_\alpha
\end{equation}
while the Moyal bracket reads
\begin{equation}
	\comm{U}{V}_\ast=i\theta\left(U_\beta\:V_\alpha-V_\beta\:U_\alpha\right).
\end{equation}
Partial derivatives are indicated by the subscript shorthand. The form of the flow equation is now largely fixed, up to the specific choice of the generator. As an example, consider the generator $\eta(\alpha,\beta)=\comm{\alpha}{H(\alpha,\beta)}_\ast$ which we will use later on. In this case the flow equation becomes
\begin{equation}
		\afg{H(\alpha,\beta,\ell)}{\ell}=\theta^2\left(H_{\beta\beta}H_\alpha-H_\beta H_{\beta\alpha}\right).
\end{equation}

The remaining problem is that of constructing the initial conditions, i.e. $H(0)$, in terms of $g$ and $h$ (or equivalently $\alpha$ and $\beta$) in such a way that the smoothness conditions are satisfied. The reader may have noticed that we have not specified how the realization of $g$ and $h$ should be constructed on ${\cal H}$. Put differently, there is no obvious rule which associates a specific basis of ${\cal H}$ with the eigenstates of $g$. It seems reasonable that this freedom may allow us to construct smooth initial conditions through an appropriate choice of basis, whereas a malicious choice could produce very poorly behaved functions. We know of no way to proceed on such general terms, and we will instead tackle this problem on a case-by-case basis. In all of these we will use the algebraic properties of operators appearing in the Hamiltonian to reduce this problem to one of representation theory.
\\

Although two operators are clearly the minimum required to construct a complete operator basis, it is also possible to introduce multiple such pairs. This would be a natural choice when ${\cal H}$ is a tensor product of Hilbert spaces ${\cal H}_i$ $(i=1,\ldots,m)$, each of which is of a high dimension. We can introduce $m$ pairs of operators $\{g_i,h_i\}$ which satisfy $h_ig_i=e^{-i\theta_i}g_ih_i$ and $h_ig_j=g_jh_i$ for all $i\neq j$. In the same way as before this leads to $m$ pairs of scalar variables $\{\alpha_i,\beta_i\}$ for which the product rule, to first order in the $\theta_i$'s, is
\begin{equation}
		U(\boldsymbol{\alpha},\boldsymbol{\beta})\ast V(\boldsymbol{\alpha},\boldsymbol{\beta})=U V+i\sum_{j=1}^m\theta_j U_{\beta_j} V_{\alpha_j}.
\end{equation}
Note that $\{\alpha_i,\beta_i\}$ are analogous to conjugate position and momenta coordinates representing the independent degrees of freedom of the system. The Moyal bracket acts like the Poisson bracket for these coordinates: 
\begin{equation}
	\comm{\alpha_i}{\beta_j}_\ast=-i\theta\delta_{ij}\ \ {\rm and}\ \ \comm{\alpha_i}{\alpha_j}_\ast=\comm{\beta_i}{\beta_j}_\ast=0.
\end{equation}

This formulation strongly suggests an analogy with semi-classical approximation schemes. Our approach to solving the flow equation is indeed very closely related to the Wigner-Weyl-Moyal \cite{Moyal,Wigner} formalism, which describes the construction of a mapping between quantum operators and functions of classical phase space coordinates. This allows for the description of a quantum system in a form formally analogous to classical dynamics. When applied to the flow equations this formalism produces results similar to those obtained before. The central approximation again involves a non-perturbative expansion, but which is now controlled by $\hbar$, and so is semi-classical in nature. Let us formalize some of these notions in the context of a single particle in $3$ dimensions. The relevant Hilbert space is ${\cal H}=L^2({\mathbb R}^3)$ and the position and momentum operators satisfy the standard commutation relations
\begin{equation}
	\comm{\hat{x}_i}{\hat{p}_j}=i\hbar\delta_{ij}\ \ \ \ i,j=1,2,3.
\end{equation}
We first introduce the characteristic operator \cite{Gardiner}
\begin{equation}
	U(\bs{t},\bs{s})=e^{i\bs{t}\cdot\hbs{p}}e^{i\bs{s}\cdot\hbs{x}}
\end{equation}
where $\bs{t}=\left(t_1,t_2,t_3\right)\in{\mathbb R}^3$, $\hbs{x}=\left(\hat{x}_1,\hat{x}_2,\hat{x}_3\right)$ and similar for $\bs{s}\in{\mathbb R}^3$ and $\hbs{p}$. Varying the arguments of $U(\bs{t},\bs{s})$ over their domains produces a set of operators analogous to ${\cal P}$ (equation \eref{setP}), where the discrete powers $n$ and $m$ correspond to the continuous labels $\bs{t}$ and $\bs{s}$. We again find both completeness and orthogonality with respect to the trace norm:
\begin{eqnarray}
	{\rm tr}\left[U(\bs{t}',\bs{s}')^\dag U(\bs{t},\bs{s})\right]&=&\int{\rm d}\bs{x}\inp{\bs{x}}{e^{-i\bs{s}'\cdot\hbs{x}}e^{-i\bs{t}'\cdot\hbs{p}}e^{i\bs{t}\cdot\hbs{p}}e^{i\bs{s}\cdot\hbs{x}}}{\bs{x}}\nonumber \\
	&=&\frac{1}{(2\pi\hbar)^3}\int{\rm d}\bs{x}\:{\rm d}\bs{p}\:e^{i\bs{p}\cdot(\bs{t}-\bs{t}')}e^{i\bs{x}\cdot(\bs{s}-\bs{s}')}\nonumber  \\
	&=&\frac{1}{\hbar^3}\delta(\bs{t}-\bs{t}')\delta(\bs{s}-\bs{s}').
\end{eqnarray}
Using this, an operator $A(\hbs{x},\hbs{p})$ can be represented as 
\begin{equation}
	A(\hbs{x},\hbs{p})=\int{\rm d}\bs{t}\:{\rm d}\bs{s}\:\tilde{A}(\bs{t},\bs{s})U(\bs{t},\bs{s})
\end{equation}
where
\begin{equation}
	\tilde{A}(\bs{t},\bs{s})=\hbar^3\:{\rm tr}\left[U(\bs{t},\bs{s})^\dag A(\hbs{x},\hbs{p})\right]
\end{equation}
is a scalar function. Now consider the product of two operators represented in this manner:
\begin{equation}
 A(\hbs{x},\hbs{p}) B(\hbs{x},\hbs{p})=\int{\rm d}\bs{t}\:{\rm d}\bs{s}\:{\rm d}\bs{t}'\:{\rm d}\bs{s}'\tilde{A}(\bs{t},\bs{s}) \tilde{B}(\bs{t}',\bs{s}')U(\bs{t}+\bs{t'},\bs{s}+\bs{s'})e^{-i\hbar\bs{t}'\cdot\bs{s}}
\end{equation}
The non-commutativity of $\hbs{x}$ and $\hbs{p}$ gives rise to the scalar factor  $e^{-i\hbar\bs{t}'\cdot\bs{s}}$, which is the only element distinguishing this product from one of regular scalar functions. We conclude, as before, that the position and momentum operators may be treated as scalar variables provided that we modify the product rule to incorporate this phase. This leads to the Moyal product
\begin{equation}
A(\bs{x},\bs{p})\ast B(\bs{x},\bs{p})= A(\bs{x},\bs{p})e^{i\hbar\:\stackrel{\leftarrow}{\partial_{\bs{x}}}\cdot\stackrel{\rightarrow}{\partial_{\bs{p}}}}B(\bs{x},\bs{p})
\end{equation}
where $\bs{x},\bs{p}\in{\mathbb R}^3$ and $\stackrel{\leftarrow}{\partial_{\bs{x}}}\cdot\stackrel{\rightarrow}{\partial_{\bs{p}}}=\sum_i\stackrel{\leftarrow}{\partial_{x_i}}\:\stackrel{\rightarrow}{\partial_{p_i}}$. Note that an expansion of the exponential is now controlled by $\hbar$ instead of $\theta\propto{\rm dim}({\cal H})^{-1}$. To leading order the Moyal bracket is given by
\begin{equation}
	\comm{A(\bs{x},\bs{p})}{B(\bs{x},\bs{p})}_\ast=i\hbar\sum_{i=1}^{3}\left(A_{x_i}\:B_{p_i}-A_{p_i}\:B_{x_i}\right),
\end{equation}
where the subscripts denote partial derivatives.

We conclude that in a semi-classical approximation the Hamiltonian and generator may be replaced by scalar functions $H(\bs{x},\bs{p},\ell)$ and $\eta(\bs{x},\bs{p},\ell)$, and that the flow equation is given in terms of the Moyal bracket by
\begin{equation}
	\afg{H(\bs{x},\bs{p},\ell)}{\ell}=\comm{\eta(\bs{x},\bs{p},\ell)}{H(\bs{x},\bs{p},\ell)}_\ast.
\end{equation}
When solved to leading order in $\hbar$ this equation describes the renormalization of the Hamiltonian within a semi-classical approximation. Further quantum corrections can be included by simply expanding the Moyal bracket to higher orders in $\hbar$.
\chapter{The Lipkin Model}

\section{Introduction}
Since its introduction in 1965 as a toy model for two shell nuclear interactions the Lipkin-Meshov-Glick model \cite{Lipkin} has served as a testing ground for new techniques in many-body physics. Here we will use it to  illustrate both the working of the flow equations and the solution methods presented in the previous chapter. While the simple structure of the model will allow  many calculations to be performed exactly, its non-trivial phase structure will provide a true test for our non-perturbative approach. 

We begin with an overview of the model, its features and the quantities we are interested in calculating. After pointing out some specific aspects of the flow equations for the Lipkin model we proceed to treat the equations using the methods developed earlier. Finally we present the results obtained from the numerical solutions of the resulting PDE's and compare them with some known results. New aspects of the model, brought to the fore by this treatment, will also be discussed.
 
\section{The model}
\label{lipkinmodel}
The Lipkin model describes $N$ fermions distributed over two $\Lambda$-fold degenerate levels separated by an energy of $\xi$. For simplicity we shall take $N$ to be even. Fermi statistics require that $N \leq 2 \Lambda$, and accordingly the thermodynamic limit should be understood as $\Lambda \rightarrow \infty$ followed by $N \rightarrow \infty$. The interaction $V$ introduces scattering of pairs between levels. Labelling the two levels by $\sigma=\pm1$, the Hamiltonian reads
\begin{equation}
 H=\frac{\xi}{2} \sum_{\sigma,p}\sigma a^\dag_{p,\sigma} a_{p,\sigma}+\frac{V}{2}\sum_{p,p',\sigma}a^\dag_{p,\sigma}a^\dag_{p',\sigma}a_{p',-\sigma}a_{p,-\sigma}
\end{equation}
where the indices $p$ and $p^\prime$ run over the level degeneracy $1\ldots\Lambda$. A spin representation for $H$ can be found by introducing the $\texttt{su}(2)$ generators
\begin{equation}
J_z=\frac{1}{2}\sum_{\sigma,p}\sigma a^\dag_{p,\sigma} a_{p,\sigma} \ \ \ {\rm and} \ \ \ J_\pm=\sum_{p}\sigma a^\dag_{p,\pm1} a_{p,\mp1}.
\end{equation}
Together with the second order Casimir operator $J^2=J_z^2+J_z+J_-J_+$, these satisfy the regular $\texttt{su}(2)$ commutation relations:
\begin{equation}
	\left[J_z,J_\pm\right]=\pm J_{\pm},\ \ \ \left[J_+,J_-\right]=2J_{z}\ \ \ {\rm and}\ \ \  \left[J^2,J_\pm\right]=\left[J^2,J_z\right]=0.
\end{equation}
 
We divide $H$ by $\xi$ and define the dimensionless coupling constant $\lambda=NV/\xi$ to obtain
\begin{equation}
	H=J_z+\frac{\lambda}{2 N}\left(J_+^2+J_-^2\right),
\end{equation}
where all energies are now expressed in units of $\xi$. The factor of $N$ in the definition of $\lambda$ brings about the $1/N$ in front of the second term, which ensures that the Hamiltonian as a whole is extensive and scales like $N$.  Since $\left[H,J^2\right]=0$ the Hamiltonian acts within irreducible representations of $\texttt{su}(2)$ where states are labelled by the eigenvalues of $J^2$ and $J_z$, i.e., $J^2\ket{j,m}=j(j+1)\ket{j,m}$ and $J_z\ket{j,m}=m\ket{j,m}$ for $m=-j,\ldots,j$.  The Hamiltonian thus assumes a block diagonal structure of sizes $2j+1$. The low-lying states occur in the multiplet $j=N/2$, and we fix $j$ at this value throughout, using the shorthand $\ket{m}\equiv\ket{j,m}$ for the basis states. The Hamiltonian can be reduced further by noting that it leaves the subspaces of states with either odd or even spin projection invariant. States belonging to one of these subspaces are referred to as having either odd or even parity. We  denote the eigenstate of $H$ with energy $E_n$ by $\ket{E_n}$, where $E_0\leq E_1\leq E_2\ldots$. When $\lambda=0$ the ground state is simply $\ket{E_0}=a^\dag_{1,-1}\ldots a^\dag_{N,-1}\ket{0}$ which is written as $\ket{E_0}=\ket{-j}$ in the spin basis. Non-zero values of $\lambda$ cause particle-hole excitations across the gap, and at $\lambda=\pm 1$ the model exhibits a phase transition from an undeformed first phase to a deformed second phase. To distinguish the two phases we use the order parameter $\Omega\equiv1+\left<J_z\right>/j$ where $\left<J_z\right>$ is the expectation value of $J_z$ in the ground state. As we will show, $\Omega$ is non-zero only within the second phase.

The phases can be characterized further by the energy gap $\Delta=E_1-E_0$ which is positive in the first phase and vanishes like $\sqrt{1-\lambda^2}$ as $\lambda$ approaches $1$. In the second phase the ground states of the odd and even parity subspaces become degenerate, causing the parity symmetry to be broken and the corresponding energy gap to vanish. Further discussion of this model and its features can be found in \cite{Lipkin}.

\section{The flow equation approach}
\label{lipkinflow}
We will follow the formulations of Sections \ref{gChoice} and \ref{etaChoice} and choose the generator as
\begin{equation}
	\eta(\ell)=\comm{J_z}{H(\ell)}.
\end{equation}
Let us consider the consequences of this choice. Firstly, since $J_z$ is non-degenerate, we expect the final Hamiltonian $H(\infty)$ to be diagonal in the known spin basis. Furthermore the eigenvalues of $H$ appear on the diagonal of $H(\infty)$ in the same order as in $J_z$, i.e. increasing from top to bottom. Secondly, note that $H$ possesses a band diagonal structure with respect to $J_z$, which plays the role of $Q$ in Section \ref{etaChoice}. Based on our earlier discussion we expect this structure to be conserved during flow, and so $H(\ell)$ will only connect states of which the spin projection differs by two.

Let $\ket{E_n,\ell}=U^\dag(\ell)\ket{E_n}$ denote the transformed eigenstates of $H$. From the ordering property of eigenvalues in $H(\infty)$ we conclude that $\ket{E_n,\infty}=\ket{-j+n}$, and so a general expectation value may be calculated through
\begin{equation}
	\inp{E_n}{O}{E_n}=\inp{E_n,\infty}{O(\infty)}{E_n,\infty}=\inp{-j+n}{O(\infty)}{-j+n}.
\end{equation}

Before continuing we mention some of the previous applications of flow equations to the Lipkin model. The first of these was by Pirner and Friman \cite{PirnerFriman}, who dealt with newly generated terms by linearizing them around their ground state expectation values. This yielded good results in the first phase but lead to divergences in the second; a common ailment of these types of approximations. Subsequent work by Mielke \cite{Mielke} relied on an ansatz for the form of $H(\ell)$'s matrix elements, while Stein \cite{Stein2} followed a bosonization approach. Dusuel and Vidal \cite{Vidal1,Vidal2} used flow equations together with the Holstein-Primakoff boson representation to compute finite-size scaling exponents for a number of physical quantities. Their approach was also based on a $1/N$ expansion. A method which produced reliable results in both phases was that of \cite{FGS}. This method also relied on the linearization of newly generated operators but, in contrast with \cite{PirnerFriman}, did so around a dynamically changing (``running") expectation value taken with respect to the flowing ground state. This ``running" expectation value could be solved for in a self consistent manner, and then used in the flow equation for $H(\ell)$. This lead to non-perturbative results for the ground state, excitation gap and order parameter in both phases. Although it described the low energy physics well, this method failed to produce the correct spectrum for the higher energy states. 

\section{Expansion method}
In order to apply the method of Section \ref{fluxExpand} we require an irreducible set of operators $\{H_0,H_1\}$ in terms of which the flowing Hamiltonian will be constructed. For the purposes of the Lipkin model we will choose $H_0=J_z$ and $H_1=\frac{1}{4j}\left(J_+^2+J_-^2\right)$, i.e. the diagonal and off-diagonal parts of the Hamiltonian respectively. The reader may well remark that this does not constitute an irreducible set on the entire Hilbert space, since the subspaces of odd and even spin projection are left invariant, as pointed out in Section \ref{lipkinmodel}. Indeed, it is only within a subspace of definite parity (i.e. odd or even spin projection) that this set is irreducible. However, we note that the flow equation will never mix these odd and even subspaces, and so the flow will proceed independently within each subspace. This ensures that our representation of $H(\ell)$ will never require an operator which connects odd and even states, and it turns out that this choice of $\{H_0,H_1\}$ is indeed sufficient. The same conclusion extends to $U(\ell)$ by equation \eref{flowEqForU}, and then to any flowing operator $O(\ell)=U^\dag(\ell)OU(\ell)$, provided that $O$ can be represented in this way.

\begin{figure}[t]
\begin{center}
	\epsfig{file=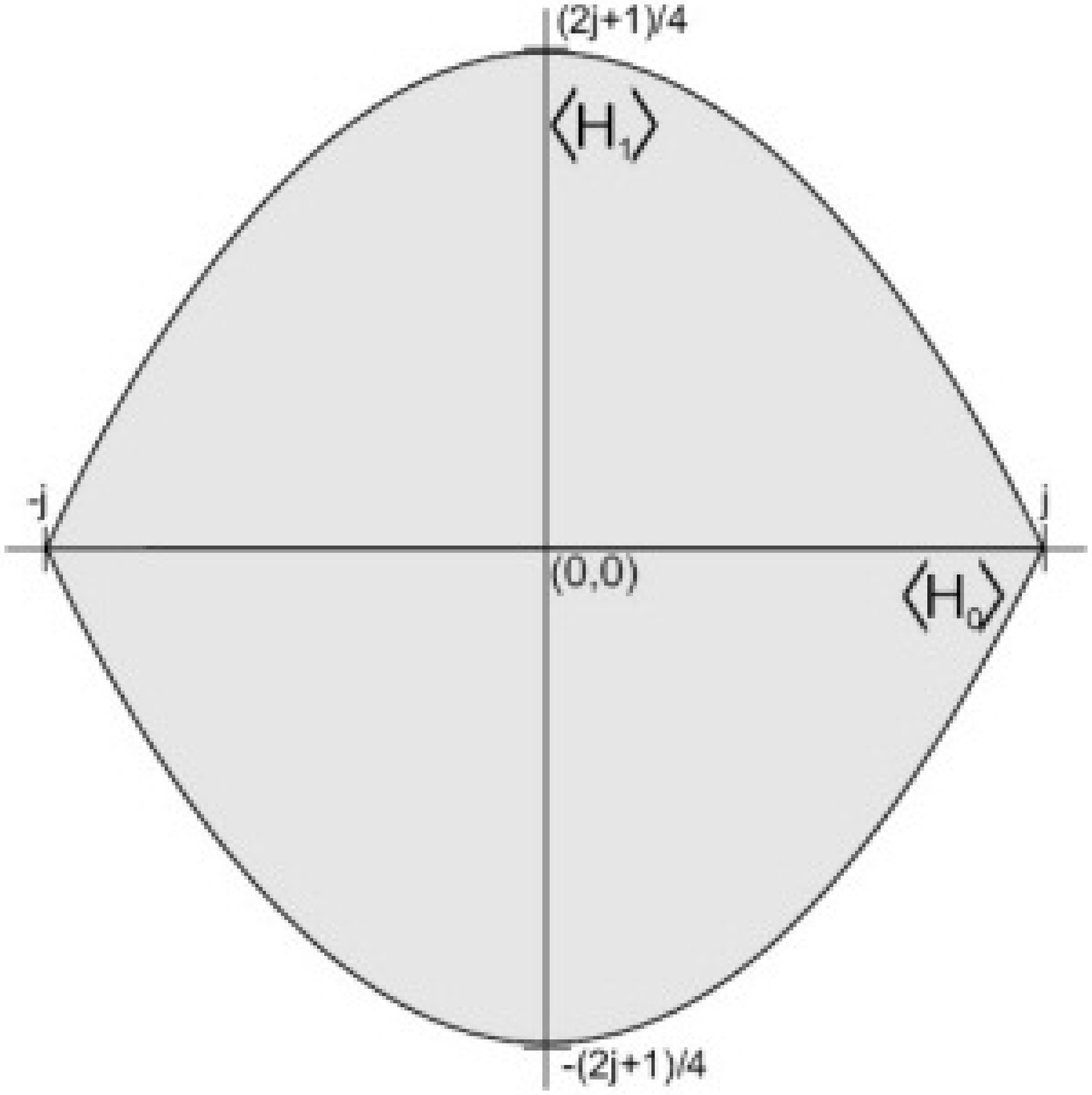,height=6.0cm,clip=,angle=0}
	\caption{The domain of $f(\ave{H_0},\ave{H_1},\ell)$.}
	\label{domainfig}
\end{center}
\end{figure}

We will calculate the averages of $H_0$ and $H_1$ with respect to the coherent state \cite{Klauder,Perelomov}
\begin{eqnarray}
\label{cohstate}
\ket{z}\equiv (1+r^2)^{-j}\exp(z J_+)\ket{-j},
\end{eqnarray}
where $z=r\exp(i\theta)\in\mathbb{C}$. See Appendix \ref{appb} for details on the properties of these states and  the method by which the averages are calculated. We find that:
\begin{equation}
	\ave{H_0}=j\left(\frac{r^2-1}{r^2+1}\right)\ \ {\rm and}\ \ \ave{H_1}=\frac{(2j-1)\,r^2\cos(2\theta)}{(1+r^2)^2}.
\end{equation}
This constitutes a mapping of the complex plane onto the domain shown in Figure \ref{domainfig}.
With these definitions in place we conclude that the flowing Lipkin Hamiltonian may be written as
\begin{eqnarray}
	H(\ell)=f+f^{10}\Delta H_0+f^{01}\Delta H_1+\frac{1}{2}f^{11}\left\{\Delta H_0,\Delta H_1\right\}+\frac{1}{2}f^{20}\Delta H_0^2+\frac{1}{2}f^{02}\Delta H_1^2+\cdots
	\label{lipkinhexp}
\end{eqnarray}
with $H_0$ and $H_1$ respectively the diagonal and off-diagonal parts of the original Hamiltonian. By definition $\Delta H_i=H_i-\inp{z}{H_i}{z}$ and so $f$ is defined on the domain pictured in Figure \ref{domainfig}. The initial condition becomes  
\begin{equation}
f(\ave{H_0},\ave{H_1},\ell=0)=\ave{H_0}+\lambda \ave{H_1}.
\end{equation}
Due to the coherent nature of the state \eref{cohstate} one might expect that terms corresponding to high order fluctuations in \eref{lipkinhexp} will contribute less significantly to $\inp{z}{H(\ell)}{z}$ than the scalar term $f$. This statement can be made precise as follows: since the flowing Hamiltonian is extensive $f$ should be proportional to $j$, as is the case with $\ave{H_0}$ and $\ave{H_1}$. To keep track of the orders of $j$ we introduce the scaleless variables $x\equiv\ave{H_0}/j$, $y\equiv\ave{H_1}/j$ and $\tilde{f}\equiv f/j$. When taking the inner-product with respect to $\ket{z}$ on both sides of \eref{lipkinhexp} the linear terms fall away and we obtain
\begin{eqnarray}
	\ave{H(\ell)}=f+\frac{1}{2}f^{11}\ave{\left\{\Delta H_0,\Delta H_1\right\}}+\frac{1}{2}f^{20}\ave{\Delta H_0^2}+\frac{1}{2}f^{02}\ave{\Delta H_1^2}+\cdots.
\end{eqnarray}
Each term is, up to a constant factor, of the form 
\begin{equation}
	f^{ik}\ave{{\rm Prod}(i+k)}=j^{1-(i+k)}\frac{\partial^{\,i+k} \tilde{f}}{\partial x^i\partial y^k}\ave{{\rm Prod}(i+k)}
\end{equation}
where ${\rm Prod}(i+k)$ denotes some arbitrary product of $i+k$ fluctuations. Using the results of Appendix \ref{appc} we see that such a term is at most of order ${\cal O}(j^{1-(i+k)}j^{\left\lfloor (i+k)/2 \right\rfloor})$. The leading order term corresponds to $i+k=0$, i.e. the scalar term $f$. We conclude that $f$ is the leading order contribution to $\inp{z}{H(\ell)}{z}$, expressed not as a function of $z$ and $z^*$, but rather of the averages $\ave{H_0}$ and $\ave{H_1}$. 

\subsection{The flow equation in the $j\rightarrow\infty$ limit.}
\label{flowinf}
We consider the flow of the Hamiltonian and an arbitrary observable $O$. Since $H(\ell)$ determines $U(\ell)$ one would expect a one-way coupling between the equations. It is assumed that $O$ is a hermitian operator constructed in terms of $H_0$ and $H_1$. Furthermore $\inp{z}{O}{z}$ must be a rational function of $j$, which ensures that when taking derivatives of $g(x,y,\ell)$ no additional factors of $j$ are generated. First we summarize the equations concerned
\begin{eqnarray}
		H(\ell)&=&f+f^{10}\Delta H_0+f^{01}\Delta H_1+\frac{1}{2}f^{11}\left\{\Delta H_0,\Delta H_1\right\}+\frac{1}{2}f^{20}\Delta H_0^2+\frac{1}{2}f^{02}\Delta H_1^2+\cdots,\label{expand2}\\
		O(\ell)&=&g+g^{10}\Delta H_0+g^{01}\Delta H_1+\frac{1}{2}g^{11}\left\{\Delta H_0,\Delta H_1\right\}+\frac{1}{2}g^{20}\Delta H_0^2+\frac{1}{2}g^{02}\Delta H_1^2+\cdots,\\
		\frac{d H(\ell)}{d\ell}&=&\comm{\comm{H_0}{H(\ell)}}{H(\ell)},\ \ \ H(0)=H_0+\lambda H_1, \label{floweqh2} \\
		\frac{d O(\ell)}{d\ell}&=&\comm{\comm{H_0}{H(\ell)}}{O(\ell)},\ \ \ O(0)=O,
\end{eqnarray}
where $f(\ave{H_0},\ave{H_1},0)=\ave{H_0}+\lambda\ave{H_1}$ and $g(\ave{H_0},\ave{H_1},0)$ is just $\inp{z}{O}{z}$ up to leading order in $j$. Next we substitute the expansion of $H(\ell)$ into the flow equation \eref{floweqh2} and take the expectation value on both sides with respect to the coherent state. Arguing as before we identify the leading order term on the left as being ${\rm \partial}f/{\rm \partial}{\ell}$. A general term on the right is of the form
\begin{equation}
	f^{in}f^{kl}\ave{\comm{\comm{\Delta H_0}{{\rm Prod}(i+n)}}{{\rm Prod}(k+l)}}.
\end{equation}
By transforming to scaleless variables and using the result of Appendix \ref{appc} it is seen to be at most of order ${\cal O}(j^{2-t+\left\lfloor(1+t)/2\right\rfloor})$, where $t=i+n+k+l$. Since the $t=0$ and $t=1$ terms are zero (they involve commutators of scalars) the leading order contributions come from the $t=2,3$ terms. These are exactly the terms found by considering the expansion of $H(\ell)$ up to second order in the fluctuations. The expectation values of the double commutators may be calculated and expressed as functions of $x$ and $y$ using the method outlined in Appendix \ref{appb}.
It is found that, due to cancellations, only four of the potential fifteen terms make leading order contributions. The corresponding expectation values are, to leading order
\begin{eqnarray}
	\ave{\comm{\comm{\Delta H_0}{\Delta H_1}}{\Delta H_0}}&=&\gamma_1(x,y)j\\
	\ave{\comm{\comm{\Delta H_0}{\Delta H_1}}{\Delta H_1}}&=&\gamma_2(x,y)j\\
	\ave{\comm{\comm{\Delta H_0}{\acomm{\Delta H_0}{\Delta H_1}}}{\Delta H_1}}&=&\gamma_3(x,y)j^2\\
	\ave{\comm{\comm{\Delta H_0}{\Delta H_1^2}}{\Delta H_0}}&=&-\gamma_3(x,y)j^2
\end{eqnarray}
where $\gamma_1=-4y$, $\gamma_2=2x(1-x^2)$ and $\gamma_3=-2+4x^2-2x^4+8y^2$. When keeping only the leading order terms on both sides the flow equation becomes
\begin{equation}
\pafg{\tilde{f}}{\ell}=\gamma_1\tilde{f}^{01}\tilde{f}^{10}+\gamma_2\tilde{f}^{01}\tilde{f}^{01}+\frac{1}{2}\gamma_3\tilde{f}^{11}\tilde{f}^{01}-\frac{1}{2}\gamma_3\tilde{f}^{02}\tilde{f}^{10},
\end{equation}
where superscripts denote derivatives to the rescaled averages $x$ and $y$. For further discussion it is convenient to use the variables $(x,q=\cos(2\theta))\in\left[-1,1\right]\times\left[-1,1\right]$, where $q$ and $y$ are related by $q=2y/(1-x^2)$. The square domain of these variables also simplifies the numerical solution significantly. We note that the arguments above can be applied, completely unchanged, to the flow equation of $O(\ell)$ as well. We obtain, now for both $H(\ell)$ and $O(\ell)$, the coupled set
\begin{eqnarray}
\pafg{\tilde{f}}{\ell}=4(q^2-1)\left(\pafg{\tilde{f}}{q}\frac{\partial^2\tilde{f}}{\partial q\partial x}-\pafg{\tilde{f}}{x}\frac{\partial^2\tilde{f}}{\partial q^2}\right)-4q\pafg{\tilde{f}}{q}\pafg{\tilde{f}}{x} \label{floweqftilde}\\
\pafg{\tilde{g}}{\ell}=4(q^2-1)\left(\pafg{\tilde{g}}{q}\frac{\partial^2\tilde{f}}{\partial q\partial x}-\pafg{\tilde{g}}{x}\frac{\partial^2\tilde{f}}{\partial q^2}\right)-4q\pafg{\tilde{f}}{q}\pafg{\tilde{g}}{x}
\end{eqnarray}
Note that, in contrast to the equation for $\tilde f$, the equation for $g$ is a linear equation that can be solved once $\tilde f$ has been obtained from \eref{floweqftilde}.
 
In Section \ref{lipkinflow} it was mentioned that $H(\ell)$ retains its band diagonal structure during flow, which means that $H_1$ only appears linearly in the representation of $H(\ell)$. This implies that $f(\ave{H_0},\ave{H_1},\ell)$ should be linear in $\ave{H_1}$, or, in the new variables, linear in $q$. When the form $\tilde{f}(x,q,\ell)=n_0(x,\ell)+n_1(x,\ell)\:q$ is substituted into \eref{floweqftilde} this is indeed seen to be the case, and we obtain a remarkably simple set of coupled PDE's for $n_0$ and $n_1$:
\begin{eqnarray}
	\frac{\partial n_0}{\partial \ell}&=&-4 n_1\frac{\partial n_1}{\partial x},\nonumber \\
	\frac{\partial n_1}{\partial \ell}&=&-4 n_1\frac{\partial n_0}{\partial x}.
	\label{simplesetcoupled}
\end{eqnarray}
The initial conditions are 
\begin{equation}
	n_0(x,0)=x\ \ \ {\rm and}\ \ \ n_1(x,0)=\frac{\lambda y}{q}=\frac{\lambda (1-x^2)}{2}.
\end{equation}

\subsection{Finding the spectrum}
\label{spectrum}
From $f=j\tilde{f}$ it is possible to obtain the entire spectrum of $H$. For this discussion it is convenient to use the original $(\ave{H_0},\ave{H_1})$ variables. Recall that in the $\ell\rightarrow\infty$ limit $H(\ell)$ flows toward a diagonal form and that the eigenvalues appear on the diagonal in the same order as in $H_0$, i.e. increasing from top to bottom. This implies that the $n^{\rm th}$ eigenvalue of $H$ is given by
\begin{equation}
	E_n=\inp{-j+n}{H(\ell=\infty)}{-j+n},
\end{equation}
where $E_0$ corresponds to the ground state energy. As $H(\ell)$ flows towards a diagonal form the terms of expansion \eref{expand2} containing $H_1$ will disappear and eventually $H(\infty)$ and $f$ will become functions only of $H_0$ and $\ave{H_0}$ respectively. The eigenvalues of $H$ are given by
\begin{equation}
	E_n=f(\ave{H_0}=-j+n,0,\ell=\infty),
\end{equation}
where $\ave{H_1}$ has been set to zero. This can be understood in two ways. Looking at equations \eref{gammabardef} and \eref{genparamh} we see that the functional dependence of $H(\ell)$ on $H_0$ is the same as that of $\bar{\Gamma}$ (now called $f$) on $\ave{H_0}$. Since taking the expectation value of $H(\infty)$ with respect to $\ket{-j+n}$ is equivalent to substituting $-j+n$ for  $H_0$, the result follows. Alternatively, consider equation \eref{fexpandflux} and note that setting $\ave{H_0}=-j+n$ makes the ($n+1)^{\rm st}$ diagonal element of $\Delta H_0$ zero. Since $H_1$ does not appear we see that when taking the inner product with $\ket{-j+n}$ only the  scalar term, $f$, will survive. 
\subsection{Calculating expectation values}
\label{expval}
Next we return to the arbitrary operator $O$ introduced earlier. The aim is to calculate $\inp{E_n}{O}{E_n}$ where $\ket{E_n}$ is the (unknown) eigenstate of $H=H_0+\lambda H_1$ corresponding to the energy $E_n$. We can formulate this calculation in terms of a flow equation by noting that
\begin{equation}
	\inp{E_n}{O}{E_n}=\inpns{E_n}{U(\ell)U^\dag(\ell) OU(\ell)U^\dag(\ell)}{E_n}=\inp{E_n,\ell}{O(\ell)}{E_n,\ell}
\end{equation}
where $\ket{E_n,\ell}=U^\dag(\ell)\ket{E_n}$, $O(\ell)=U^\dag(\ell)OU(\ell)$ and $U(\ell)$ is the unitary operator associated with the flow of $H(\ell)$. This equation holds for all $\ell$, and particularly in the $\ell\rightarrow\infty$ limit. Since $\ket{E_n,\infty}=\ket{-j+n}$ it follows that 
\begin{equation}
	\inp{E_n}{O}{E_n}=\inp{-j+n}{O(\infty)}{-j+n},
\end{equation}
and we conclude that the expectation value of $O$ in the $\ket{E_n}$ eigenstate of $H$ is simply the $(n+1)^{\rm th}$ diagonal element of $O(\infty)$ in the $H_0$ basis. Furthermore $\inp{z}{O(\ell)}{z}$ has the character of a generating function in the sense that
\begin{equation}
	\left.\frac{\partial^{n+m}\inp{z}{O(\ell)}{z}}{{\partial z^*}^n\partial z^m}\right|_{z=z^*=0}\propto\inp{-j+n}{O(\ell)}{-j+m},
\end{equation}
thus knowing $g$, which is just $\inp{z}{O(\ell)}{z}$ up to leading order in $j$, is sufficient to obtain all the matrix elements of $O(\ell)$ with high accuracy. However, unless an analytic solution for $g$ is known, we are limited to numerical calculations for the low lying states. In particular, the ground state expectation value is found by setting $z=0$. 

\section{Moyal bracket method}
\label{lipkinMoyal}
In Section \ref{methodMoyal} the Moyal Bracket method was used to obtain a simple, but general, realization of the flow equation as a partial differential equation in the variables $\alpha$, $\beta$ and $\ell$. For the present case this equation becomes
\begin{equation}
\afg{H(\alpha,\beta,\ell)}{\ell}=\comm{\comm{J_z(\alpha,\beta)}{H(\alpha,\beta,\ell)}_\ast}{H(\alpha,\beta,\ell)}_\ast,
\end{equation}
where $\comm{U}{V}_\ast=i\theta\left(U_\beta\:V_\alpha-V_\beta\:U_\alpha\right)$ is the Moyal bracket to leading order in $\theta$. Here $\theta=2\pi/(N+1)=2\pi/(2j+1)$ and so an expansion in $\theta$ is indeed controlled by the size of the system. As was mentioned earlier the model specific information enters through the initial condition, which we need to construct in terms of $\alpha$ and $\beta$. We will not perform this construction for the Hamiltonian directly, but rather follow the more general route of constructing an irreducible representation of $\texttt{su}(2)$, in terms of which both the Hamiltonian and observables can be readily obtained.

We wish to find three functions $J_z(\alpha,\beta)$, $J_+(\alpha,\beta)$ and $J_-(\alpha,\beta)$ which satisfy the $\texttt{su}(2)$ commutation relations with respect to the Moyal bracket $\comm{\cdot}{\cdot}_\ast$. Note that the Moyal formalism has allowed an essentially algebraic problem, that of constructing a specific representation, to be reduced to that of solving a set of differential equations. We also remind ourselves that, since we use the Moyal bracket in a first order approximation, our representation can only be expected to be correct up to this same order. 

We begin the construction by making the following ansatz
\begin{equation}
	J_+=e^{i\beta}f(\alpha),\ \ J_-=e^{-i\beta}f(\alpha)\ \ {\rm and}\ \ J_z=p(\alpha)
\end{equation}
which clearly requires the representation to be unitary. This ansatz is based on the interpretation of $h=e^{i\beta}$ as a ladder operator which connects states labelled by the eigenvalues of $g=e^{i\alpha}$. Substituting these forms into the required commutation relations $\comm{J_z}{J_\pm}_\ast=\pm J_\pm$ and $\comm{J_+}{J_-}_\ast=2J_z$ produces the set
\begin{equation}
	\theta\afg{}{\alpha}p(\alpha)=1\ \ {\rm and}\ \ -\theta\afg{}{\alpha}f^2(\alpha)=2p(\alpha)
\end{equation}
which is easily solved to obtain
\begin{equation}
	p(\alpha)=\frac{\alpha}{\theta}+a_1\ \ {\rm and}\ \ f^2(\alpha)=-\frac{\alpha^2}{\theta^2}-2a_1\frac{\alpha}{\theta}+a_2.
\end{equation}
Here $a_1$ and $a_2$ are integration constants that we fix by requiring that the second order Casimir operator assumes a constant value corresponding to the $j$-irrep:
\begin{equation}
	J^2=j(j+1)=\frac{1}{2}(J_+\ast J_-+J_-\ast J_+)+J_z\ast J_z=a_2+a_1^2.
\end{equation}
We can satisfy this constraint to leading order in $j$ by setting $a_2=0$ and $a_1=-j$. Finally we arrive at
\begin{equation}
J_+=e^{i\beta}\sqrt{\frac{2j\alpha}{\theta}-\frac{\alpha^2}{\theta^2}},\ \ J_-=e^{-i\beta}\sqrt{\frac{2j\alpha}{\theta}-\frac{\alpha^2}{\theta^2}}\ \ {\rm and}\ \ J_z=\frac{\alpha}{\theta}-j.
\end{equation}
Before continuing we define some more suitable variables. Consider the operator $\bar{g}=-i\log(g)/\theta$ which, according to equation \eref{gEV}, has eigenvalues $0,1,\ldots,2j$ and is represented by $\alpha/\theta$. This suggests that the natural domain of $\alpha/\theta$ is $[0,2j]$. We will use the scaleless variable $x=\alpha/(j \theta)-1\in[-1,1]$ in what follows. The representation now becomes
\begin{equation}
J_+=je^{i\beta}\sqrt{1-x^2},\ \ J_-=je^{-i\beta}\sqrt{1-x^2}\ \ {\rm and}\ \ J_z=jx.
\label{repSU2}
\end{equation}

We remark that this construction is by no means unique, although it is expected that other examples are related to this one by a similarity transformation. For example, complex non-unitary representation such as 
\begin{equation}
J_+=ie^{i\beta}\left(j(x-1)-1\right),\ \ J_-=ie^{-i\beta}\left(j(x+1)+1\right)\ \ {\rm and}\ \ J_z=jx.	
\label{repSU2NU}
\end{equation}
may be constructed. In fact, this is an exact representation to all orders in $\theta$, since $x$ (and thus $\alpha$) only occurs linearly and the Moyal bracket truncates after the first derivative. This illustrates the important role of the association between the eigenstates of $g$ and the particular basis of the Hilbert space. For our purposes the unitary representation \eref{repSU2} will be sufficient, although it is straightforward to check that the same results can be obtained using \eref{repSU2NU}.

By substituting the representation \eref{repSU2} into the Lipkin model Hamiltonian, and being careful to use the Moyal product in calculating the squares of $J_\pm$, we obtain the initial condition, to leading order, as:
\begin{equation}
	H(x,\beta,0)=jx+\frac{j\lambda}{2}(1-x^2)\cos(2\beta).
\end{equation}
The flow equation reads
\begin{equation}
	\pafg{H}{\ell}=\frac{1}{j}\left(H_{\beta\beta}H_x-H_\beta H_{\beta x}\right).
\end{equation}
Again it is expected that the band diagonality of $H(\ell)$ will be manifested as a constraint on the form of the solutions of the flow equation. We note that scattering between states of which the spin projection differ by two is associated with the $\cos(2\beta)$ term also appearing in the initial condition. Motivated by this we try the form $H(x,\beta,\ell)=j(n_0(x,\ell)+n_1(x,\ell)\cos(2\beta))$. Note that a factor of $j$, responsible for the extensivity of the Hamiltonian, has been factored out. Upon substituting this form into the flow equation we obtain 
\begin{equation}
	\pafg{n_0}{\ell}=-4n_1\pafg{n_1}{x}\ \ \ {\rm and}\ \ \ \pafg{n_1}{\ell}=-4n_1\pafg{n_0}{x},
	\label{moyaln0n1}
\end{equation}
which agrees with the equations of \eref{simplesetcoupled} in the previous section. For this form the flow of an observable is given by
\begin{equation}
	\pafg{O}{\ell}=2\sin(2\beta)\pafg{n_1}{x}\pafg{O}{\beta}-4\cos(2\beta)n_1\pafg{O}{x}.
	\label{moyalaux}
\end{equation}

We have seen that the governing equations obtained using the Moyal bracket method agrees with those of the previous section, although the interpretation of the constituents do differ. Next we turn to the matter of extracting the spectrum and expectation values from the solutions of these equations. As this procedure is largely similar to that of the previous section we will remain brief. First, note that through the representation \eref{repSU2NU} we may consider $n_0(x,\infty)$, and thus $H(\infty)$ itself, to be a function only of $J_z$. We arrive at the familiar result that \mbox{$E_n=jn_0(x=-1+n/j)$} for $n=0,1,\ldots,2j$. Expectation values are given by the diagonal elements of the flowed hermitian observable $O(x,\beta,\infty)$. The forms of $J_\pm(\alpha,\beta)$ suggest that in general $O(\ell)$ may be written as 
\begin{equation}
	O(x,\beta,\ell)=\sum_{n=0}^\infty f_n\left(x,\ell\right)\cos(2n\beta).
	\label{fourier}
\end{equation}
where $\cos(2n\beta)$ corresponds to an off-diagonal term proportional to $J^{2n}_++J^{2n}_-$. (We only expect even powers since there is no mixing between the odd and even subspaces.) Importantly the $f_0(x,\infty)$ term contains the desired information about the diagonal entries. We can isolate $f_0$ by integrating over $\beta$, which projects out the off-diagonal terms. Thus, in summary,
\begin{eqnarray}
\inp{E_n}{O}{E_n}&=&\inp{-j+n}{O(\infty)}{-j+n}\nonumber \\
&=&f_0(x=-1+n/j,\ell=\infty)\nonumber \\
&=&\frac{1}{2\pi}\int_0^{2\pi}{\rm d}\beta\;O(x=-1+n/j,\beta,\ell=\infty).
\label{usefourier}
\end{eqnarray}

\subsection{Probing the structure of eigenstates}
\label{eigenstates}
The ability to calculate expectation values for a large class of operators enables us to probe the structure of the eigenstates in a variety of ways. For this purpose we consider diagonal operators of the form $O(f)=\sum_{m=-j}^jf(m/j)\ket{m}\bra{m}$ where $f$ is a smooth, $j$-independent function defined on $[-1,1]$. Since $O(f)=f(J_z/j)$, the initial condition for the flow equation is simply $O(f,\alpha,\beta,\ell=0)=f(\alpha)$. Suppose $\ket{E_n}=\sum_{m=-j}^jc_m\ket{m}$ is the eigenstate under consideration, in which case
\begin{equation}
\inp{E_n}{O(f)}{E_n}=\sum_{m=-j}^jf\left(m/j\right)\left|c_m\right|^2=\sum_{m=-j}^jf\left(m/j\right)\left|\braket{m}{E_n}\right|^2.
\end{equation}
This leads to the interpretation of the expectation value as the average of the expansion coefficients squared $\left|\braket{m}{E_n}\right|^2$, weighted by the function $f(x)$. Ideally we would like to choose $O(f)$ as  the projection operator onto some basis state $\ket{m}$, as this would provide the most direct way of calculating the contributions of individual states. However, the projection operator does not fall within the class of operators corresponding to smooth initial conditions since there is no continuous function $f(x)$ for which $O(f)=\ket{m}\bra{m}$ in the large $j$ limit. Instead we choose $f(x,\bar{x})=\exp(-\gamma(\bar{x}-x)^2)$ where $\bar{x}\in[-1,1]$ and $\gamma>>1$. This weight function focuses on the contribution of those basis states $\ket{m}$ for which $m/j$ lies in a narrow region centered around $\bar{x}$. Considered as a function of $\bar{x}$, it is expected that $\ave{O(f(x,\bar{x}))}$ would approximate $\left|\alpha_{j\bar{x}}\right|^2$ up to a constant factor, provided that the latter varies slowly on the scale of $1/\sqrt{\gamma}$ in the region of $\bar{x}$. Clearly $\gamma$ controls the accuracy of this method, and also determines the scale on which the structure of the eigenstates can be resolved. 

\section{Numerical results}
In this section we analyze the results obtained by solving the equations derived in the previous two sections. We will mainly use the terminology of the Moyal bracket approach in what follows. See \cite{Kriel} for a treatment in terms of the expansion method. For compactness the $\ell$ argument of functions will occasionally be suppressed, in which case $\ell=\infty$ should be assumed. 
\subsection{Spectrum and expectation values}
\label{numres}
\begin{figure}[t]
\begin{narrow}{-0.35in}{0in}
\begin{tabular}{cc}
	\epsfig{file=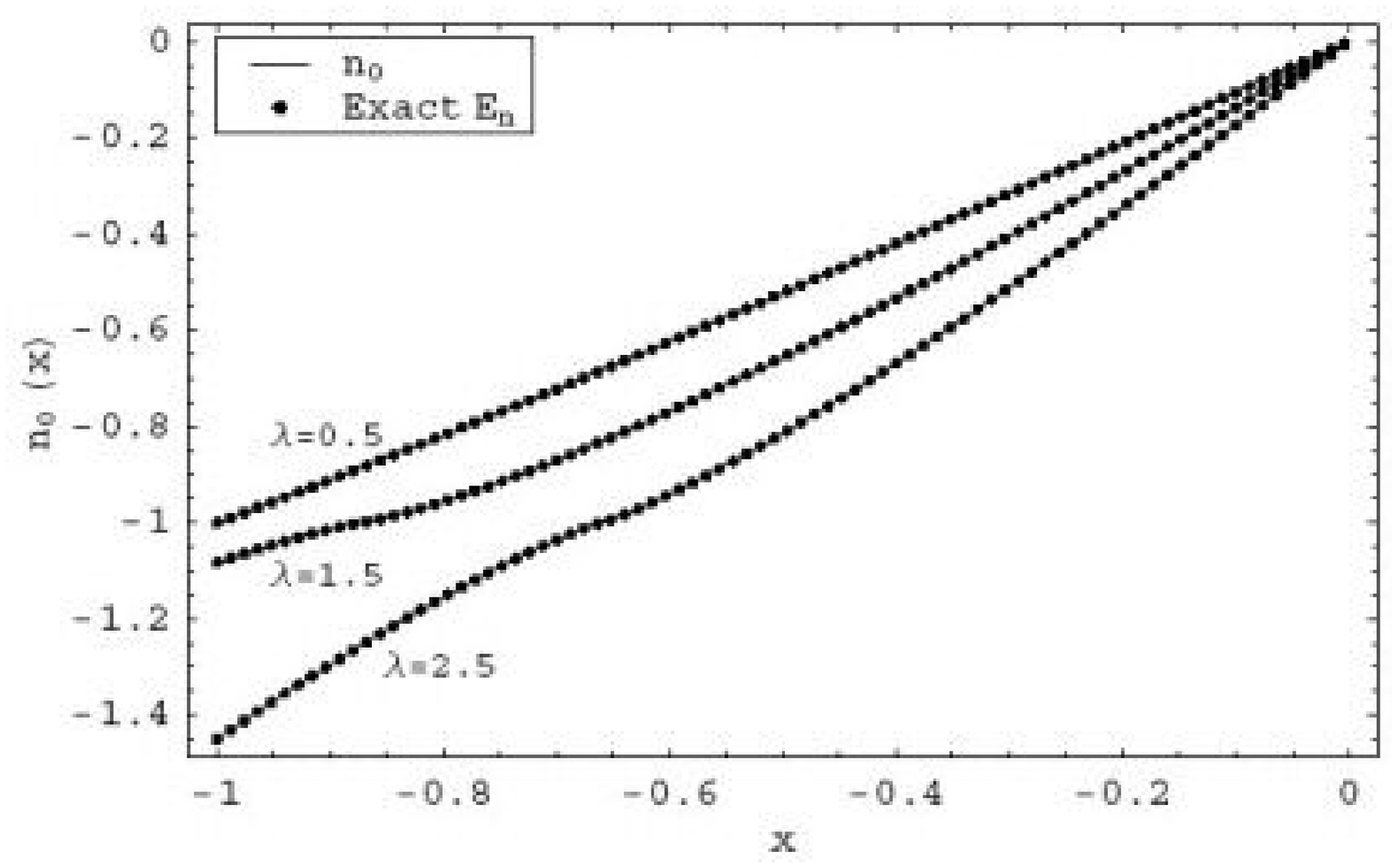,height=5cm,clip=,angle=0} &
	\epsfig{file=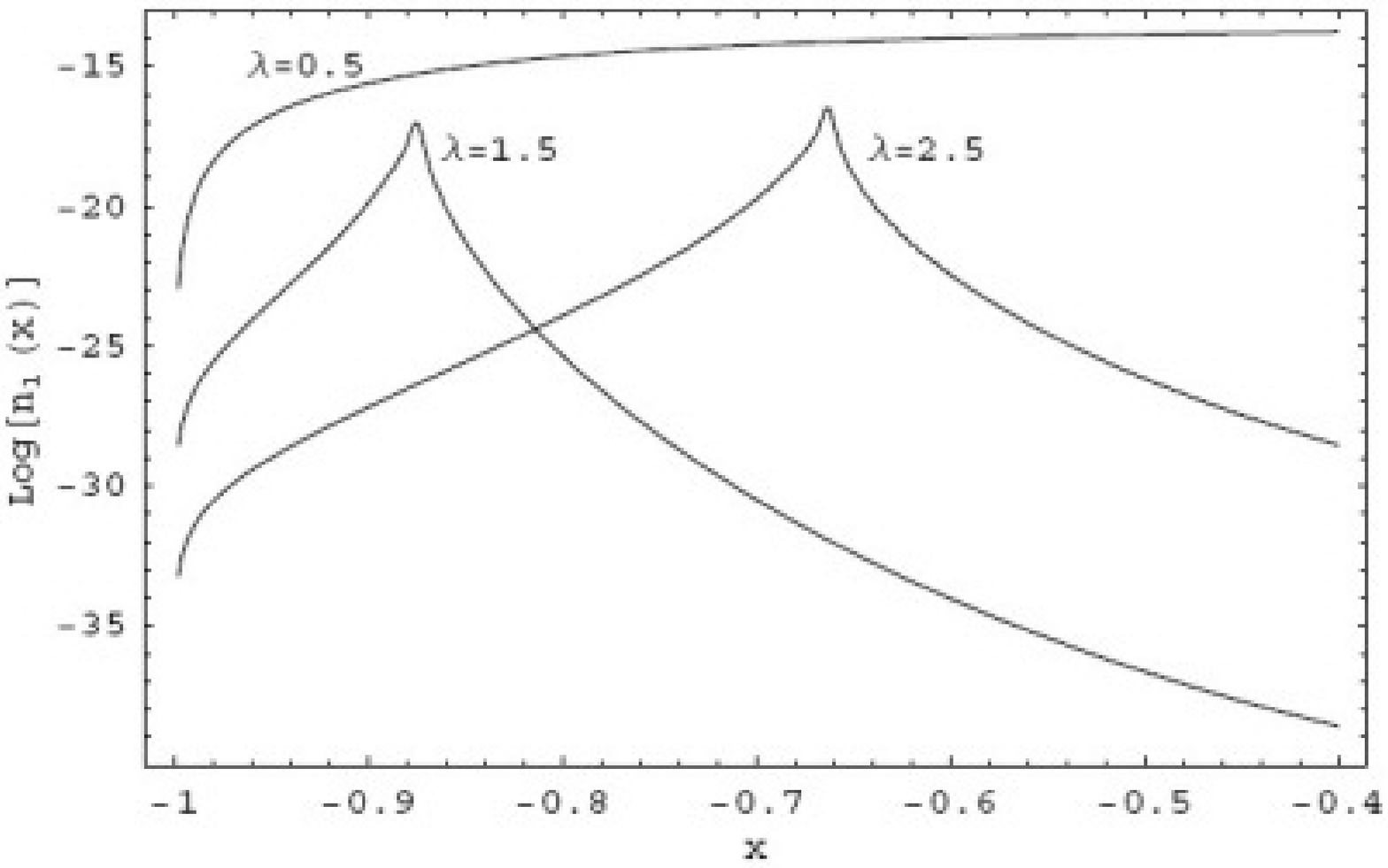,height=5cm,clip=,angle=0} \\
 	(a) & (b)\\
	\vspace{0.05cm} &\\
 \epsfig{file=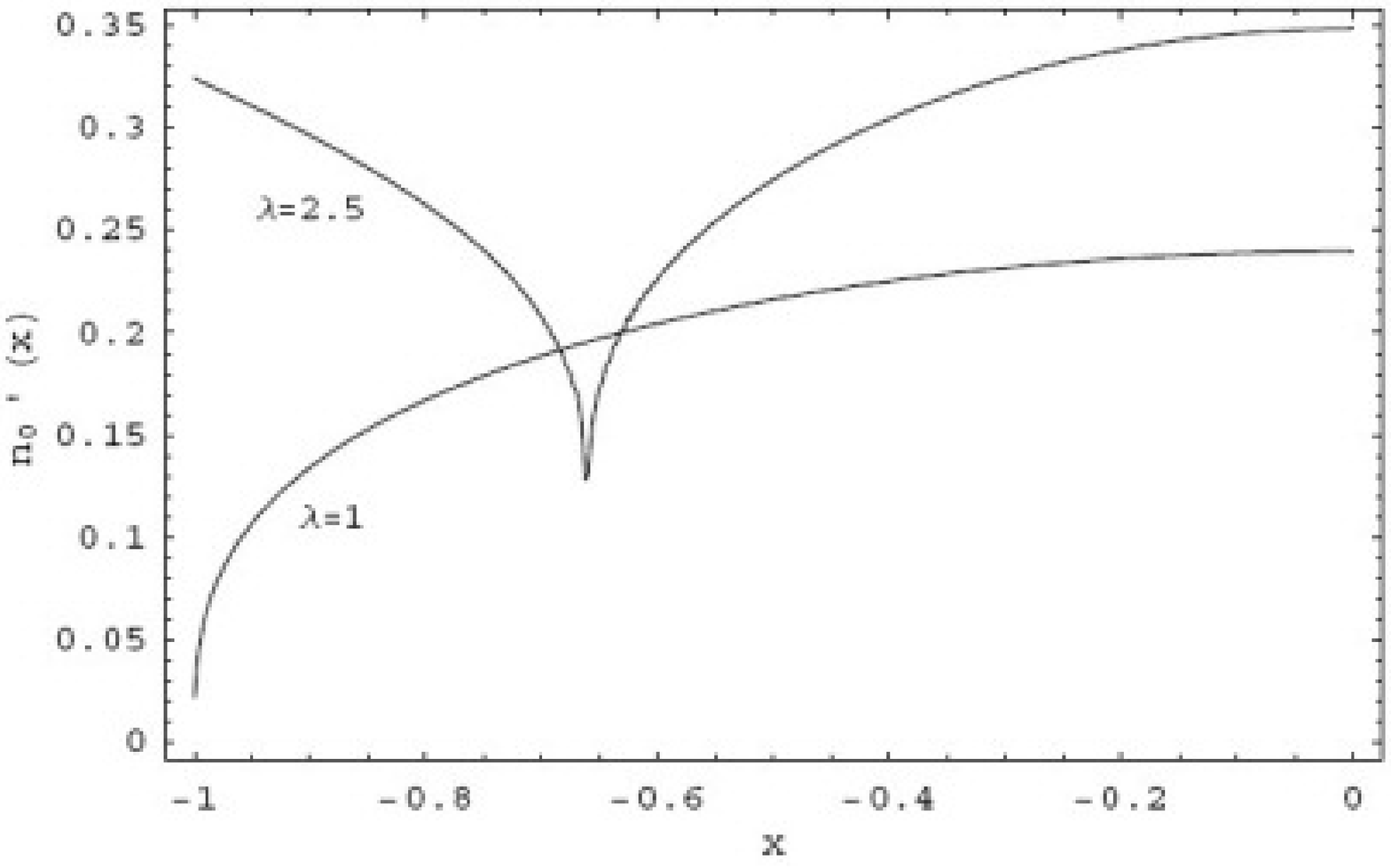,height=5cm,clip=,angle=0}  &
 \epsfig{file=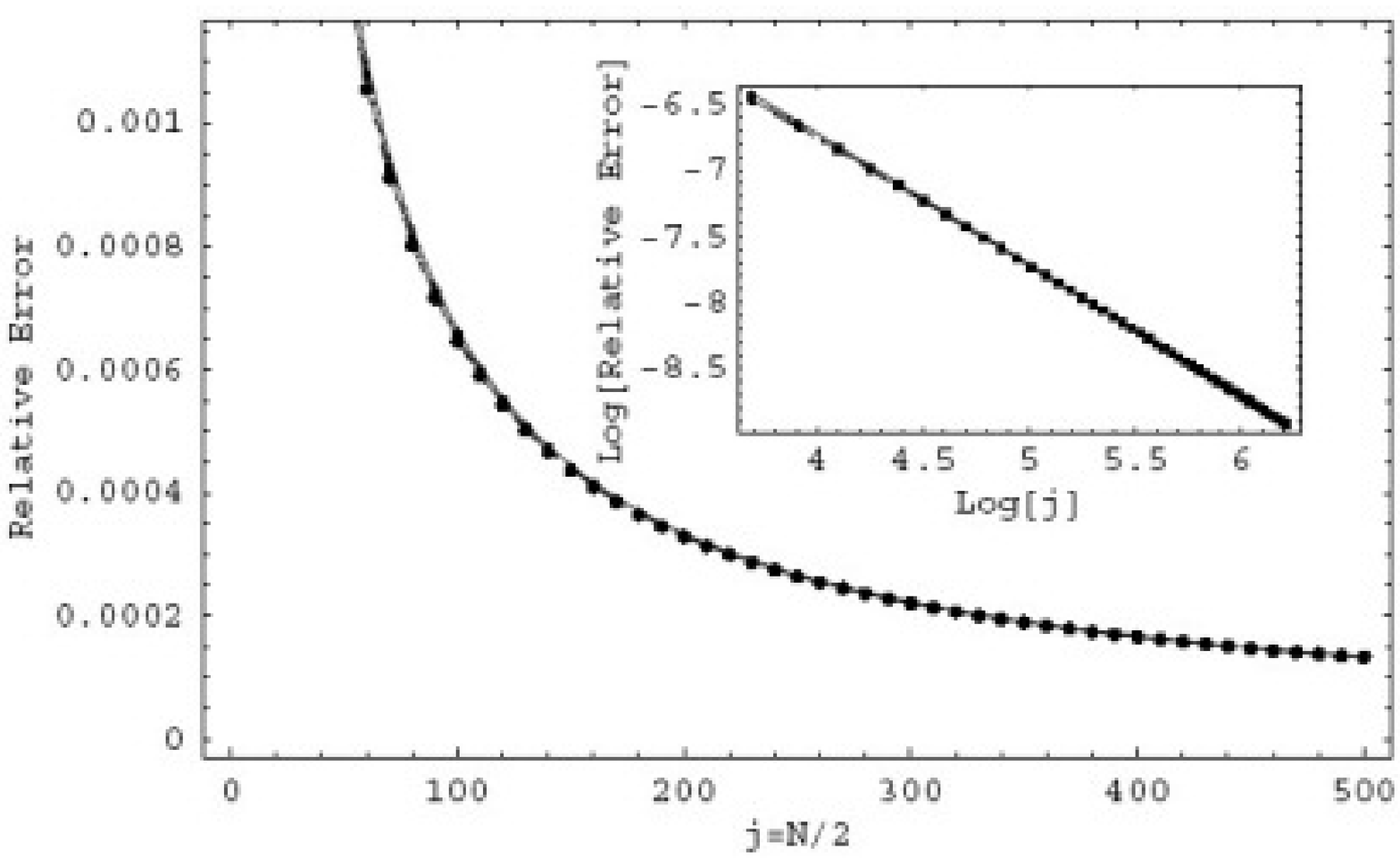,height=5cm,clip=,angle=0} \\
	(c) & (d)
\end{tabular}
\end{narrow}
\caption[The solutions of the Lipkin model flow equation \eref{moyaln0n1}.]{(a) and (b) The solutions to equation \eref{moyaln0n1} at $\ell\approx100$. The dots in (a) represent the exact spectrum for $j=1000$. For clarity only every twelfth eigenvalue is shown. (c) The derivative of $n_0$ with respect to the state label $x$. This is related to the gaps in the spectrum by equation \eref{gapderiv}.  (d) The relative error in the first five states as a function of $j$ for $\lambda=0.5$}
\label{figset3}

\end{figure}

\begin{figure}[th]
\begin{narrow}{-0.35in}{0in}
\begin{tabular}{cc}
	\epsfig{file=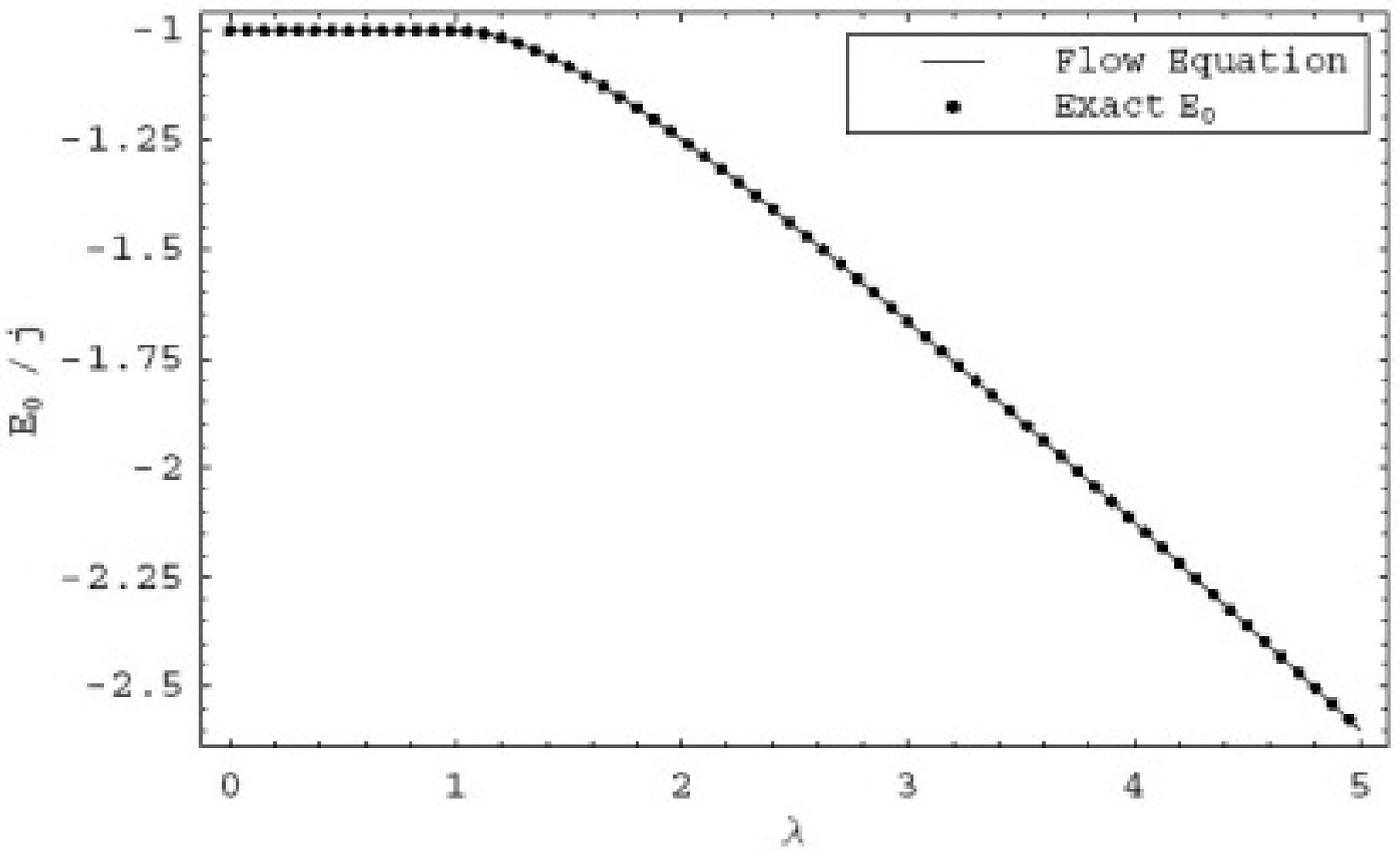,height=5cm,clip=,angle=0} &
	\epsfig{file=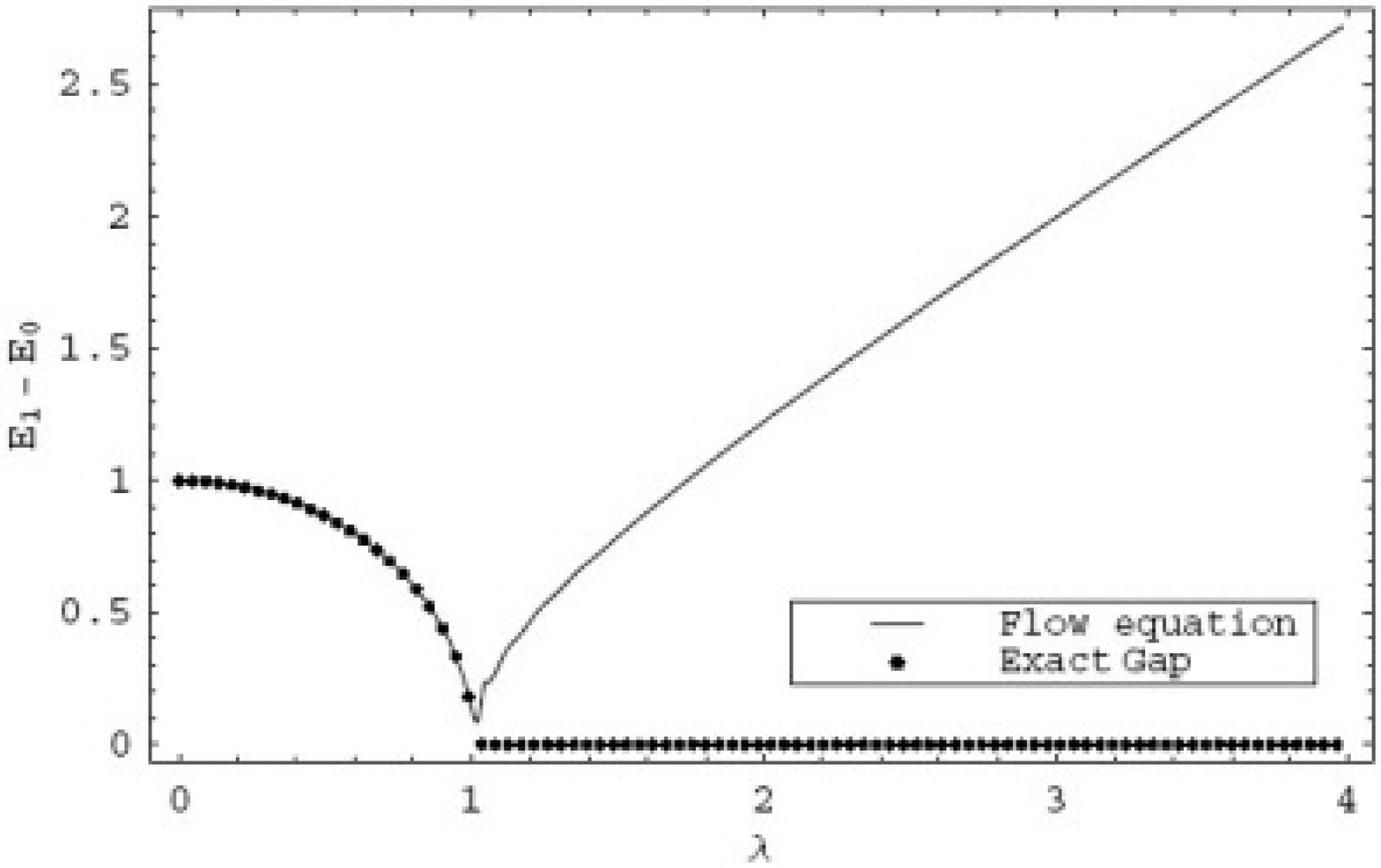,height=5cm,clip=,angle=0} \\
 	(a) & (b)\\
	\vspace{0.05cm} &\\
 \epsfig{file=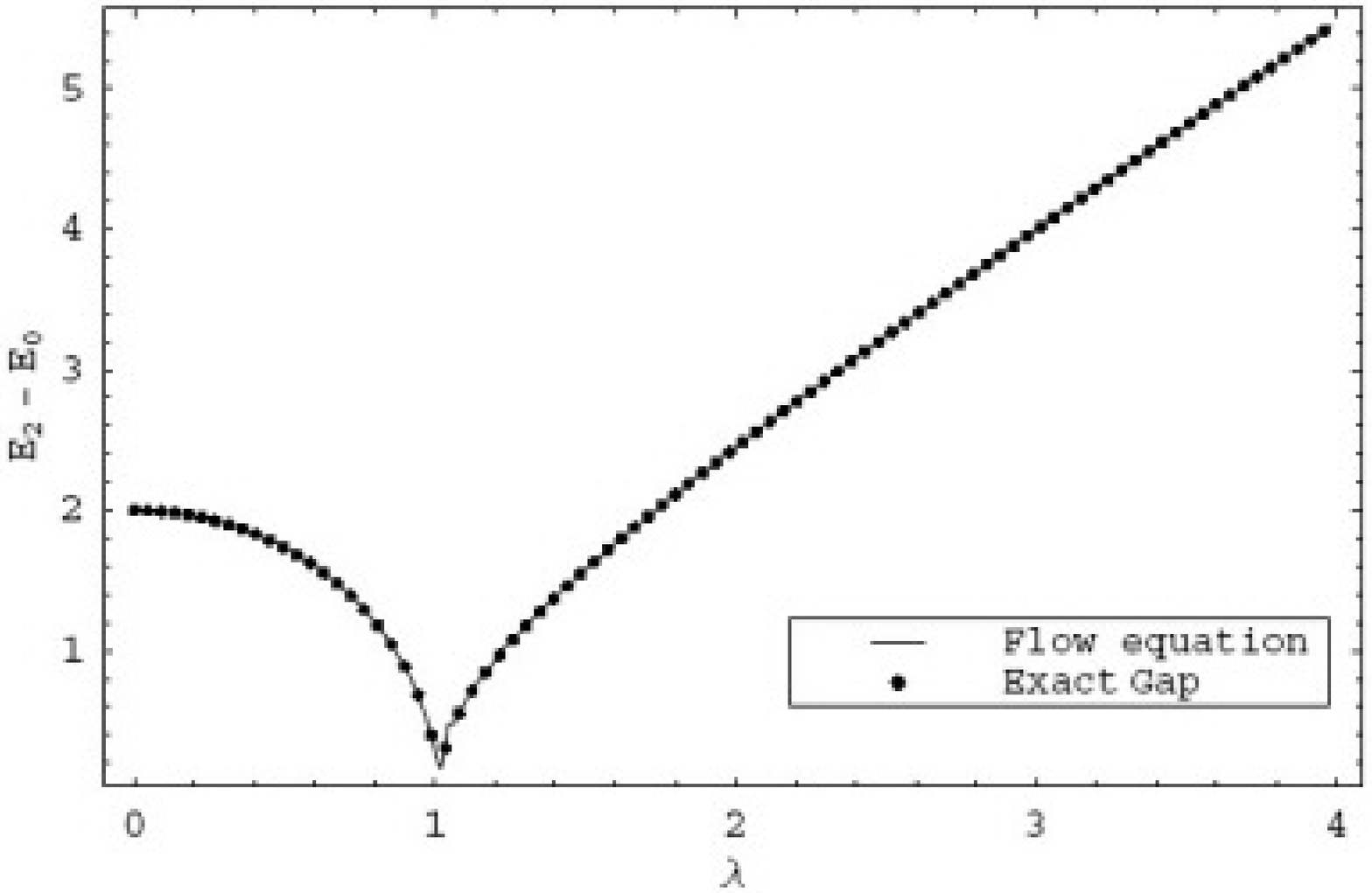,height=5cm,clip=,angle=0}  &
 \epsfig{file=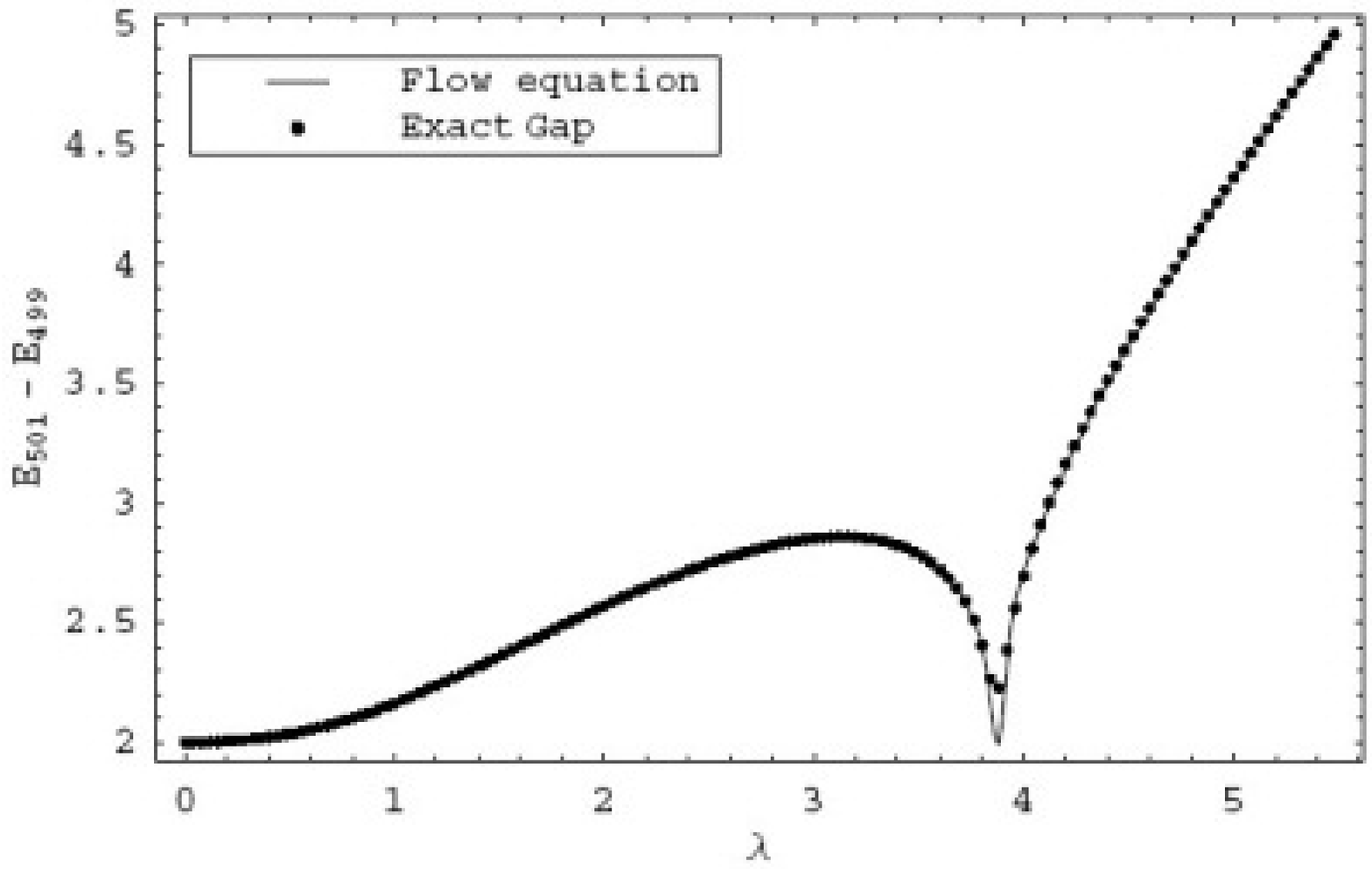,height=5cm,clip=,angle=0} \\
	(c) & (d)
\end{tabular}
\end{narrow}
	\caption[Eigenvalues and gaps between successive states as functions of $\lambda$.]{Energy levels and gaps as functions of the coupling constant. (a) The ground state energy.  (b) The gap between the first excited state and ground state. (c) The gap between the second excited state and ground state. (d) The gap between $E_{501}$ and $E_{499}$ for $j=1000$.}
	\label{figset4}
\end{figure}

\begin{figure}[th]
\begin{narrow}{-0.28in}{0in}
\begin{tabular}{cc}
	\epsfig{file=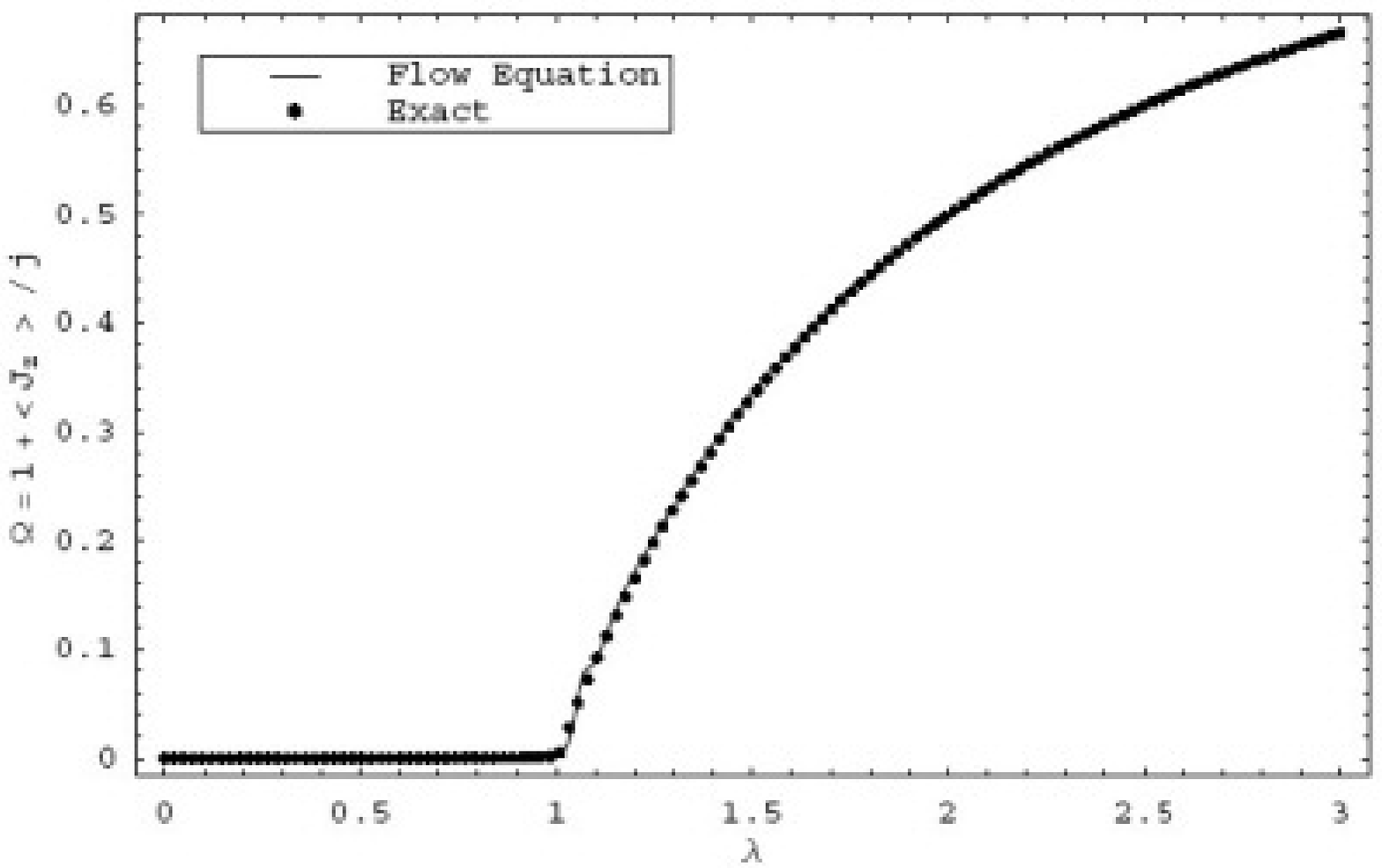,height=5cm,clip=,angle=0} &
	\epsfig{file=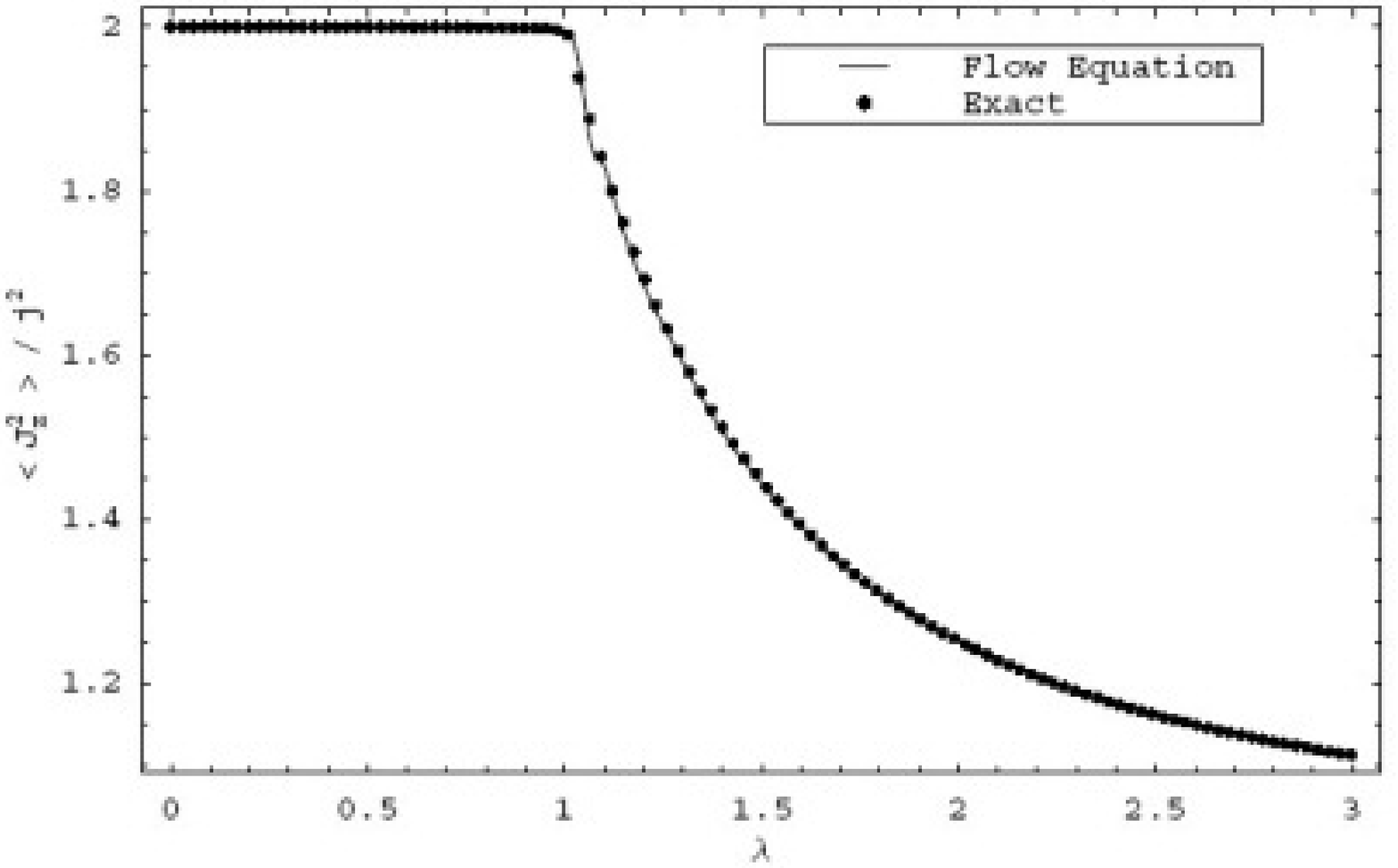,height=5cm,clip=,angle=0} \\
 	(a) & (b)\\
	\vspace{0.05cm} &\\
 \epsfig{file=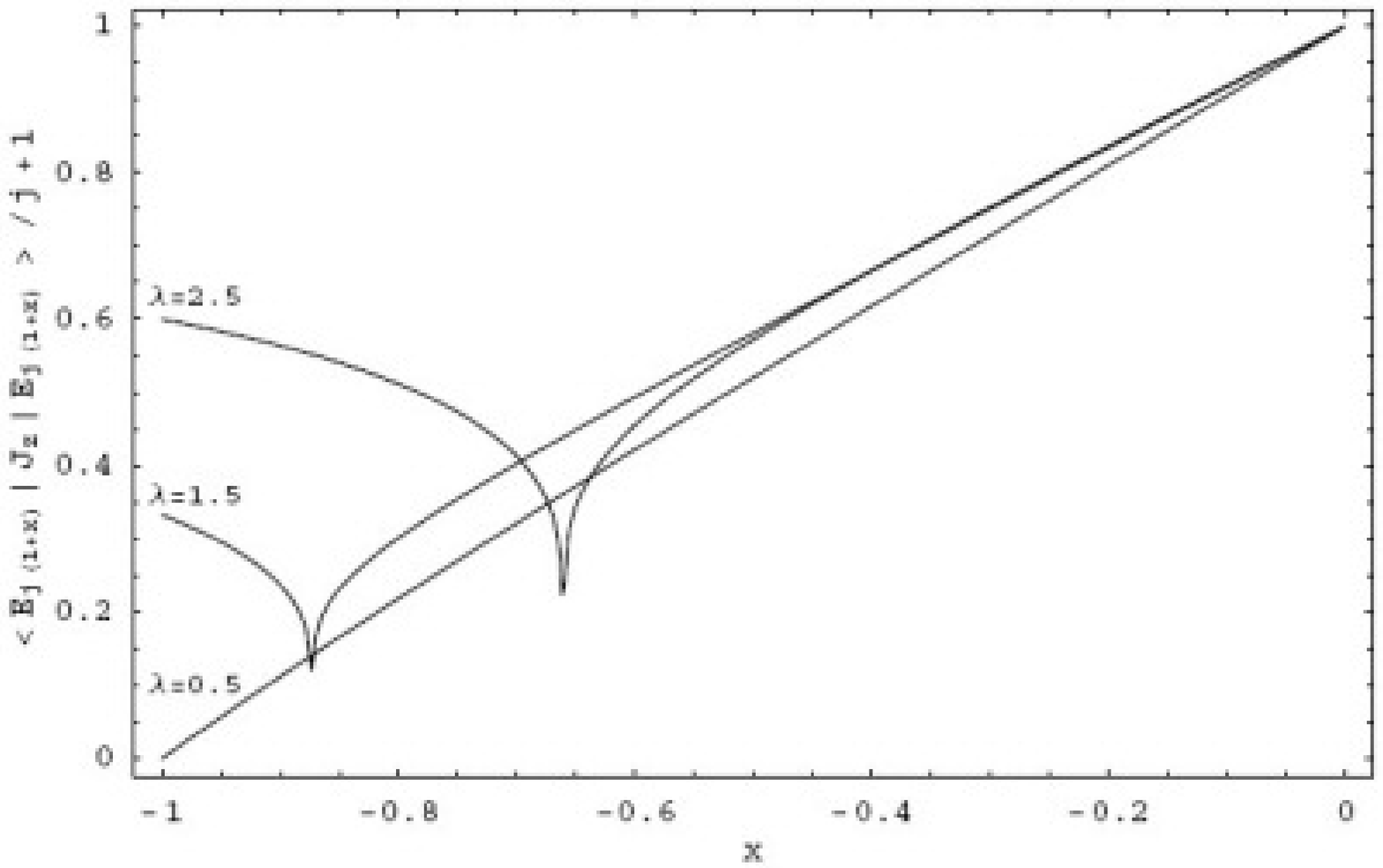,height=5cm,clip=,angle=0}  &
 \epsfig{file=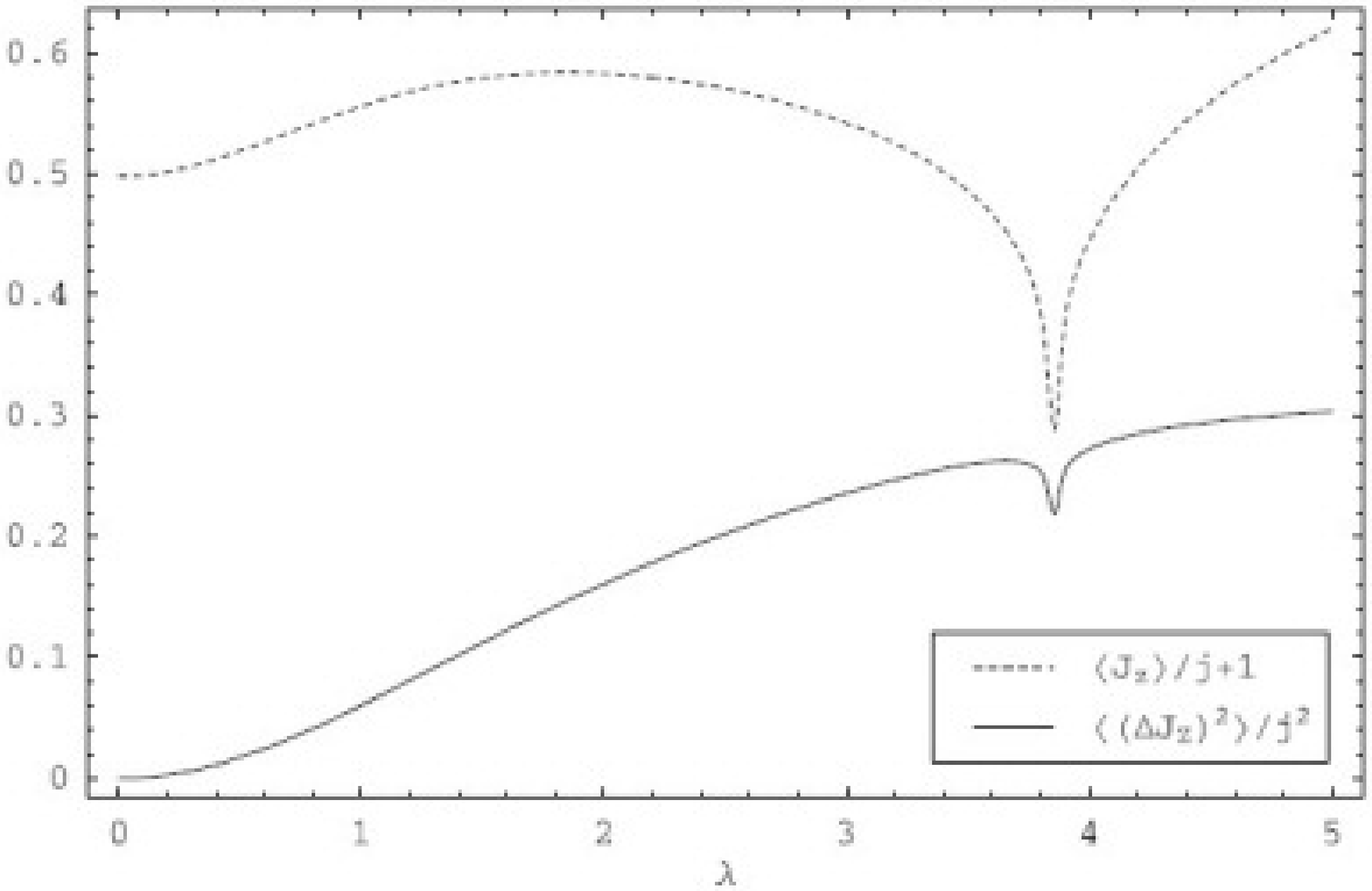,height=5cm,clip=,angle=0} \\
	(c) & (d)
\end{tabular}
\end{narrow}
	\caption[Expectation values as functions of $\lambda$.]{(a) The order parameter $\Omega=1+\ave{J_z}/j$ as a function of $\lambda$. (b) $\ave{J^2_z}/j^2$ in the ground state as a function of $\lambda$. (c) Expectation values of $J_z$ as functions of the state label $x$ for different values of the coupling. (d) $\ave{J_z}+1$ and $\ave{(\Delta J_z)^2}$ in $\ket{E_{250}}$ for $j=1000$ as a function of $\lambda$. The sharp local minimum occur at the point where $x(\lambda)=-0.5$. }
	\label{figset5}
\end{figure}

\begin{figure}[th]
\begin{narrow}{-0.35in}{0in}
\begin{tabular}{cc}
	\epsfig{file=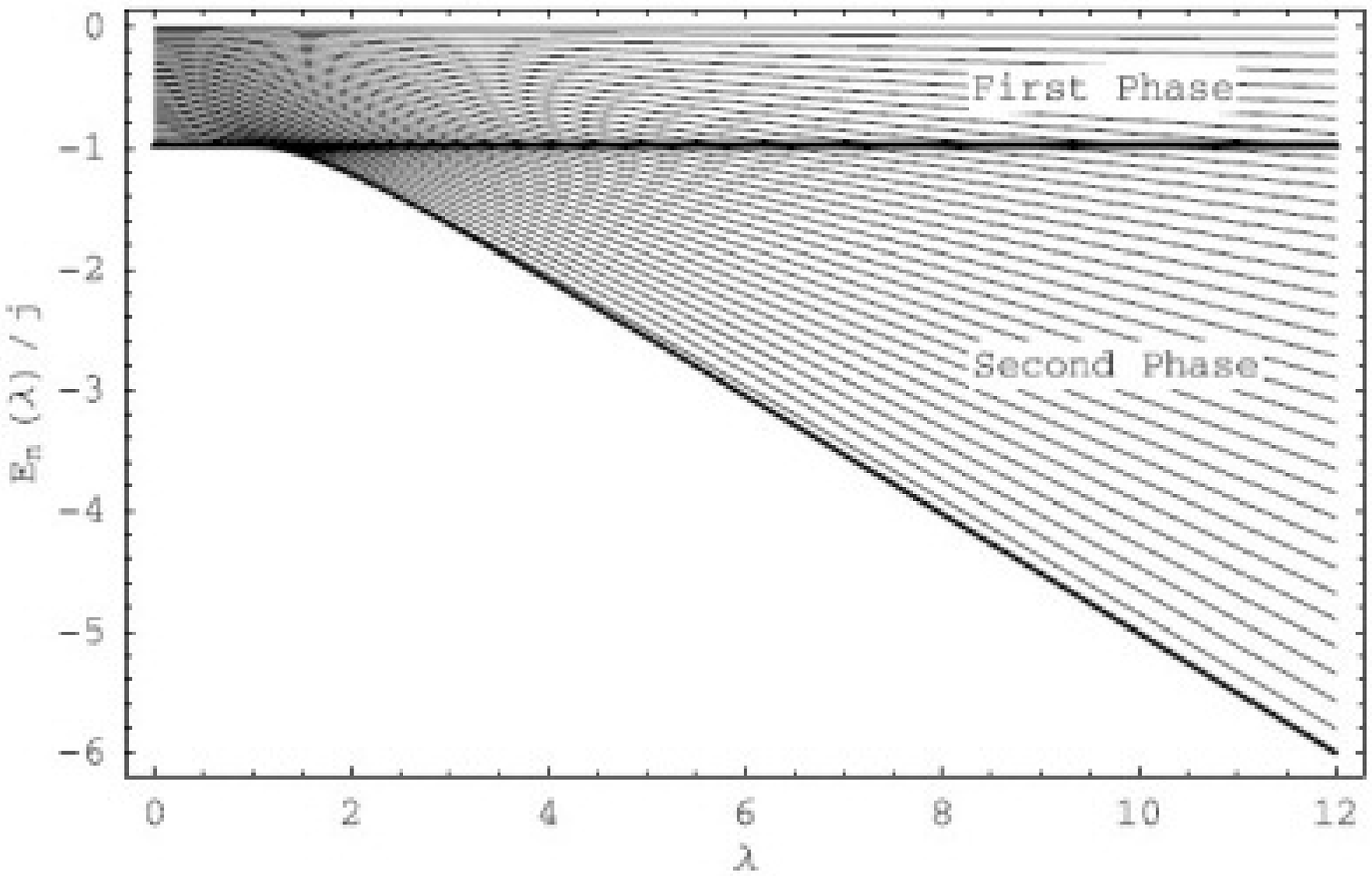,height=5cm,clip=,angle=0} &
	\epsfig{file=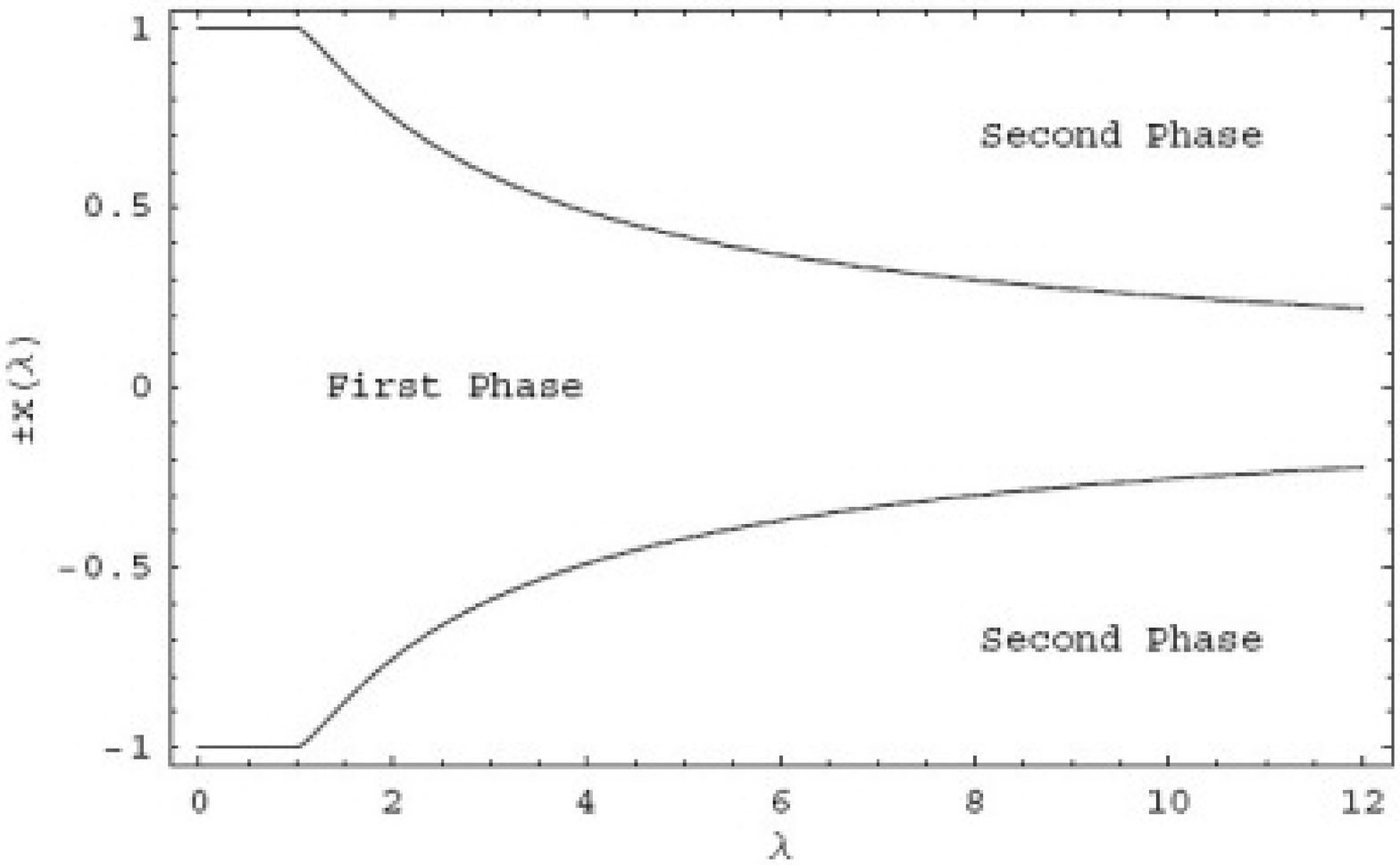,height=5cm,clip=,angle=0} \\
 	(a) & (b)\\
	\vspace{0.05cm} & \\ \epsfig{file=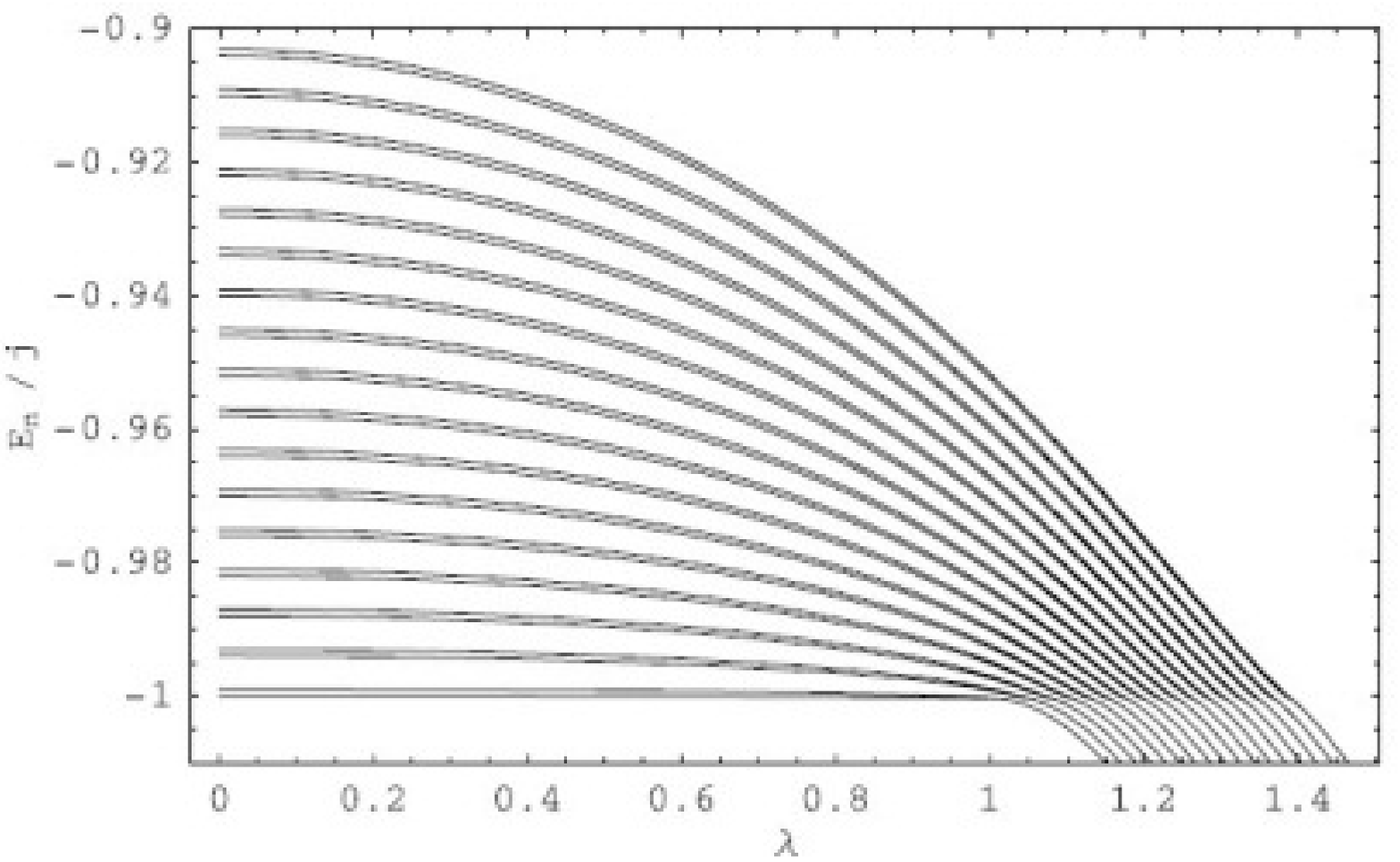,height=5cm,clip=,angle=0}  &
 \epsfig{file=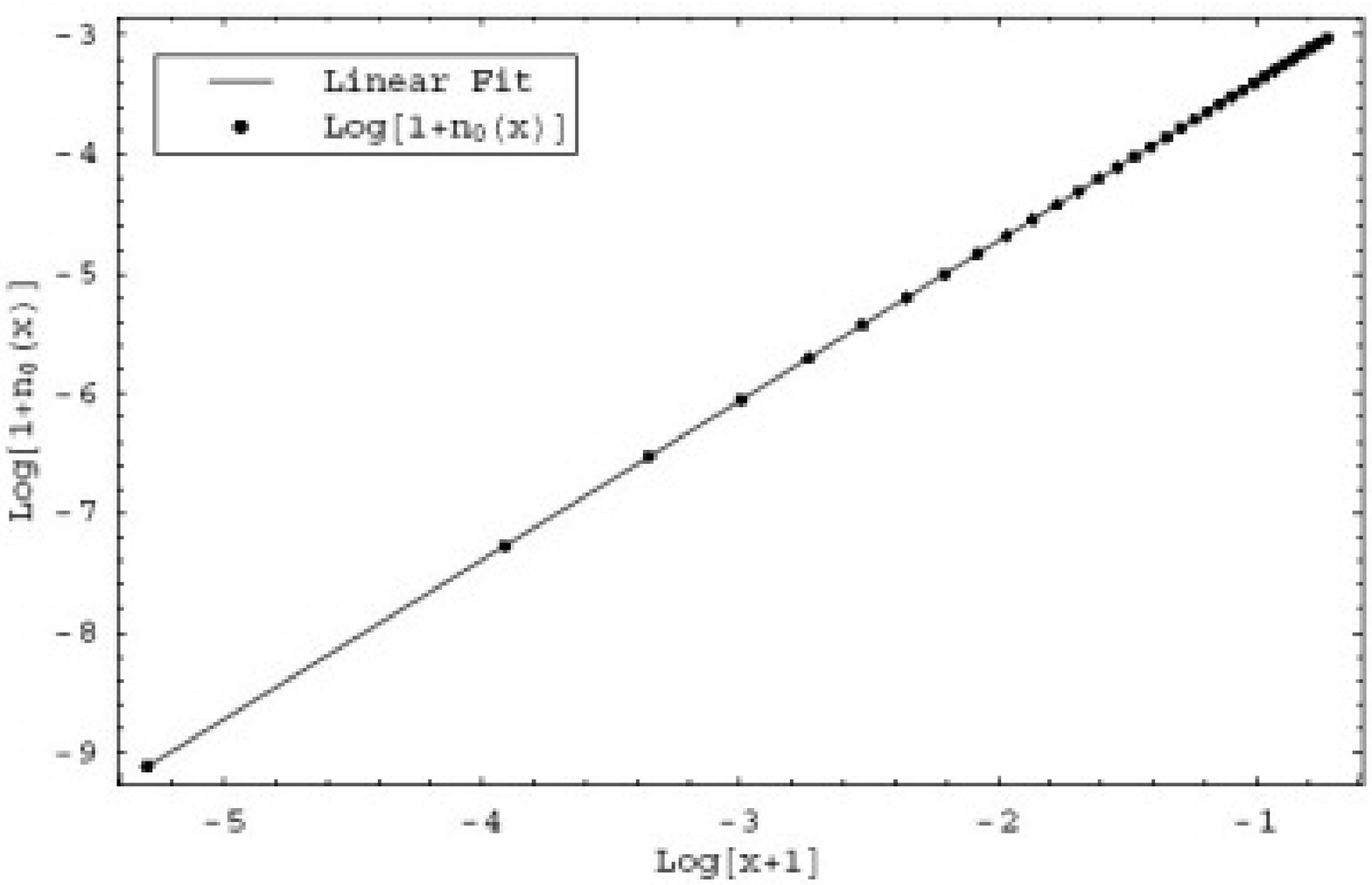,height=5cm,clip=,angle=0} \\
	(c) & (d)
\end{tabular}
\end{narrow}
	\caption[Aspects of the phase structure of the Lipkin model.]{(a) The negative half of the spectrum as a function of $\lambda$. Every fourteenth eigenvalue is shown. (b) $x(\lambda)$ versus $\lambda$. The relevant points in both the positive and negative halves of the spectrum are shown. (c) Energies of pairs of consecutive states of differing parity as function of $\lambda$. Every third pair of the first 300 states are shown. (d) A log-log plot at $\lambda=1$ of $n_0(x)+1$ versus $x+1$, which is equivalent to $E_n$ versus $n$. The slope of the linear fit is $1.33207$; within $0.1\%$ of $4/3$.}
	\label{figset6}
\end{figure}

First, we consider some structural properties of $n_0(\ell)$ and $n_1(\ell)$. It is known that the matrix elements of the Lipkin Hamiltonian possess the symmetry $\inp{m}{H}{m}=-\inp{-m}{H}{-m}$, which implies that the spectrum is anti-symmetric around $E_j=0$. This symmetry is respected by the flow equation and manifests itself through an invariance of equations \eref{moyaln0n1} under the substitutions $n_0\rightarrow-n_0$, $n_1\rightarrow n_1$ and $x\rightarrow -x$. This implies that $n_0(x,\ell)=-n_0(-x,\ell)$ and $n_1(x,\ell)=n_1(-x,\ell)$, and so we may restrict ourselves to the interval $[-1,0]$ in the $x$ dimension. This will be seen to correspond to the negative half of the spectrum, from which the entire spectrum can easily be obtained. We also note that $n_1(\pm1,0)=0$, and that this remains the case at finite $\ell$ since $\partial n_1/\partial \ell$ is proportional to $n_1$. Applying the same argument to $n_0$ seems to suggest that no flow occurs for $n_0$ at $x=-1$ and that $n_0(-1,\ell)=-1$ for all $\ell$. This conclusion is, however, incorrect since it is found that in the second phase $\partial n_1/\partial x$ develops a square root singularity at $x=-1$, which allows $n_0(-1,\ell)$ to flow away from $-1$. Numerically this can be handled easily by solving for $m(x,\ell)\equiv n_1^2(x,\ell)$, instead of $n_1$. We made use of the well established fourth order Runge-Kutta method \cite{Burden} to integrate the PDE's to sufficiently large $\ell$-values. 

It was shown that the eigenvalues of $H$ are given by $E_n=j n_0(-1+n/j)$. Furthermore, for sufficiently large $j$ it holds that
\begin{eqnarray}
E_{n+1}-E_n&=&j\left[n_0\left(-1+\frac{n+1}{j}\right)-n_0\left(-1+\frac{n}{j}\right)\right]\nonumber\\
	&\approx&\left.\pafg{n_0}{x}\right|_{x=-1+n/j}.
	\label{gapderiv}
\end{eqnarray}
Figures \ref{figset3} (a) and (b) show $n_0$ and $\log(n_1)$ at $\ell\approx100$ for three values of $\lambda$. At this point in the flow $n_1$, which represents the off-diagonal part of $H(\ell)$, is already of the order of $10^{-20}$, and may be neglected completely. As a comparison with exact results we calculated the spectrum for $j=1000$ using direct diagonalization and plotted the pairs $(x=-1+n/j,E^{\rm exact}_n/j)$ for $n=0,\ldots,j$ as dots. We observe an excellent correspondence for all states and in both phases, with an average error of about $0.05\%$. This small discrepancy can be attributed to numerical errors and finite size effects, since we are comparing exact results obtained at finite $j$ with those of the flow equation which was derived in the $j\rightarrow\infty$ limit. This is again illustrated in Figure \ref{figset3} (d) which shows the relative error in the first five eigenvalues as a functions of $j$ for $\lambda=0.5$. The lines fall almost exactly on one another, so no legend is given. The log-log inset shows a set of straight lines with gradients equal to one, clearly illustrating the $1/j$ behaviour of the errors. 

Next we consider the dependence of the eigenvalues on the coupling $\lambda$. Figure \ref{figset4} (a) shows the ground state energy as a function of $\lambda$ together with the exact result for $j=1000$. We see that $E_0$ is fixed at $-j$ in the first phase and begins to decrease linearly with $\lambda$ at large coupling. The phase transition in the Lipkin model is known to be of second order, and is characterized by a large number of avoided level crossings \cite{Sachdev}. This brings about complex non-analytic behaviour in the gaps between energy levels. The non-perturbative treatment of the flow equation enables us to reproduce much of this behaviour correctly. As examples we display the gaps $E_1-E_0$, $E_2-E_0$ and $E_{501}-E_{500}$ in Figure \ref{figset4}. One quantity that is not reproduced correctly are the gaps between successive states belonging to subspaces of differing parity. (Recall that subspaces of odd and even spin projection (parity) are not mixed by the Hamiltonian, as was shown in Section \ref{lipkinmodel}.) As a definite case consider the gap $E_1-E_0$ which is known to vanish like $1/j$ in the second phase, while the flow equation produces a gap which grows linearly with $\lambda$. This can be attributed to the fact that the gaps are not extensive quantities and so they depend on higher order corrections in $1/j$, which were neglected in our derivation. 

The function $n_1(x,\ell)$ found by solving equations \eref{moyaln0n1} can now be substituted into equation \eref{moyalaux} to obtain the flow equation for a general observable $O$. On a technical note, we found it advantageous to use expansion \eref{fourier} to write the flow equation for $O(x,\beta,\ell)$ as a coupled set of $1+1$ dimensional equations for the $f_n$'s, rather than treating it as a general $2+1$ dimensional PDE. First we consider the flow of $J_z$, which corresponds to the initial condition $O(x,\beta,0)=jx$. Figure \ref{figset5} (a) shows the results obtained for the order parameter together with exact results for $j=1000$. We again find excellent agreement in both phases. Other observables can be considered by a simple modification of the initial conditions. For example $O(x,\beta,0)=(jx)^2$ produces the second moment of $J_z$ in the ground state, as shown in Figure \ref{figset5} (b). The solution to equation \eref{moyalaux} provides us with expectation values corresponding to excited states as well, which we find by evaluating $f_0$, as defined in \eref{fourier}, at the values $x=-1+n/j$ for $n=1,2,\ldots$. Figure \ref{figset5} (c) shows $f_0$ for $O=J_z$ at different values of the coupling strength. 

The non-perturbative application of the flow equation method to the Lipkin model has provided some interesting new insights into aspects of the phase transition. We will investigate some of these next. First we fix some notation. In the large $j$ limit we may label states with the continuous label $x\in[-1,1]$ through the association $x\leftrightarrow\ket{E_{(1+x)j}}$. In this way a state is labelled according to its fractional position in the spectrum, for example $x=-1$ always corresponds to the ground state, while $x=-1/2$ denotes the state lying one quarter way up the spectrum. Now consider the behaviour of $n_0(x)$ and $n_0'(x)$ depicted in Figures \ref{figset3} (a) and (c) respectively. We note that for $\lambda>1$ there always exists a value of $x<0$, denoted by $x(\lambda)$, where the derivative of $n_0(x)$ very nearly vanishes. Furthermore, $x(\lambda)$ must be a point of inflection of $n_0(x)$ since the derivative of $n_0(x)$ to $x$ cannot change sign. If ${\rm d}n_0(x)/{\rm d}x$ were to become negative the flow equation would become unstable and cause $n_1(x)$ to grow exponentially, contradicting the results of Section \ref{gChoice} concerning the form of $H(\infty)$. This explains why we only observe a plateau at $x(\lambda)$, and no more drastic behaviour. At $\lambda=1$ this happens precisely at the ground state, i.e. $x(1)=-1$, while in the first phase no such point exists. In \cite{Heiss2} it was found that the energies of the low-lying states obey the scaling law 
\begin{equation}
	E_n+j\propto n^{4/3}/N^{1/3}=N(x+1)^{4/3}
\end{equation}
at $\lambda=1$. This can be confirmed using the flow equations by considering the behaviour of $n_0(x)$ close to $x=-1$. Figure \ref{figset6} (d) shows a log-log plot of $n_0(x)+1$ versus $x+1$ together with a linear fit which reproduces the power of $4/3$ to within $0.1\%$.  Earlier a direct link was established between the gaps separating successive eigenvalues and the derivative of $n_0$. This suggests that $x(\lambda)$ corresponds to a point in the spectrum with a very high density of states, brought about by a large number of avoided level crossings. Interestingly this point always occurs at the same absolute energy, namely $E_{(1+x(\lambda))j}=-j$, although this corresponds to increasingly highly excited energies relative to the ground state. We believe that in the thermodynamic limit, contrary to Figure \ref{figset3} (c), the derivative of $n_0$, and the corresponding gap, should vanish completely at this point. This has been confirmed in \cite{Heiss2} where it was shown that the gap is given by
\begin{equation}
	\Delta E=\frac{2\pi\sqrt{\lambda^2-1}}{\ln N}
\end{equation}
where $N=2j$. That our solution does not reflect this can be attributed to numerics, as this concerns a single point which is effectively ``invisible" to the finite discretization used in the numerical method. The off-diagonal part of the Hamiltonian, represented by $n_1$, also exhibits striking behaviour at $x(\lambda)$. In fact, upon returning to the flow equation \eref{moyaln0n1} we make the interesting observation that at the point where the derivative of $n_0(x)$ vanishes, $n_1(x)$ is not forced to flow to zero, but may in fact attain a non-trivial fixed point value.  This is shown in Figure \ref{figset3} (b) which shows $\log(n_1(x))$ as a function of $x$. One clearly sees a sharp peak at the point where the derivative of $n_0(x)$ nearly vanishes. The peak only occurs for $\lambda>1$ and moves to the right as $\lambda$ is increased.  This is also consistent with the known result \cite{Wegner} that an off-diagonal element $m_{ij}$ of $H(\ell)$ decays roughly as $\exp(-(E_{i-1}-E_{j-1})^2 \ell)$ at large $\ell$. This suggests a connection between quantum phase transitions, the corresponding disappearance of an energy scale (gap) \cite{Sachdev} in the thermodynamic limit and the absence of decoupling in the Hamiltonian, also in the thermodynamic limit.

The occurrence of this point in the second phase lends itself to the interesting interpretation of a ``quantum phase transition" at higher energies. Indeed, apart from possessing some notable properties itself, it separates regions of the spectrum with markedly different characteristics. It is well known that states alternate between odd and even parity as one moves up in the spectrum. In the first phase these odd-even pairs are separated by a finite gap. In the second phase one finds a degeneracy between these successive odd and even states developing below the $x(\lambda)$ point. In particular this implies a degenerate ground state with broken symmetry in the second phase, although, as we have seen, this description may be applied to all states below $x(\lambda)$. Figure \ref{figset6} (c) shows a subset of eigenvalues as a function of the coupling, clearly illustrating this coalescing of pairs. Turning to Figure \ref{figset5} (d), which shows the expectation value $\inp{E_{(1+x)j}}{J_z}{E_{(1+x)j}}$ as a function of the state label $x$, we see a sharp local minimum occurring at $x(\lambda)$ which separates the two phases. At the point $x(\lambda)$ itself the corresponding state $\ket{E_{(1+x(\lambda))j}}$ is characterized by sharp localization around the $\ket{j,-j}$ basis state, also noted by \cite{Heiss2}. This is clearly illustrated by Figure \ref{figset5} (d) which shows a pronounced decrease in the spread of the state $\ket{E_{j/4}}$ in the $J_z$ basis occurring at the point where $x(\lambda)=-0.5$. We end this section with two ``phase diagrams" which we hope will further clarify the discussion above. Figure \ref{figset6} (a) shows the $\lambda$ dependency of a subset of the negative, even eigenvalues. For $\lambda\leq1$ the eigenvalues are confined between $0$ and $-j$. As $\lambda$ increases first the ground state (shown in bold) and then the excited states begin to cross the $E=-j$ phase boundary until eventually, in the large coupling limit, only the $\ket{E_j=0}$ state retains its first phase character. For finite $j$ successive eigenvalues show avoided level crossings on the phase boundary.  In the thermodynamic limit one would, however, expect that successive eigenvalues will coalesce as they cross the phase boundary; signaling a vanishing gap. A similar diagram, based in the label $x$ rather than the energy itself, is shown in Figure \ref{figset6} (b). Keeping in mind the symmetries $E_n=-E_{2j-n}$ and $\inp{E_{n}}{J_z}{E_{n}}=-\inp{E_{2j-n}}{J_z}{E_{2j-n}}$ this discussion can easily be adapted to apply to the positive half of the spectrum as well.\\
\begin{figure}[t]
\begin{center}
	\epsfig{file=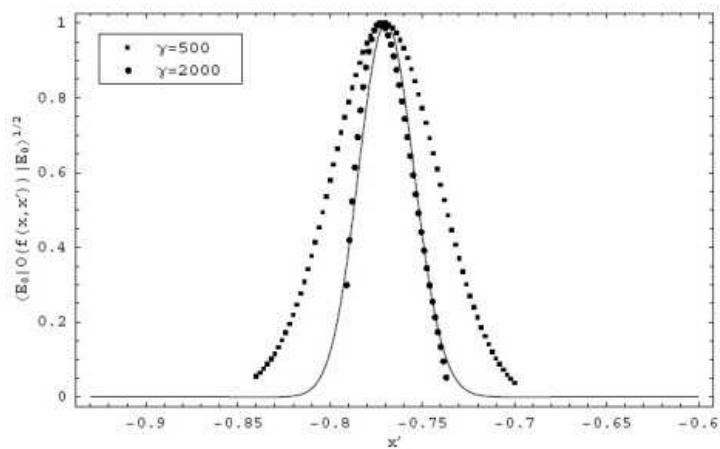,height=6cm,clip=,angle=0}\\
	\caption[Results obtained for the ground state wavefunction using the flow equation.]{The square root of the ground state expectation value of $O(f(x,x'))$ at $\lambda=1.3$ and for a range of $x'$ values. The solid line is the exact modulus of the ground state wavefunction in the $J_z$ basis for $j=4000$. All results have been normalized to have the same maximum value.}
	\label{eigstruc}
\end{center}
\end{figure}

These findings, published in \cite{Kriel}, have stimulated further investigation of 
these phenomenon in the contexts of exceptional points \cite{Heiss} and semi-classical approximations \cite{Heiss2}. These have shed light on the origins of the high density of states and localization occurring at $x(\lambda)$, as well as the mechanism  responsible for the degeneracy between states of differing parity. 

\subsection{Structure of the eigenstates}
\label{eigenstatenumeric}
In Section \ref{eigenstates} we outlined a strategy whereby the structure of the eigenstates could be probed using the expectation values of a class of specially constructed operators. In this manner it is possible to approximate the modulus of the expansion coefficients of the state in the $J_z$ basis. Here we present the numeric results of this procedure using the operators $O(f(x,\bar{x}))$ with $f(x,\bar{x})=\exp(-\gamma(\bar{x}-x)^2)$. We will consider the ground state at $\lambda=1.3$ and hope to find  that $\inp{E_0}{f(x,\bar{x})}{E_0}\approx\left|\braket{m=\bar{x}j}{E_0}\right|^2$. Figure \ref{eigstruc} shows that this is indeed the case. As expected larger values of $\gamma$ produce a more accurate reproduction of the form of the absolute wavefunction.

\section{Solution in the local approximation}
\label{firstPhaseSol}
The numeric treatment of the previous section, although very effective as a computational tool, obscures some of the more subtle properties of the flow equation. In this regard an analytic approach provides some valuable insights, and we consider one such treatment next.

At $\lambda=0$ the eigenstates of the Lipkin Hamiltonian are simply the $J_z$ basis states. Of course, this is no longer the case at non-zero coupling, although we expect that, in the first phase at least, the eigenstates will remain localized in the spin basis. The unitary transformation  diagonalising $H$ will reflect this by mainly mixing states of which the spin projections differ only slightly. This is manifested in the flow equations through the locality of the evolution of $n_{0,1}(x,\ell)$. By this we mean that the flow at a point $x=x_1$ is very weakly affected by the values of $n_{0,1}(x,\ell)$ at points far from $x_1$. We first consider the low lying states which can be investigated by  solving $n_{0,1}(x,\ell)$ within a neighbourhood of $x=-1$. Since $x=J_z/j$ (from representation \eref{repSU2}) this is equivalent to restricting the calculation to states for which $\left(\left<J_z\right>+j\right)/j<<1$; an approximation commonly used in this context. We begin by linearizing the initial condition about the point $x=-1$, which gives
\begin{equation}
	n_0(x,0)=-1+(x+1)\ \ \ {\rm and}\ \ \ n_1(x,0)=\lambda (1+x).
\end{equation}
The form of $n_0(x,0)$ is chosen to coincide with the parametrization of $H(\ell)$ we introduce next. From equation \eref{moyaln0n1} it is clear that the linearity of $n_{0,1}$ is preserved during flow, which allows for the simple parametrization of $H(\ell)$ as
\begin{equation}
	H(x,\beta,\ell)/j=-1+a_0(\ell)(1+x)+a_1(\ell)(1+x)\cos(2 \beta),
\end{equation}
where $a_0(0)=1$ and $a_1(0)=\lambda$. The flow of the coefficients is given by
\begin{equation}
	\afg{a_0}{\ell}=-a_1 a_1\ \ \ {\rm and}\ \ \ \afg{a_1}{\ell}=-a_1 a_0
\end{equation}
where $\ell$ has been rescaled by a factor of four. These equations leave $a_0(\ell)^2-a_1(\ell)^2=1-\lambda^2$ invariant, and since $a_1(\infty)=0$ we conclude that $H(\infty)=-j+\sqrt{1-\lambda^2}(j+J_z)$. This correctly predicts the characteristic square root behaviour of the gap $E_1-E_0=\sqrt{1-\lambda^2}$, as well as the low lying spectrum within a harmonic approximation.\\

The natural question arising here is whether this approach can be extended to allow for the treatment of the highly excited states as well. Indeed, if the flow at a point $c\in[-1,1]$ exhibits this locality property it seems reasonable that a local solution about $c$ would provide a good result for the corresponding energy eigenvalue $E_{j(c+1)}$. We will show how this local solution can be found analytically, and that by combining the solutions at various points an approximation for $n_0(x,\infty)$ may be constructed. The first step is again the linearization of $H(0)$, now about an arbitrary point $c\in[-1,1]$. The linearity of $H(\ell)$ in $(x-c)$ is conserved during flow, allowing for the parametrization
\begin{equation}
	H(\ell)/j=a_0(\ell)+a_1(\ell)\left(x-c\right)+\left[a_2(\ell)+a_3(\ell)\left(x-c\right)\right]\cos(2 \beta),
\end{equation}
where the flow of the coefficients are given by
\begin{equation}
	\afg{a_0}{\ell}=-a_3 a_2,\ \ \ \ \ \ \ \afg{a_1}{\ell}=-a^2_3,\ \ \ \ \ \ \ \afg{a_2}{\ell}=-a_1 a_2,\ \ \ \ \ \ \ \afg{a_3}{\ell}=-a_1 a_3.
\end{equation}
The initial conditions are $a_0(0)=c$, $a_1(0)=1$, $a_2(0)=\lambda(1-c^2)/2$ and $a_3(0)=-c\lambda$. These equations leave $I=a_1^2-a_3^2$ invariant, and since $a_3(\infty)=0$ we conclude that $a_1(\infty)=\sqrt{1-c^2\lambda^2}$. The presence of $c$ in this equation is significant, as it indicates that the validity of the local approximation is a function both of the coupling and the specific point under consideration. The stability condition of the local solution is
\begin{equation}
	-\frac{1}{\lambda}\leq c \leq \frac{1}{\lambda},
\end{equation}
which holds for all $c\in[-1,1]$ in the first phase. In the second phase the stable domain is restricted to the interval $[-1/\lambda,1/\lambda]$. This is reminiscent of the discussion in the previous section concerning the transition  appearing at higher energies in the second phase. It was seen that there exists points $\pm x(\lambda)$ which separate regions of the spectrum for which the states exhibit either first or second phase behaviour. In fact, all the states lying within $[-1/\lambda,1/\lambda]$ have a first phase character, as $x(\lambda)$ is always found to be greater than $1/\lambda$. This suggests a connection between the phase structure, the properties of eigenstates, and the locality, or lack thereof, exhibited by the flow equation. 

Continuing with the derivation, we find that the flow equations can be solved exactly:
\begin{eqnarray}
	a_0(\ell)&=&\alpha_3\:\left(\coth{(\alpha_1\ell+\alpha_2)}-\coth{(\alpha_2)}\right)+c,\ \ \ \ \ 
	a_1(\ell)=\alpha_1\:\coth{(\alpha_1\ell+\alpha_2)}\nonumber\\
	a_2(\ell)&=&\alpha_3\:{\rm cosech}{(\alpha_1\ell+\alpha_2)},\ \ \ \ \ 
	a_3(\ell)=\alpha_1\:{\rm cosech}{(\alpha_1\ell+\alpha_2)}.
\end{eqnarray}
The $\alpha_i$'s are integration constants that can be fixed as follows:
\begin{eqnarray}
	a_1(\infty)=\sqrt{1-c^2\lambda^2}\ \ &\Longrightarrow&\ \ \alpha_1=\sqrt{1-c^2\lambda^2}\nonumber\\
 \nonumber\\
\frac{a_1(0)}{a_3(0)}=\cosh(\alpha_2)=\frac{-1}{c\lambda}\ \ &\Longrightarrow&\ \ \alpha_2={\rm arccosh}{\left[\frac{-1}{c\lambda}\right]}\\
\nonumber \\
\frac{\alpha_3}{\alpha_1}=\frac{a_2(0)}{a_3(0)}=\frac{c^2-1}{2c}\ \ &\Longrightarrow&\ \ \alpha_3=\frac{(c^2-1)\sqrt{1-c^2\lambda^2}}{2c}. \nonumber
\end{eqnarray}
The value of $H(\infty)/j$ at $x=c$ is expected to provide a good approximation for the exact solution $n_0(c,\infty)$. We find that
\begin{equation}
	\left.\frac{H(\infty)}{j}\right|_{x=c}=a_0(\infty)=c+\frac{(1-c^2)\left(1-\sqrt{1-c^2\lambda^2}\right)}{2c}.
	\label{local1stOrder}
\end{equation}

\begin{figure}[t]
\begin{narrow}{-0.35in}{0in}
\begin{tabular}{cc}
	\epsfig{file=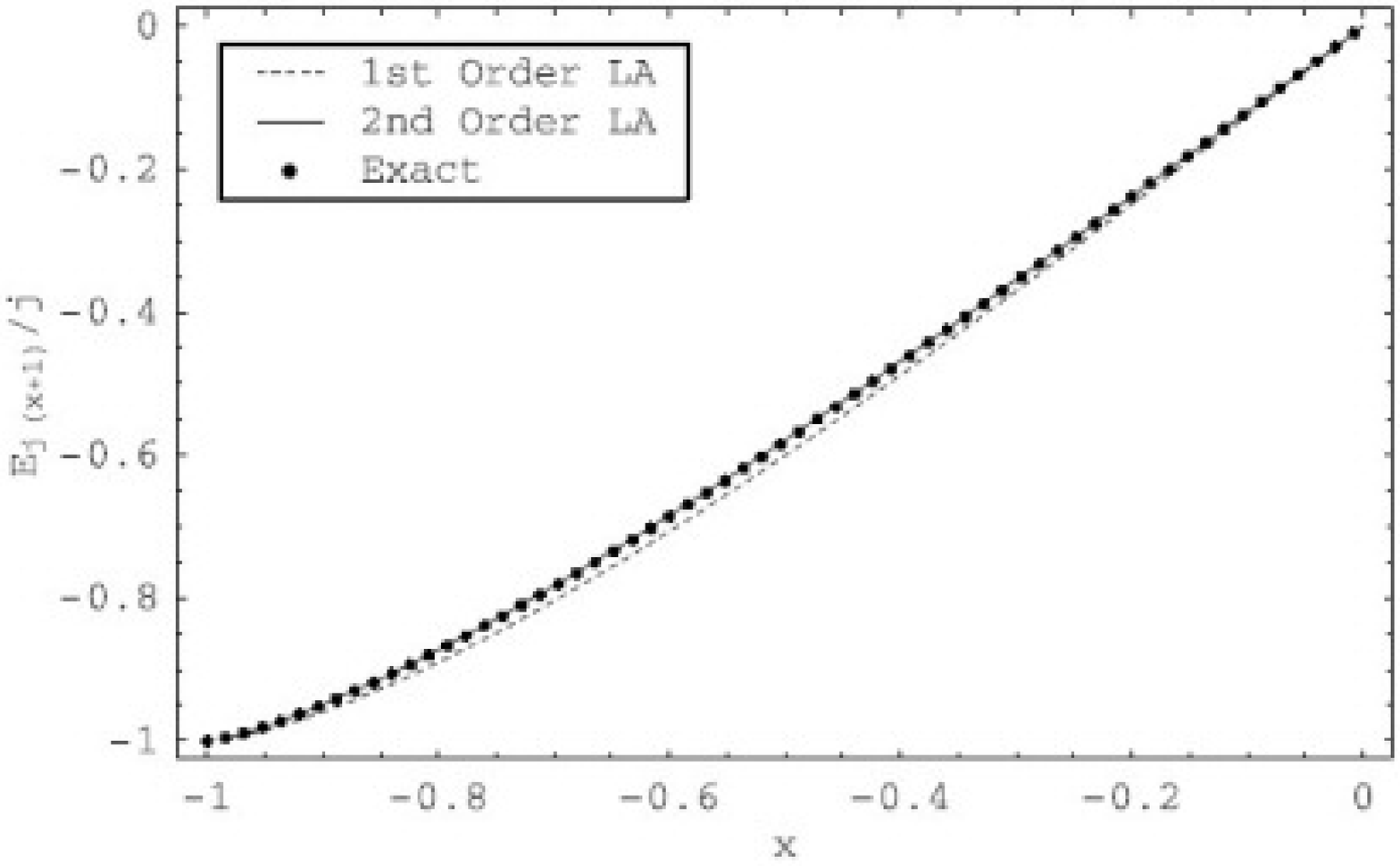,height=5cm,clip=,angle=0} &
	\epsfig{file=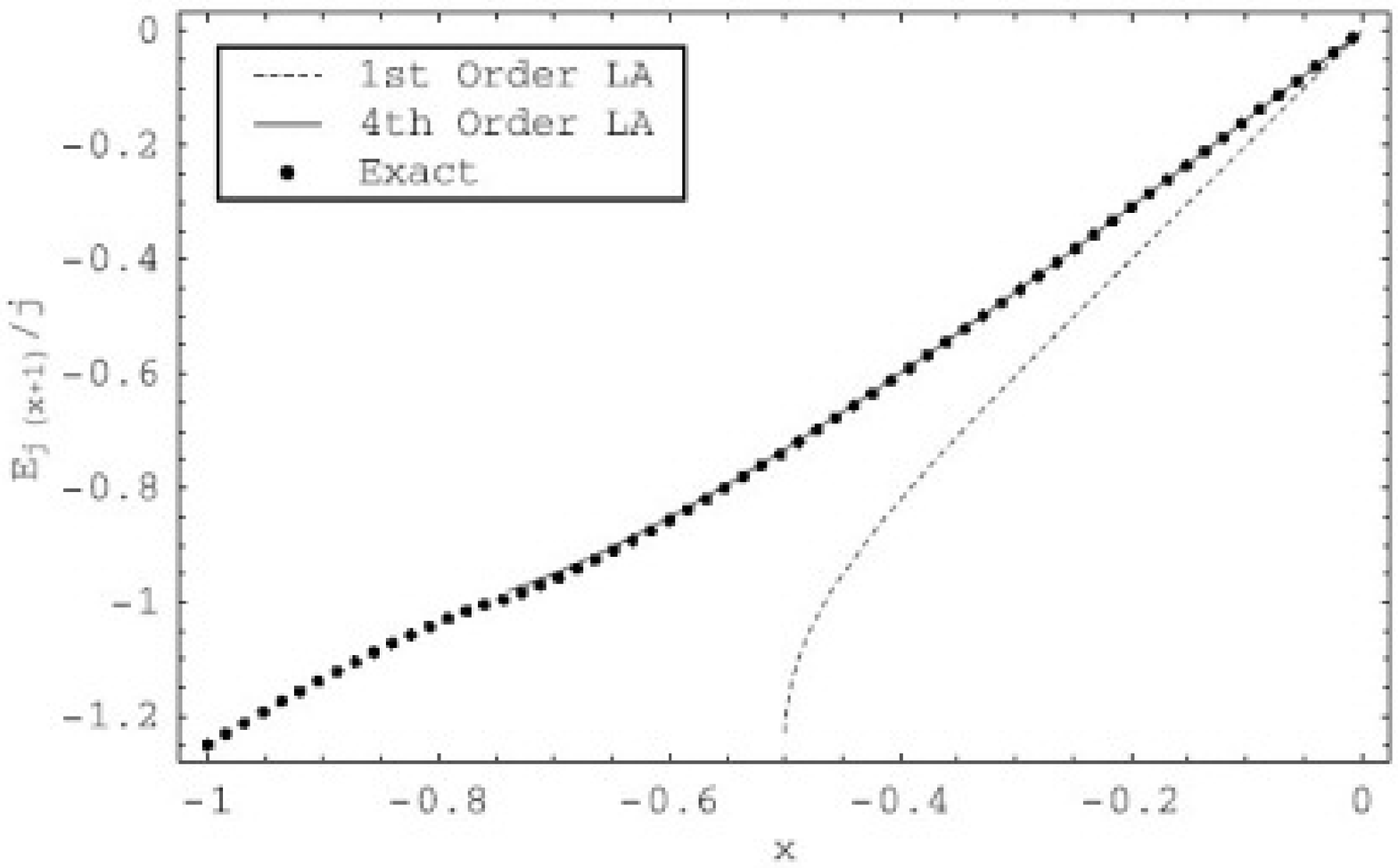,height=5cm,clip=,angle=0} \\
 	(a) & (b)\\
	\end{tabular}
\end{narrow}
	\caption[Results of the local approximation]{The results of the local approximation (LA) for (a) $\lambda=1$ and (b) $\lambda=2$ together with the exact results for $j=1000$. The dashed line corresponds to the first order solution given in \eref{local1stOrder}. Higher order local solutions, obtained numerically, are also shown.}
	\label{figset7}
\end{figure}

Figure \ref{figset7} shows this analytic result together with the exact eigenvalues. While it reproduces the spectrum of the first phase well, the local solution fares progressively worse at stronger coupling. This does not signal a lack of locality in the flow, but simply reflects the low order to which the flow equation was solved. The inclusion of higher order terms $\sim(x-c)^n$ in the solution leads to much improved results, as is clear from the figure. Note that the point of breakdown for the fourth order solution in (b) has moved closer to the actual phase boundary at $x(\lambda=2)\approx-0.75$ which separates regions of first and second phase character. 

In conclusion, we have seen that the local approximation provides non-perturbative results for states which possess a first phase character. In particular, the linear case can be solved analytically and reproduces the entire spectrum of the normal phase to good accuracy; the first such result that we are aware of. 
\chapter{The Dicke Model}
\section{Introduction}
Since its introduction in 1954 in the study of collective phenomenon in quantum optics, the Dicke model \cite{Dicke} has received considerable attention in a wide range of fields, including that of quantum chaos and quantum phase transitions. We refer the reader to \cite{Emary} for an extensive list of references to these and related studies. Here we present a flow equation treatment of the Dicke model. Other boson and spin-boson problems have been considered in this context \cite{Kleff,Kehrein}, and a special case of the Dicke model was treated perturbatively in \cite{Contreras}. Vidal and Dusuel \cite{Vidal3} have calculated finite-size scaling exponents for various quantities in the context of the Dicke model using flow equations. Compared to the Lipkin model this case presents a greater challenge to the flow equation approach, mainly due to the presence of two independent degrees of freedom. In order to keep the equations manageable we implement a two step procedure, first bringing the Hamiltonian into a block diagonal form before diagonalizing it completely. Unfortunately the complexity of the resulting PDE still prevents us from performing the first step exactly, and so a perturbative approach will be followed.

We begin with an overview of the model and its structure relevant to the flow equation treatment. After introducing the necessary variables and representations we digress briefly to first treat a special case analytically before proceeding with the derivation of the full flow equation. Using this equation we will construct an effective form of Hamiltonian in which certain degrees of freedom have been decoupled. Finally the different approaches to diagonalizing this effective form are introduced and compared using numerical results. 

\section{The model}
\label{DickeModelOverview}
The Dicke model describes the interaction of $N$ two-level atoms with a number of bosonic fields via a dipole interaction. For simplicity we will take $N$ to be even; the odd case requires only minor modifications. Following \cite{Emary}, we consider one such bosonic mode with frequency $\omega_1$ and coupling strength $\lambda$. The corresponding bosonic creation and annihilation operators are $b^\dag$ and $b$. Associated with each atom are the spin-1/2 operators $\left\{s^{(i)}_+,s^{(i)}_-,s^{(i)}_z\right\}$ which obey the standard $\texttt{su}(2)$ commutation relations. All $N$ atoms have equal level splitting $\omega_0$. The Dicke Hamiltonian reads
\begin{equation}
	H=\omega_0\sum_{i=1}^Ns_z^{(i)}+\omega_1b^\dag b+\frac{\lambda}{\sqrt{N}}\sum_{i=1}^N\left(s_+^{(i)}+s_-^{(i)}\right)\left(b^\dag+b\right).
\end{equation}
The $1/\sqrt{N}$ factor ensures that the Hamiltonian remains extensive when the bosonic mode is macroscopically occupied, i.e. when $\ave{b^\dag b}\sim N$. By introducing the collective spin operators $J_z=\sum_{i=1}^Ns_z^{(i)}$ and $J_\pm=\sum_{i=1}^Ns_\pm^{(i)}$ we obtain the simplified form
\begin{equation}
	H=\omega_0J_z+\omega_1b^\dag b+\frac{\lambda}{\sqrt{N}}\left(J_++J_-\right)\left(b^\dag+b\right).
	\label{dickeH}
\end{equation}
\begin{figure}[t]
\begin{center}
\epsfig{file=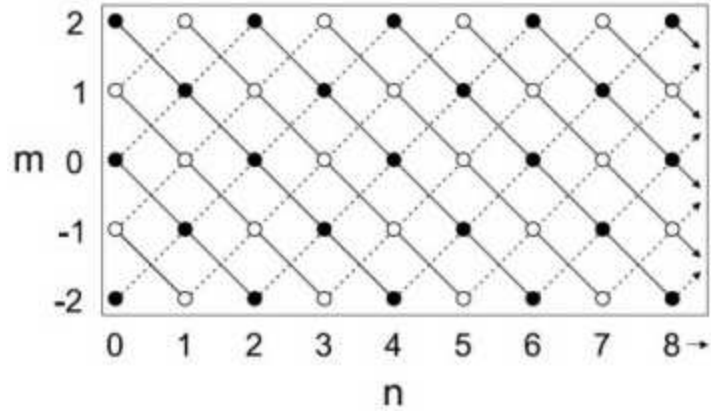,height=5.5cm,clip=,angle=0}
\caption[A schematic representation of the basis states spanning the Dicke model Hilbert space.]{A schematic representation adapted from \cite{Emary} of the basis states \newline $\{\ket{m,n}\ :\ m=-j,\ldots,j;\ n=0,1,2,\ldots\}$ for $j=2$. Filled and hollow circles correspond to even and odd parity states respectively. Solid lines connect states belonging to the same $Q$-sector while dashed lines run between states connected by the $J_+b^\dag+J_-b$ term of the Hamiltonian.}
\label{dickeBasisPlot}
\end{center}
\end{figure}
A natural basis for the Hilbert space ${\cal H}$ is given by the eigenstates of the total collective spin operator $J^2=J_z^2+J_z+J_-J_+$ together with $J_z$ and the boson number operator $\hat{n}=b^\dag b$:
\begin{equation}
	{\cal H}={\rm span}\left\{\ket{j,m,n}\ :\ j=0,\ldots,N/2;\ \ m=-j,\ldots,j;\ \ n=0,1,2,\ldots\right\}.
\end{equation}
Here $J^2\ket{j,m,n}=j(j+1)\ket{j,m,n}$, $J_z\ket{j,m,n}=m\ket{j,m,n}$ and $\hat{n}\ket{j,m,n}=n\ket{j,m,n}$. Since $\comm{H}{J^2}=0$ the Hamiltonian does not mix different $j$-sectors of the total spin representation. The relevant sector for the low-lying states is $j=N/2$, and we fix $j$ at this value throughout, dropping the $j$ label in the basis states. We will focus on the resonant case where $\omega_0=\omega_1=1$. As before, other cases simply correspond to different initial conditions for the equations we will derive. For these parameter values the model undergoes a quantum phase transition at $\lambda_c=1/2$ from the normal to the so-called super-radiant phase. The second phase is characterized by the macroscopic occupation of the bosonic mode in the ground state, i.e. $\ave{\hat{n}}\sim j$.\\

Next we consider the structure of $H$ relevant to our application of the flow equations. Central to this discussion is the operator $Q=J_z+\hat{n}+j$. With each distinct eigenvalue  $q=0,1,2,\ldots$ of $Q$ we associate the corresponding subspace of eigenstates 
\begin{equation}
	Q_q={\rm span}\left\{\ket{k,q-k-j}\ :\ k=-j,-j+1,\ldots,{\rm min}(j,q-j)\right\},
\end{equation}
which we refer to as a $Q$-sector. The dimension of $Q_q$ increases linearly from a minimum of one at $q=0$ to a maximum value of $2j+1$ at $q=2 j$, from where it remains constant. A schematic representation of the basis states and $Q$-sectors appears in Figure \ref{dickeBasisPlot}. Grouping the terms in the Hamiltonian according to equation \eref{hform} leads to
\begin{equation}
H=\underbrace{J_z+\hat{n}+\frac{\lambda}{\sqrt{2j}}\left(J_+b+J_-b^\dag\right)}_{T_0}+\underbrace{\frac{\lambda}{\sqrt{2j}}\left(J_+b^\dag+J_-b\right)}_{T_2+T_{-2}},
\label{dickeH2}
\end{equation}
which makes it clear that $H$ possesses a band block diagonal structure with respect to $Q$, as defined in Section \ref{etaChoice}. Since $H$ only changes $Q$ by two it leaves the subspaces corresponding to odd or even $Q$ invariant. This implies a parity symmetry, similar to that of the Lipkin model, which may be compactly expressed in terms of the operator $\Pi=\exp{[i\pi Q]}$ as $\comm{H}{\Pi}=0$. This symmetry gets broken in the second phase where the ground states of the odd and even sectors become degenerate. 

Early studies \cite{Lieb,Wang} of the model's thermodynamic properties were performed in the rotating wave approximation (RWA) which amounts to dropping the $T_2+T_{-2}$ term. We will consider this case in more detail later on, as it reappears as the first order contribution to the solution of the flow equation. 

\section{The flow equation approach}
We have seen that the Hamiltonian contains interactions responsible for scattering both between different $Q$-sectors and within a fixed sector. A complete diagonalization of $H$ would require a transformation which eliminates both these interactions. Although the flow equation is certainly capable of generating such a transformation in a single application, it is conceptually clearer and technically much simpler to perform this diagonalization in two separate steps. As suggested in the previous section we will first eliminate  interactions which connect different $Q$-sectors through the structure preserving flow generated by $\eta(\ell)=\comm{Q}{H(\ell)}$. As before this is done in the thermodynamic or large $j$ limit. A physical interpretation of $Q$ is that of a counting operator for the elementary excitations or energy quanta present in the system. The flow will produce an effective Hamiltonian $H_{eff}\equiv H(\infty)$ which conserves the number of these excitations. We expect there to be new terms, both diagonal and off-diagonal, appearing in $H_{eff}$. Apart from the provision that they conserve $Q$ there is no obvious constraint on these newly generated interactions, and so no band (block) diagonal structure need be present within the $Q$-sectors.

Having brought $H$ into a block diagonal form there are two ways in which to proceed. We may construct, for large but finite $j$, the matrices corresponding to each $Q$-sector and then apply direct diagonalization. Although still numerically intensive this is a greatly reduced problem compared to diagonalizing the original fully interacting Hamiltonian, as the submatrices are at most of size $(2 j+1)\times (2j+1)$. Alternatively we may  remain at infinite $j$ and apply the flow equations to each sector separately using, for example, the generator $\eta(\ell)=\comm{J_z}{H(\ell)}$. For a fixed value of $Q$ the problem now becomes effectively one dimensional as we may eliminate either the bosonic or spin degree of freedom in favor of $Q$, which acts as a scalar parameter. Both these methods will be demonstrated later on. 

Finally, a remark on the ordering of eigenvalues. When applying the flow equations to each sector in the second step we expect complete diagonalization and that the eigenvalues will appear on the diagonal in increasing order \cite{Brockett}. For the first step, which ends in a block diagonal form, the situation is no longer so clear, as the ordering proof in \cite{Brockett} is only valid for the case of complete diagonalization. In short, it is generally unknown in which sector a certain eigenstate of $H$ will be found when we diagonalize $H_{eff}$. In the first phase we expect the same locality encountered in the Lipkin model (Section \ref{firstPhaseSol}) and so the low-lying states should belong to the sectors with $Q\ll j$. In numerical investigations for small $j$ values we have observed this ordering in both phases. In particular, the ground state is mapped to the single basis state of the $Q=0$ subspace. As this ordering is a non-perturbative phenomenon we do not expect to observe it in our perturbative treatment.

\section{Flow equations in the $j\rightarrow\infty$ limit}
In the subsections that follow we derive the flow equation for the first step of the diagonalization process using the Moyal bracket method. We begin by introducing the relevant variables and representations.
\subsection{Variables and representations}
To account for the model's two independent degrees of freedom we introduce two pairs of operators $g,h$ and $g',h'$ as described in Section \ref{methodMoyal}. These satisfy the exchange relations
\begin{equation}
	hg=e^{-i\theta}gh\ \ \ {\rm and}\ \ \  h'g'=e^{-i\theta'}g'h'
\end{equation}
while operators coming from different pairs commute. The pair $g,h$ is used to represent the spin degree of freedom through the representation constructed in Section \ref{lipkinMoyal} in terms of $x_s=\alpha/(j \theta)-1\in[-1,1]$ and $\beta$:
\begin{equation}
J_+=je^{i\beta}\sqrt{1-x_s^2},\ \ J_-=je^{-i\beta}\sqrt{1-x_s^2}\ \ {\rm and}\ \ J_z=jx_s.
\end{equation}
In similar fashion we wish to construct a representation for the boson algebra $\{b^\dag,b\}$ in terms of $g'$ and $h'$. Two new issues, not encountered in the $\texttt{su}(2)$ case, arise here. Firstly, it is well known that only infinite dimensional representations of the boson algebra exist, whereas $g'$ and $h'$ are finite dimensional. The second point concerns the scale we should associate with the bosonic operators. If $b^\dag$  and $b$ were naively assumed to be scaleless with respect to $j$ (or $N$), only the $J_z$ term in the Hamiltonian would survive when working to leading order in $j$. We will address these issues in the context of our proposed approximate representation
\begin{equation}
	b={h'}^\dag \sqrt{\bar{g}'},\ \ \ \ b^\dag=\sqrt{\bar{g}'}h'\ \ \ {\rm and}\ \ \  b^\dag b=\hat{n}=\bar{g}'
	\label{bosonRep1}
\end{equation}
where $\bar{g}'=-i\log(g')/\theta'={\rm diag}(0,1,2,\ldots,D'-1)$ and $D'$ is the dimension of the space. Denoting the eigenstates of $\bar{g}'=\hat{n}$ by $\ket{n}\ n=0,1,\ldots,D'-1$ we see that $b\ket{n}=\sqrt{n}\ket{n-1}$ and $b^\dag\ket{n}=\sqrt{n+1}\ket{n+1}$ for $0\leq n<D'-1$. (See equations \eref{gEV} and \eref{gLadder}.) The creation operator $b^\dag$ maps the highest state $\ket{D'-1}$ to zero, since $h'\ket{D'-1}=\ket{0}$. This amounts to a truncation of the boson Fock-space, and the operators in \eref{bosonRep1} agree with the truncated forms of the exact infinite-dimensional operators. Proceeding as before, we treat $g'$ and $h'$ as scalars and define $\alpha'$ and $\beta'$ through $g'=e^{i\alpha'}$ and $h'=e^{i\beta'}$. The representation now becomes 
\begin{equation}
	b=e^{-i \beta'}\sqrt{\alpha'/\theta'+1},\ \ \ \  b^\dag=e^{i \beta'}\sqrt{\alpha'/\theta'}\ \ \ {\rm and}\ \ \ b^\dag b=\alpha'/\theta'
\end{equation}
which satisfies, to leading order in $\theta'$, the desired commutation relation with respect to the bosonic Moyal bracket $\comm{b}{b^\dag}_\ast=1$. For the moment we set $D'=uj+1$, and return to the role of $u$ later. The second issue concerns the scale of $j$ to be associated with $b$ and $b^\dag$. This is really a question of which values of $\alpha'/\theta'$ are relevant to the low energy physics of the system. In the Dicke model it is known \cite{Emary} that ground state occupation of the bosonic mode is microscopic, i.e. $\ave{\hat{n}}\sim {\cal O}(j^0)$, in the first phase and macroscopic, i.e. $\ave{\hat{n}}\sim {\cal O}(j)$, in the second. We can describe both these cases by considering $b^\dag b=\alpha'/\theta'\propto j$. A natural choice of variables is $x_b=\alpha'/(j \theta')\in[0,u]$ which is scaleless and analogous to $x_s$ of the $\texttt{su}(2)$ case. The $j$-scale now becomes explicit in both the representation
\begin{equation}
	b=\sqrt{j}e^{-i \beta'}\sqrt{x_b+1/j},\ \ \ \ b^\dag=\sqrt{j}e^{i \beta'}\sqrt{x_b}\ \ \ {\rm and}\ \ \ b^\dag b=jx_b
\end{equation}
and bosonic Moyal bracket
\begin{equation}
	\comm{U(x_b,\beta')}{V(x_b,\beta')}_\ast=i\left(U_{\beta'} V_{x_b}-V_{\beta'} U_{x_b}\right)/j.
\end{equation}
We also observe that the dimension $D'=uj+1$, which controls the Fock-space cutoff, does not appear explicitly. For practical purposes we may safely assume $u$ to be much larger than the interval of $x_b$ under consideration.\\

Combining the two representations we obtain the initial condition to leading order in $j$ as
\begin{equation}
H(x_s,\beta,x_b,\beta',\ell=0)=j\left[x_s+x_b+\lambda\sqrt{2x_b(1-x_s^2)}\cos{(\beta-\beta')}+\lambda\sqrt{2x_b(1-x_s^2)}\cos{(\beta+\beta')}\right],
\label{initkond}
\end{equation}
where the terms are ordered as in equation \eref{dickeH2}. 
\subsection{Solution in the local approximation}
\label{dickeLocal}
Before proceeding with the derivation of the general flow equation, which will involve significant behind-the-scenes numeric and symbolic computation, let us consider a case amenable to an analytic treatment. Using the same locality argument put forth for the Lipkin model in Section \ref{firstPhaseSol} we assert that for the low-lying states only the flow within a small region around $x_s=-1$ is relevant, and that the evolution of $H$ in this region is governed by its local properties. This is equivalent to the assumption that $\ave{J_z+j}\sim j^{0}$ for the low-lying states. A similar approximation is made in the bosonization treatment of \cite{Emary}, the results of which will be reproduced here using the flow equations. In practice we implement this approximation by simply replacing $\sqrt{1-x_s^2}$ by $\sqrt{2}\sqrt{1+x_s}$ in the initial condition \eref{initkond}; the result of a Taylor expansion to leading order in $x_s+1$. First we consider the flow generated by $\eta(\ell)=\comm{Q}{H(\ell)}$. As with the Lipkin model the local approximation leads to a very simple form for $H(\ell)$, with only two new terms, proportional to $b^2+(b^\dag)^2$ and $J_+^2+J_-^2$, being generated. As required the band block diagonal structure of $H$ is conserved. The flowing Hamiltonian may be parametrized as
\begin{eqnarray}
H(\ell)/j&=&-1+a_0(\ell)\left(x_s+x_b+1\right)+2\:a_1(\ell)\sqrt{x_b(1+x_s)}\cos{(\beta-\beta')}\label{localHform}\\
&&+2\:a_2(\ell)\sqrt{x_b(1+x_s)}\cos{(\beta+\beta')}+a_3(\ell)\left(x_b \cos{(2 \beta')}+(x_s+1) \cos{(2 \beta)}\right)\nonumber
\end{eqnarray}
where $a_i(\ell)$ $i=0,1,2,3$ are scalar coefficients. In operator language this amount to
\begin{eqnarray}
H(\ell)&=&-j+a_0(\ell)\left(j+J_z+\hat{n}\right)+a_1(\ell)\left[J_+b+J_-b^\dag\right]/\sqrt{j}
\\&&+a_2(\ell)\left[J_+b^\dag+J_-b\right]/\sqrt{j}+a_3(\ell)\left[((b^\dag)^2+b^2)+(J_+^2+J_-^2)/j\right]/2.\nonumber
\end{eqnarray}
Observables linear in $J_z$ and $\hat{n}$ may be parametrized in a similar fashion. We will consider $O=Q-j=J_z+\hat{n}$ for which $O(\ell)$ is of the form \eref{localHform} with the flowing coefficients denoted by $\bar{a}_i(\ell)$ for $i=0,1,2,3$.

Substituting these forms into the flow equations \eref{fullFE1} and \eref{fullFEO} and then matching the coefficients on both sides yield
\begin{equation}
		\afg{a_0}{\ell}=-a_2^2-a_3^2,\ \ \ \ \ \ \afg{a_1}{\ell}=-2a_2a_3,\ \ \ \ \ \ \afg{a_2}{\ell}=-a_0a_2-a_1a_3,\ \ \ \ \ \ \afg{a_3}{\ell}=-a_1a_2-a_0a_3
\end{equation}
and
\begin{equation}
		\afg{\bar{a}_0}{\ell}=-a_2 \bar{a}_2-a_3 \bar{a}_3,\ \ \ \ \ \afg{\bar{a}_1}{\ell}=-a_2\bar{a}_3-a_3\bar{a}_2,\ \ \ \ \ \afg{\bar{a}_2}{\ell}=-\bar{a}_0a_2-\bar{a}_1a_3,\ \ \ \ \ \afg{\bar{a}_3}{\ell}=-\bar{a}_1a_2-\bar{a}_0a_3.
\end{equation}
The initial conditions for $H(\ell)$ and $O(\ell)$ are $a_0(0)=1$, $a_1(0)=a_2(0)=\lambda$, $a_3(0)=0$ and 
$\bar{a}_0=1$, $\bar{a}_1=\bar{a}_2=\bar{a}_3=0$ respectively. 

First we consider $H(\ell)$. In the $\ell\rightarrow\infty$ limit $a_2$ and $a_3$ is expected to vanish, while $a_0$ and $a_1$ will assume new, renormalized values. The initial and final values of the $a_i$'s may be related using the flow invariants $I_1=a_0^2+a_1^2-a_2^2-a_3^2$ and $I_2=a_0a_1-a_2a_3$. By combining these with the initial conditions we find that
\begin{equation}
	a_0^2(\infty)+a_1^2(\infty)=1\ \ \ \ \ {\rm and}\ \ \ \ \ a_0(\infty)a_1(\infty)=\lambda.
	\label{hypercircle}
\end{equation}
The renormalized values of $a_0$ and $a_1$ correspond to a point where the circle and hyperbole described by these equations intersect. For $\lambda>0.5$ no such point exists, and the solution breaks down. For $\lambda\in[0,0.5]$ the fixed point is found to be
\begin{equation}
	a_0(\infty)=\left(\sqrt{1+2\lambda}+\sqrt{1-2\lambda}\right)/2\ \ \ \ \ {\rm and}\ \ \ \ \  a_1(\infty)=\left(\sqrt{1+2\lambda}-\sqrt{1-2\lambda}\right)/2.
\end{equation}
\begin{figure}[t]
\begin{center}
\begin{tabular}{ccc}
	\epsfig{file=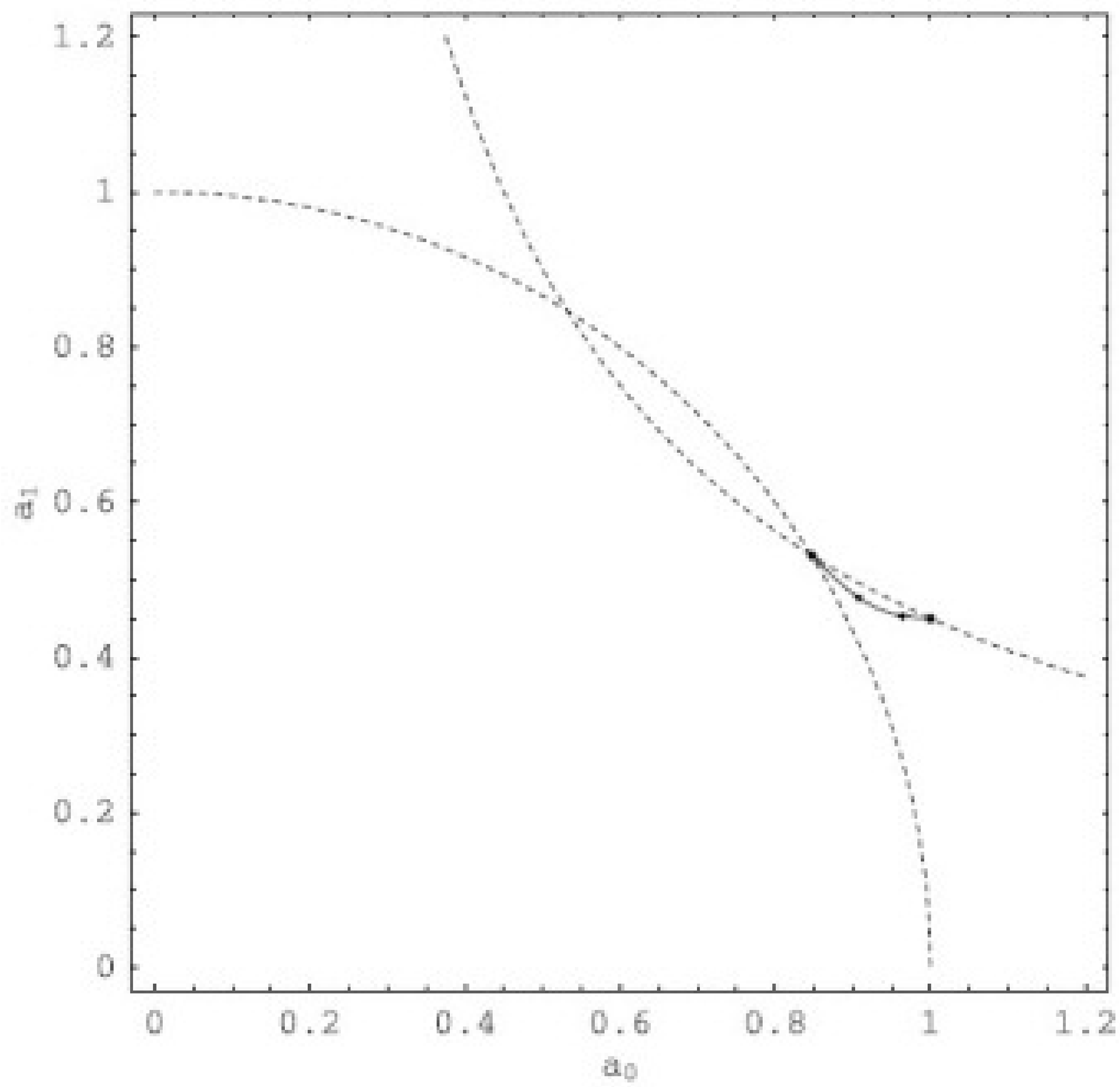,height=6.5cm,clip=,angle=0} & \makebox[0.5cm]{ }&
	\epsfig{file=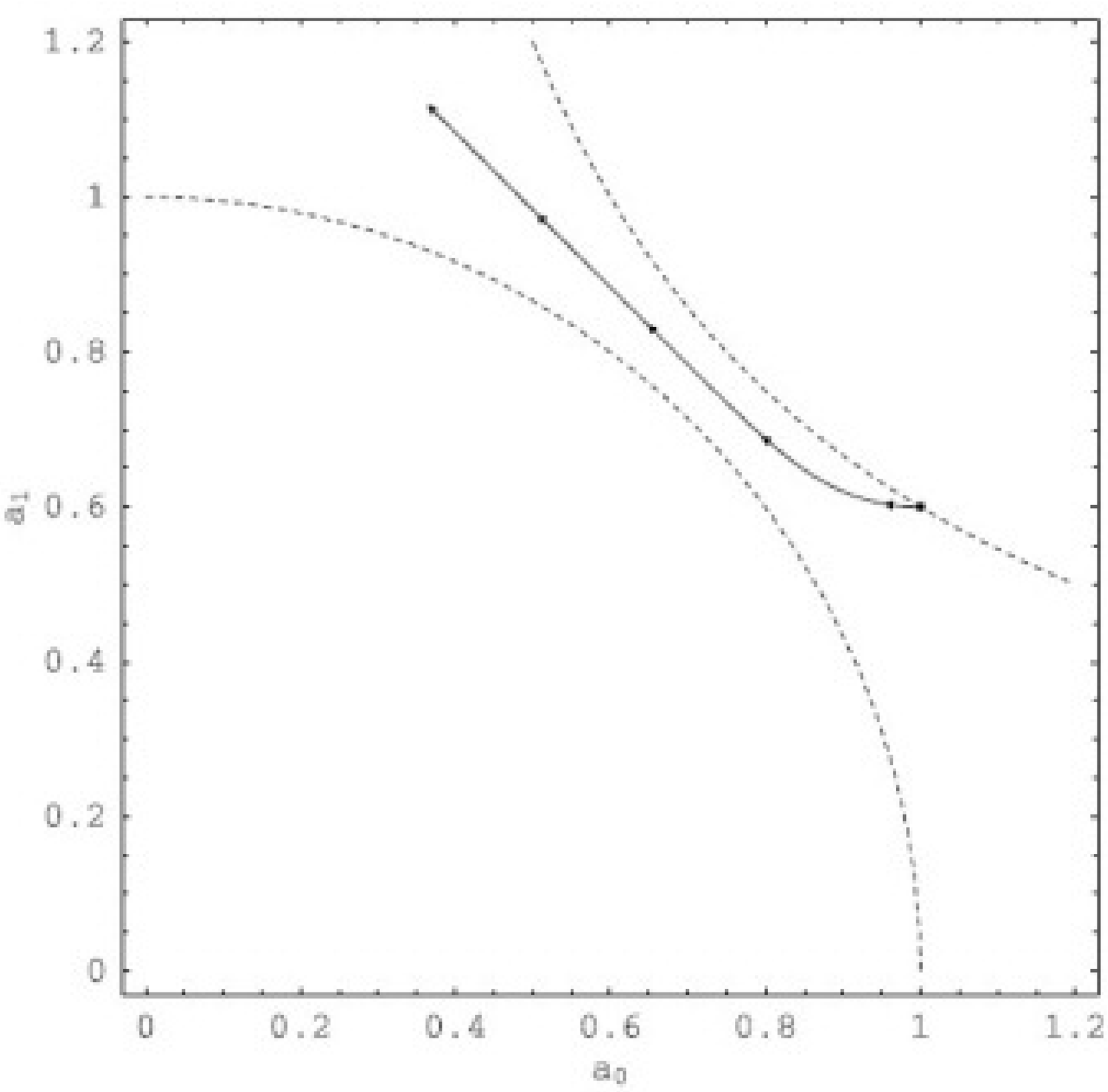,height=6.5cm,clip=,angle=0} \\
	(a) &   & (b)
	\end{tabular}
	\caption[Flow of the coupling constants $a_0(\ell)$ and $a_1(\ell)$ in the local approximation.]{The flow of $(a_0(\ell),a_1(\ell))$ for (a) $\lambda=0.45$ and (b) $\lambda=0.6$. The dashed lines represent the circle and hyperbole described by equations \eref{hypercircle}. In (b) the flow is only shown up to some finite value of $\ell$, as no convergence takes place.}
	\label{a0a1flow}
\end{center}
\end{figure}
This is illustrated in Figure \ref{a0a1flow} which shows the flow of $(a_0(\ell),a_1(\ell))$ for $\lambda=0.45$ and $\lambda=0.6$. Next we consider the flow of $O(\ell)$. To fix the final values of the $\bar{a}_i$ coefficients we require four invariants, some of which must involve the coefficients of $H(\ell)$. One such set is
\begin{eqnarray}
	&&\bar{I}_1=\bar{a}_0^2+\bar{a}_1^2-\bar{a}_2^2-\bar{a}_3^2,\ \ \ \ \ \bar{I}_2=\bar{a}_0\bar{a}_1-\bar{a}_2\bar{a}_3,\ \ \ \ \ \bar{I}_3=\bar{a}_0 a_0+\bar{a}_1 a_1-\bar{a}_2 a_2-\bar{a}_3 a_3 \nonumber\\
	 &&{\rm and}\ \ \ \ \ \bar{I}_4=\bar{a}_0 a_1+\bar{a}_1 a_0-\bar{a}_2 a_3-\bar{a}_3 a_2.
\end{eqnarray}
Combining $\bar{I}_3$ and $\bar{I}_4$ with the initial conditions lead to
\begin{eqnarray}
	a_0(\infty) \bar{a}_0(\infty)+a_1(\infty)\bar{a}_1(\infty)=a_0(0)=1 \nonumber\\
	a_1(\infty) \bar{a}_0(\infty)+a_0(\infty)\bar{a}_1(\infty)=a_1(0)=\lambda
\end{eqnarray}
and so
\begin{eqnarray}
\bar{a}_0(\infty)=\frac{1}{2}\left(\frac{1+\lambda}{\sqrt{1+2\lambda}}+\frac{1-\lambda}{\sqrt{1-2\lambda}}\right) \nonumber \\
\bar{a}_1(\infty)=\frac{1}{2}\left(\frac{1+\lambda}{\sqrt{1+2\lambda}}-\frac{1-\lambda}{\sqrt{1-2\lambda}}\right).
\end{eqnarray}
The values of $\bar{a}_2(\infty)$ and $\bar{a}_3(\infty)$ can now be solved for using $\bar{I}_1$ and $\bar{I}_2$. Note that for the purposes of calculating expectation values with respect to the eigenstates of $H$ only $\bar{a}_0(\infty)$ and $\bar{a}_1(\infty)$ are relevant. This follows from the observation that $\bar{a}_2$ and $\bar{a}_3$ correspond to terms that change $Q$ by two, whereas the eigenstates of $H_{eff}$ are also eigenstates of $Q$. For simplicity we drop these ``between-sector" scattering terms in what follows, keeping only those terms relevant to the calculation of expectation values.\\

The successful construction of the effective $Q$-preserving Hamiltonian $H_{eff}=H(\infty)$ and the observable $O_{eff}=O(\infty)$ marks the end of the first step. We now apply the flow equation a second time, using the generator $\eta(\ell)=\comm{J_z}{H_{eff}(\ell)}$. When restricted to a single $Q$-sector $J_z$ is clearly non-degenerate, and so we expect complete diagonalization in the final Hamiltonian. The local approximation again allows for a simple parametrization
\begin{equation}
	H_{eff}(\ell)/j=-1+b_0(\ell) (x_s+1)+b_1(\ell)x_b+2b_2(\ell)\sqrt{x_b(1+x_s)}\cos{(\beta-\beta')},
\end{equation}
where $b_0(0)=b_1(0)=a_0(\infty)$ and $b_2=a_1(\infty)$. The flow equations for the coefficients are
\begin{equation}
	\afg{b_0}{\ell}=2b_2^2,\ \ \ \ \ \afg{b_1}{\ell}=-2b_2^2\ \ \ \ {\rm and}\ \ \ \ \ \afg{b_2}{\ell}=-(b_0-b_1)b_2
\end{equation}
which leave $I'_1=b_0+b_1$ and $I'_2=b^2_0+b^2_1+2b_2^2$ invariant. Combining these with the initial conditions and using the fact that $b_2(\infty)=0$ leads to
\begin{eqnarray}
	b_0(\infty)&=&a_0(\infty)+a_1(\infty)=\sqrt{1+2\lambda} \nonumber\\
	 b_1(\infty)&=&a_0(\infty)-a_1(\infty)=\sqrt{1-2\lambda}.
\end{eqnarray}
Turning to $O_{eff}(\ell)$ we find that its parametrized form is again similar to that of $H_{eff}(\ell)$, and we denote the flowing coefficients by $\bar{b}_i(\ell)$. The flow equations are found to be
\begin{equation}
	\afg{\bar{b}_0}{\ell}=2b_2 \bar{b}_2,\ \ \ \ \ \afg{\bar{b}_1}{\ell}=-2b_2 \bar{b}_2\ \ \ \ {\rm and}\ \ \ \ \ \afg{\bar{b}_2}{\ell}=-(\bar{b}_0-\bar{b}_1)b_2
\end{equation}
which leave $\bar{I}'_1=\bar{b}_0+\bar{b}_1$ and $\bar{I}'_2=\bar{b}_0b_0+\bar{b}_1b_1+2\bar{b}_2b_2$ invariant. Proceeding as before we find that $\bar{b}_0(\infty)=(1+\lambda)/\sqrt{1+2\lambda}$ and  $\bar{b}_1(\infty)=(1-\lambda)/\sqrt{1-2\lambda}$. Since the eigenstates of $H_{eff}(\infty)$ are also eigenstates of $J_z$ we drop the $\bar{b}_2$ term.

In summary, we have seen that within the local approximation the two step diagonalization procedure can be performed exactly. The final Hamiltonian and transformed observable $O=J_z+\hat{n}$ is 
\begin{eqnarray}
H_{eff}(\infty)&=&-j+\left(j+J_z\right)\sqrt{1+2\lambda}+\hat{n}\sqrt{1-2\lambda} \nonumber \\
O_{eff}(\infty)&=&-j+\left(j+J_z\right)\frac{1+\lambda}{\sqrt{1+2\lambda}}+\hat{n}\frac{1-\lambda}{\sqrt{1-2\lambda}}+\ldots
\end{eqnarray}
which agrees with the result of \cite{Emary} obtained using bosonization. This clearly constitutes a harmonic approximation of the spectrum in terms of two oscillators with frequencies $\sqrt{1+2\lambda}$ and $\sqrt{1-2\lambda}$. Only the parts of $O_{eff}$ which commute with both $Q$ and $J_z$ are shown.  Note that the gap between the ground state $\ket{E_0}=\ket{-j,0}$ and first excited state $\ket{E_1}=\ket{-j,1}$ exhibits the characteristic square root behaviour $\Delta=\sqrt{1-2\lambda}$. 

\subsection{Flow equation for the Hamiltonian}
We proceed with the derivation of the general flow equation for the Hamiltonian, beginning with a more in depth study of the structure of $H(\ell)$. In Section \ref{etaChoice} it was shown that the flow generated by $\eta(\ell)=\comm{Q}{H(\ell)}$ will preserve the band block diagonal structure present in $H(0)$, thus restricting the terms in $H(\ell)$ to those changing $Q$ by either zero or two\footnote{In this equation, and in some that follow, we treat operators as commuting scalars. Since reordering can only bring about ${\cal O}(1/j)$ corrections this is sufficient for our purposes, and it simplifies the notation considerably.} 
\begin{equation}
	H(\ell)=f_0(J_z,\hat{n},\ell)+f_1(J_z,\hat{n},\ell)\left(J_+b^\dag+J_-b\right)+\sum_{\substack{k=1 \\ l\in\{k,k\pm2\}, k+l>2}}^\infty f_{k,l}(J_z,\hat{n},\ell)\left((J_+)^k (b)^l+(J_-)^k (b^\dag)^l\right).
	\label{dickeHLank}
\end{equation}
A simpler and much more convenient form can be found by introducing the operators $T_q=J_+b+J_-b^\dag$ and $${\cal F}_2=\left\{ T_s=J_+^2+J_-^2,\;T_b=(b^\dag)^2+b^2,\;T_{sb}=J_+b^\dag+J_-b\right\}.$$ We interpret $T_q$ as the fundamental $Q$-preserving interaction while the elements of ${\cal F}_2$ constitute the fundamental interactions which change $Q$ by two. As shown in Appendix \ref{Appdecomp} we can rewrite all the operators appearing in \eref{dickeHLank} in terms op $J_z$, $\hat{n}$, $T_q$ and the elements of ${\cal F}_2$ which only appear linearly. For example:
\begin{eqnarray}
	 J_+b^3+J_-(b^\dag)^3&=&T_qT_b-\hat{n}T_{sb}\nonumber\\
	 (J_+)^2b^2+(J_-)^2(b^\dag)^2&=&T_q^2+2\hat{n}(J_z^2-j^2)\nonumber\\	(J_+)^4b^2+(J_-)^4(b^\dag)^2&=&\left(\left(J_z^2-j^2\right)\hat{n}+T_q^2\right)T_s+\left(J_z^2-j^2\right)T_qT_{sb}.
\end{eqnarray}
This results in the form
\begin{equation}
H(\ell)=j\left[F_0(J_z,\hat{n},T_q,\ell)+F_1(J_z,\hat{n},T_q,\ell)T_s/j^2+F_2(J_z,\hat{n},T_q,\ell)T_b/j+F_3(J_z,\hat{n},T_q,\ell)T_{sb}/j^{3/2}\right].
\label{dickeHKort}
\end{equation}
The factor of $j$ responsible for the extensivity of $H$ has been factored out explicitly, and so each $F_i$ is a scaleless function. The initial condition is given by 
\begin{equation}
	F_0=x_s+x_b+\frac{\lambda}{\sqrt{2}}x_q,\ \ \ \ F_1=F_2=0\ \ \ \ {\rm and}\ \ \ \ F_3=\frac{\lambda}{\sqrt{2}}
\end{equation}
where $x_q=T_q/j^{3/2}$. We proceed by inserting this form into the flow equation
\begin{equation}
\afg{H}{\ell}=\comm{\comm{Q}{H}_\ast}{H}_\ast
\label{fullFE1}
\end{equation}
where $\comm{\cdot}{\cdot}_\ast$ is the full Moyal bracket
\begin{equation}
	\comm{U}{V}_\ast=i\left(U_\beta V_{x_s}-V_\beta U_{x_s}+U_{\beta'} V_{x_b}-V_{\beta'} U_{x_b}\right)/j.
	\label{fullFE2}
\end{equation}
The resulting algebra is easily handled using a symbolic processor such as \textit{Mathematica}. Matching the coefficients of $T_s$, $T_b$ and $T_{sb}$ on both sides of \eref{fullFE1} produces a set of equations of the form
\begin{equation}
\pafg{F_i}{\ell}=F^{(v)}_l\Lambda^{(i)}_{(v,l),(v',l')}F^{(v')}_{l'}\ \ \ i=0,1,2,3
\label{dickeFEH}
\end{equation}
where we sum over repeated indices. The superscripts of the $F_i$'s denote derivatives to the scaleless variables $x_s=J_z/j$, $x_b=\hat{n}/j$ and $x_q=T_q/j^{3/2}$. The $v, v'$ indices run over the set $\left\{0,x_s,x_b,x_q\right\}$ where $0$ corresponds to no derivative being taken. Governing the flow of $F_i$ is the $16\times16$ matrix $\Lambda^{(i)}_{(v,l),(v',l')}$ of which the entries are simple polynomials of the rescaled variables. The non-zero entries of these, typically sparse matrices appear in Appendix \ref{Appflowcoef}. Due to the obvious complexity of these non-perturbative equations we have been unsuccessful in finding an exact numerical solution. A perturbative solution can be found, and we present this in Section \ref{dickePSol}.

\subsection{Flow equation for an observable}
Next we derive the flow equation for an observable $O$. This case is complicated by the lack of structure present in $O(\ell)$ at non-zero $\ell$. For example, we do not generally expect band block diagonality to be present, and so $O(\ell)$ may contain interactions connecting distant $Q$-sectors. In general we are confronted by the form
\begin{equation}
O(\ell)=\sum_{\substack{\Delta=0 \\ \Delta\ {\rm even}}}^\infty\sum_{p=0}^{\Delta}\bar{F}_{(\Delta,p)}(J_z,\hat{n},T_q,\ell)\;\frac{T_{(\Delta,p)}}{j^{(\Delta+p)/2}}
	\label{dickeOKort}
\end{equation}
where $T_{(\Delta,p)}=(J_+)^p(b^\dag)^{\Delta-p}+(J_-)^p(b)^{\Delta-p}$ and $T_{(0,0)}\equiv1$. The $\Delta$ label indicates the amount by which the term changes $Q$, and similar for $p$ with respect to $J_z$. We define the index set ${\cal I}=\left\{(\Delta,p)\ :\ \Delta=0,1,2,\ldots;\ p=0,\ldots,\Delta\right\}$. Inserting this form into the flow equation
\begin{equation}
\afg{O}{\ell}=\comm{\comm{Q}{H}_\ast}{O}_\ast
\label{fullFEO}
\end{equation}
yields the coupled set
\begin{equation}
\pafg{\bar{F}_i}{\ell}=F^{(v)}_l\bar{\Lambda}^{(i)}_{(v,l),(v',l')}\bar{F}^{(v')}_{l'}\ \ \ \ i\in{\cal I}.
\label{dickeFEO}
\end{equation}
We sum over the repeated indices $l\in\{0,1,2,3\}$, $l'\in{\cal I}$ and $v,v'\in\left\{0,x_s,x_b,x_q\right\}$. $\bar{\Lambda}^{(i)}_{(v,l),(v',l')}$ is a $16\times\infty$ dimensional matrix containing polynomial functions of the scaleless variables. Note that at $\ell=\infty$ only the $F_{(0,0)}$ function is of interest for the calculation of expectation values. This follows from the observation that eigenvalues of $H(\infty)$ are also eigenvalues of $Q$, and that $F_{(0,0)}$ is the only term leaving $Q$ invariant.

Again it is clear that a direct numerical approach is intractable, and we instead try to find perturbative solutions.

\section{Perturbative solutions}
\label{dickePSol}
We wish to construct solutions to equations \eref{dickeFEH} and \eref{dickeFEO} of the forms
\begin{equation}
	H(\ell)=\sum_{p=0}^{p_{max}}\lambda^pH^{(p)}(\ell)\ \ \ {\rm and}\ \ \ O(\ell)=\sum_{p=0}^{p_{max}}\lambda^pO^{(p)}(\ell).
\end{equation}
The forms given in \eref{dickeHKort} and \eref{dickeOKort} are still valid, as they apply separately to each $H^{(p)}$ and $O^{(p)}$. When working to finite order in $\lambda$ the functional dependence of $H(\ell)$ and $O(\ell)$ on $\{x_b,x_s,x_q\}$ are constrained to be polynomial and of finite order. This also limits the type of interactions that can be generated in $O(\ell)$ since terms for which $\Delta=2n$ is at least of order $\lambda^n$, and so only $\bar{F}_{(\Delta,k)}$ with $\Delta\leq\left\lfloor p_{max}/2\right\rfloor$ are relevant. This is true at $\ell=0$, provided that $O\sim \lambda^0$ and $\comm{O}{Q}=0$, and continues to hold at $\ell>0$ since $\eta(\ell)$ is always at least of order $\lambda$ and contains only $\Delta=2$ terms. As all the relevant functions are simple polynomials we may proceed by constructing a set of coupled differential equations for the scalar coefficients appearing in these polynomials as functions of $\ell$. The extensive  algebraic manipulations involved in this step can be automated for arbitrary $p_{max}$ using \textit{Mathematica}. The resulting set of ordinary differential equations is then solved using the standard Runge-Kutta algorithm \cite{Burden}. Solutions were obtained for $H$ up to $p_{max}=20$ and for the observables $J_z$ and $\hat{n}$ up to $p_{max}=18$. These transformed operators will be denoted by $H_{eff}=H(\infty)$, $\hat{n}_{eff}=\hat{n}(\infty)$ and $(J_z)_{eff}=J_z(\infty)$. We only state the results up to fourth order here:
\begin{eqnarray}
	H^{(0)}(\infty)/j&=&\hat{n}+J_z\nonumber\\
	H^{(1)}(\infty)/j&=&T_q/\sqrt{2}\nonumber\\
	H^{(2)}(\infty)/j&=&\left(J_z^2+2\hat{n}J_z-1\right)/4\nonumber\\
	H^{(3)}(\infty)/j&=&-T_q\left(\hat{n}+4J_z\right)/(8 \sqrt{2})\nonumber\\
	H^{(4)}(\infty)/j&=&\left(7 T_q^2 + 20 J_z - 2 \hat{n} + 38 \hat{n}J_z^2+20 J_z^3\right)/64,
	\label{Hpseries}
\end{eqnarray}
\textit{where, to aid interpretation, we temporarily abuse notation by writing} $J_z,\hat{n},T_q$ \textit{for the scaleless variables} $x_s,x_b,x_q$. The rationality of the coefficients, apart from the $\sqrt{2}$ factors, have been verified to very high numerical accuracy. In similar fashion the observables are given by
\begin{eqnarray}
	J_z^{(0)}(\infty)/j&=&J_z+\ldots \nonumber\\
	J_z^{(1)}(\infty)/j&=&0+\ldots \nonumber\\
	J_z^{(2)}(\infty)/j&=&(1-2\hat{n}J_z-J_z^2)/4+\ldots \nonumber\\
	J_z^{(3)}(\infty)/j&=&(3\hat{n}+16J_z)T_q/(32 \sqrt{2})+\ldots \nonumber\\
	J_z^{(4)}(\infty)/j&=&\left(-18\hat{n} - 51T_q^2 - 228J_z - 12\hat{n}^2J_z + 282\hat{n}J_z^2 + 228J_z^3\right)/384+\ldots
	\label{Jzpseries}
\end{eqnarray}
and
\begin{eqnarray}
	\hat{n}^{(0)}(\infty)/j&=&\hat{n}+\ldots \nonumber\\
	\hat{n}^{(1)}(\infty)/j&=&0+\ldots \nonumber\\
	\hat{n}^{(2)}(\infty)/j&=&(1-2\hat{n}J_z-J_z^2)/4+\ldots \nonumber\\
	\hat{n}^{(3)}(\infty)/j&=&(5\hat{n}+16J_z)T_q/(32 \sqrt{2})+\ldots \nonumber\\
	\hat{n}^{(4)}(\infty)/j&=&\left(18\hat{n}-69T_q^2 - 132J_z - 12\hat{n}^2J_z + 
 438\hat{n}J_z^2 + 132J_z^3\right)/384+\ldots
 \label{Npseries}
\end{eqnarray}
where only the $Q$-preserving terms relevant to the calculation of expectation values are shown. 
Looking at the results for $H_{eff}$ we see the ``in-sector" scattering interaction $T_q$ appearing in the higher order terms. As the power to which $T_q$ can occur is limited by the perturbation order a band diagonal structure is present in each $Q$-sector with respect to $J_z$ (or equivalently $\hat{n}$). We expect $H_{eff}=H(\infty)$  to provide a good description of the Dicke model at weak coupling, and possibly throughout the first phase. At strong coupling there is no guarantee of the accuracy or stability of $H_{eff}$, even when working to arbitrarily high orders in the perturbation. In fact, from the terms given in \eref{Hpseries} two things become apparent which signal a breakdown of this approximation at $\lambda>0.5$. Firstly, the matrix element of the one-dimensional $Q=0$ subspace is found to be $\inp{-j,0}{H_{eff}}{-j,0}=-j$ for all $\lambda$ and orders of the perturbation considered. As is known from non-perturbative numerical studies the flow equation maps the ground state to precisely this $\ket{-j,0}$ state. However, it is also known that the ground state energy is only equal to $-j$ in the first phase, and that it decreases linearly with $\lambda$ in the second \cite{Emary}, contradicting the prediction of $H_{eff}$. Secondly $H_{eff}$ is found to become unstable at large coupling for perturbation orders higher than one. For example, consider the expectation value of $H_{eff}$ to second order with respect to a simple variational state: $\inp{-j,n}{H_{eff}}{-j,n}=-j+\left(1-\lambda^2/2\right)\hat{n}$. We note that for $\lambda>\sqrt{2}$ this expectation value is unbounded from below. Together with the variational principle this implies that the spectrum of $H_{eff}$ becomes unbounded from below at some $\lambda\leq\sqrt{2}$. For the fourth order case this instability occurs at a $\lambda\leq\sqrt{2(\sqrt{11}-1)/5}\approx0.963$. We have observed that this point of breakdown  continues to move closer to the critical point $\lambda_c=0.5$ as higher order corrections are included. This agrees with the general notion that a perturbation series, even when summed up completely, may produce divergent results beyond the series' radius of convergence.

Finally we point out that to first order the effective Hamiltonian is 
\begin{equation}
	H_{eff}=J_z+\hat{n}+\frac{\lambda}{\sqrt{2}}\left(J_+b+J_-b^\dag\right),
\end{equation}
which is equivalent to the RWA approximation in which the model was originally studied \cite{Lieb,Wang}. To this order we also observe that the $Q$-preserving parts of  $J_z$ and $\hat{n}$ are left unchanged by the transformation generated by the flow equation. 

\section{Diagonalizing $H_{eff}$}
\label{dickeFD}
The flow equation treatment of the preceding sections has brought the Dicke Hamiltonian into a block diagonal form by eliminating interactions which connect different $Q$-sectors. What remains is to diagonalize $H_{eff}$ within each $Q$-sector at a time. 

One approach, valid for large but finite $j$, is to apply direct diagonalization to each $Q$-block  (submatrix) of $H_{eff}$. This is done by first constructing the matrix representations of $J_z$, $\hat{n}$ and $T_q$ for the $Q$-sector under consideration. Finding the desired submatrix of $H_{eff}$ is simply a matter of replacing $x_s$, $x_b$ and $x_q$ by the properly scaled matrix representations of $J_z/j$, $\hat{n}/j$ and $T_q/j^{3/2}$ respectively. One need not be concerned about the precise ordering of operators in products, as different orderings only bring about ${\cal O}(1/j)$ corrections. The result of these substitutions is a matrix $M$, the size of which ranges from $1\times1$ to $(2j+1)\times(2j+1)$ depending on the value of $Q$. Generally $M$ will not be Hermitian, and so we consider the symmetrized form $(M+M^T)/2$.  Results of the subsequent numeric diagonalization appear in the next section. 

Alternatively we may remain in the large $j$ limit and apply the flow equation again, this time non-perturbatively, to diagonalize each submatrix. Within each $Q$-sector the problem is effectively  one-dimensional as we may eliminate either the bosonic or spin degree of freedom in favor of the scalar parameter $Q$. We choose to work with the spin degree of freedom, which has the natural basis
\begin{equation}
	{\cal B}_Q=\left\{\ket{m,Q}\ :\ m=-j,\ldots,{\rm min}(j,Q-j)\right\}.
\end{equation}
By writing $\hat{n}=Q-J_z-j$ and $$T_q=J_+b+J_-b^\dag=\sqrt{Q-J_z-j}\;J_-+\sqrt{Q-J_z-j+1}\;J_+$$ the effective Hamiltonian can be rewritten purely in terms of spin operators.\\

Let us clarify this last step. Since $m$ and $Q$ together fix the number of bosons there is no $n$ label in ${\cal B}_Q$. In fact, we completely eliminate the bosonic degree of freedom through the correspondence
\begin{equation}
	J_+b\leftrightarrow\sqrt{Q-J_z-j+1}J_+\ \ \ \ {\rm and}\ \ \ \ J_-b^\dag\leftrightarrow\sqrt{Q-J_z-j}J_-
\end{equation}
where the operators on the right act on ${\cal B}_Q$ and are equivalent to the operators on the left acting within a fixed $Q$-sector of the original tensor product space $\ket{m}\otimes\ket{n}$.\\

For concreteness we consider the fourth order case for the rest of the section. The results can be easily extended to higher orders. The form of the spin-only Hamiltonian is $H_{eff}=N_0(J_z)+N_1(J_z)\left(J_++J_-\right)+N_2(J_z)\left(J^2_++J^2_-\right)$, which clearly possesses a double band diagonal structure with respect to $J_z$. To preserve this form during flow we use the generator of Section \ref{etaChoice}: 
\begin{equation}
	\eta(\ell)=N_1(J_z,\ell)\left(J_+-J_-\right)+N_2(J_z,\ell)\left(J^2_+-J^2_-\right).
\end{equation}
For the numerical treatment of the flow equation it is convenient to transform to the variables of the $\texttt{su}(2)$ representation in terms of which $H_{eff}(\ell)$ and $\eta(\ell)$ become
\begin{eqnarray}
	H_{eff}(\ell)&=&j\left(n_0(x,\ell)+n_1(x,\ell)\cos(\beta)+n_2(x,\ell)\cos(2\beta)\right),\\
	\eta(\ell)&=&ij\left(n_1(x,\ell)\sin(\beta)+n_2(x,\ell)\sin(2\beta)\right),
\end{eqnarray}
where we write $x$ for $x_s$ and the initial conditions are
\begin{eqnarray}
	n_0(x,0)&=&q-1-(1-2(q-1)x+x^2)\lambda^2/4 + 
 ((1+x)^2(8x-3)+q(3-13x^2))\lambda^4/16 \nonumber, \\
 n_1(x,0)&=&\sqrt{q-1-x}\sqrt{1-x^2}(8\lambda- 
 (q-1+3x)\lambda^3)/\sqrt{32},\\
 n_2(x,0)&=&-7(q-1-x)(x^2-1)\lambda^4/32. \nonumber 
\end{eqnarray}
Here $q=Q/j$, which is now a continuous scalar parameter labelling the relevant sector. The flow of these functions are given by
\begin{eqnarray}
	\pafg{n_0}{\ell}&=&-n_1\pafg{n_1}{x}-2n_2\pafg{n_2}{x}, \nonumber \\
	\pafg{n_1}{\ell}&=&-2n_2\pafg{n_1}{x}-n_1\left(\pafg{n_0}{x}+\pafg{n_2}{x}\right), \label{2ndStepFEH} \\
	\pafg{n_2}{\ell}&=&-2n_2\pafg{n_0}{x}. \nonumber
\end{eqnarray}
The corresponding flow equation for an observable $O(\ell)$ is
\begin{equation}
	\pafg{O}{\ell}=\left(\pafg{n_1}{x}\sin{\beta}+\pafg{n_2}{x}\sin{2 \beta}\right)\pafg{O}{\beta}-\left(n_1\cos{\beta}+n_2\cos{2 \beta}\right)\pafg{O}{x},
	\label{2ndStepFEO}
\end{equation}
which we treat using a similar approach to that of equations \eref{fourier} and \eref{usefourier}. In the $\ell\rightarrow\infty$ limit we expect $n_1$ and $n_2$ to vanish, as they represent the off-diagonal parts of the flowing Hamiltonian. Using the correspondence between $x$ and $J_z$ we may consider $n_0(\infty)$, and thus $H(\infty)$ itself, to be a function of $J_z$. It follows that the eigenvalues of the particular $Q$-sector are given by $E_n=n_0(x=-1+n/j,\ell=\infty)$ for $n=0,1,\ldots,{\rm min}(2j,Q)$.

There is a subtle issue regarding the domain of these functions that still needs to be addressed. Looking back at the definition of the ${\cal B}_Q$ basis we see that the maximum value of the spin label $m$ is ${\rm min}(j,Q-j)$ which is less than $j$ when $Q<2j$. Correspondingly we need to restrict the domain of the $n_i$ functions to reflect this. Since $J_z\propto x$ the relevant domain is $x\in\left[-1,{\rm min}(1,q-1)\right]$. Indeed, outside this region the initial conditions are complex, causing the flow to become unstable as $\ell\rightarrow\infty$. 
\newpage
\section{Numeric results}
During the course of this chapter we have encountered a number of different treatments of the Dicke model. In this section we present the results obtained using these methods and compare their ranges of applicability. First we summarize the different approaches:
\begin{itemize}
	\item {\bf Direct Diagonalization}: By introducing a cut-off in the maximum value of $Q$ we can numerically diagonalize the full Dicke Hamiltonian on a truncated Hilbert space. This generally leads to very large matrices, especially in the second phase where an extensive number of $Q$-sectors must be included in order to describe the low-lying states. For sufficiently large values of the cut-off this method does provide very accurate results for the low-lying states, and we will use these as a benchmark for the flow equation results. 	
	\item {\bf Local Approximation}: This analytic method was derived in Section \ref{dickeLocal} and constitutes a harmonic approximation of the first phase, low-lying spectrum in terms of two oscillators.	
	\item {\bf Direct Diagonalization of $H_{eff}$}: In Section \ref{dickePSol} we derived the effective $Q$-preserving Hamiltonian $H_{eff}$ in a perturbative approximation. For finite $j$ the submatrices comprising the block-diagonal structure of $H_{eff}$ may be constructed and diagonalized numerically. Since these submatrices are maximally of size $(2j+1)\times(2j+1)$ this is a greatly reduced problem compared to diagonalizing the original Hamiltonian in some large truncated space.
	\item {\bf Flow Equation Diagonalization of $H_{eff}$}: As described in Section \ref{dickeFD} the individual $Q$-sectors may be diagonalized in the $j\rightarrow\infty$ limit using a flow equation equipped with an appropriate generator. The continuous variable $q=Q/j\in[0,\infty)$ appearing in the initial condition labels the relevant $Q$-sector.
	\item {\bf The Rotating Wave Approximation (RWA)}: Within this approximation the Dicke Hamiltonian is equivalent to the first order form of $H_{eff}$. In this sense the RWA is already included in the two previous cases. However, due to its prevalence in the literature we will treat it separately as a reference case.
\end{itemize}
\begin{figure}[ht]
\begin{narrow}{-0.44in}{0in}
\begin{tabular}{cc}
	\epsfig{file=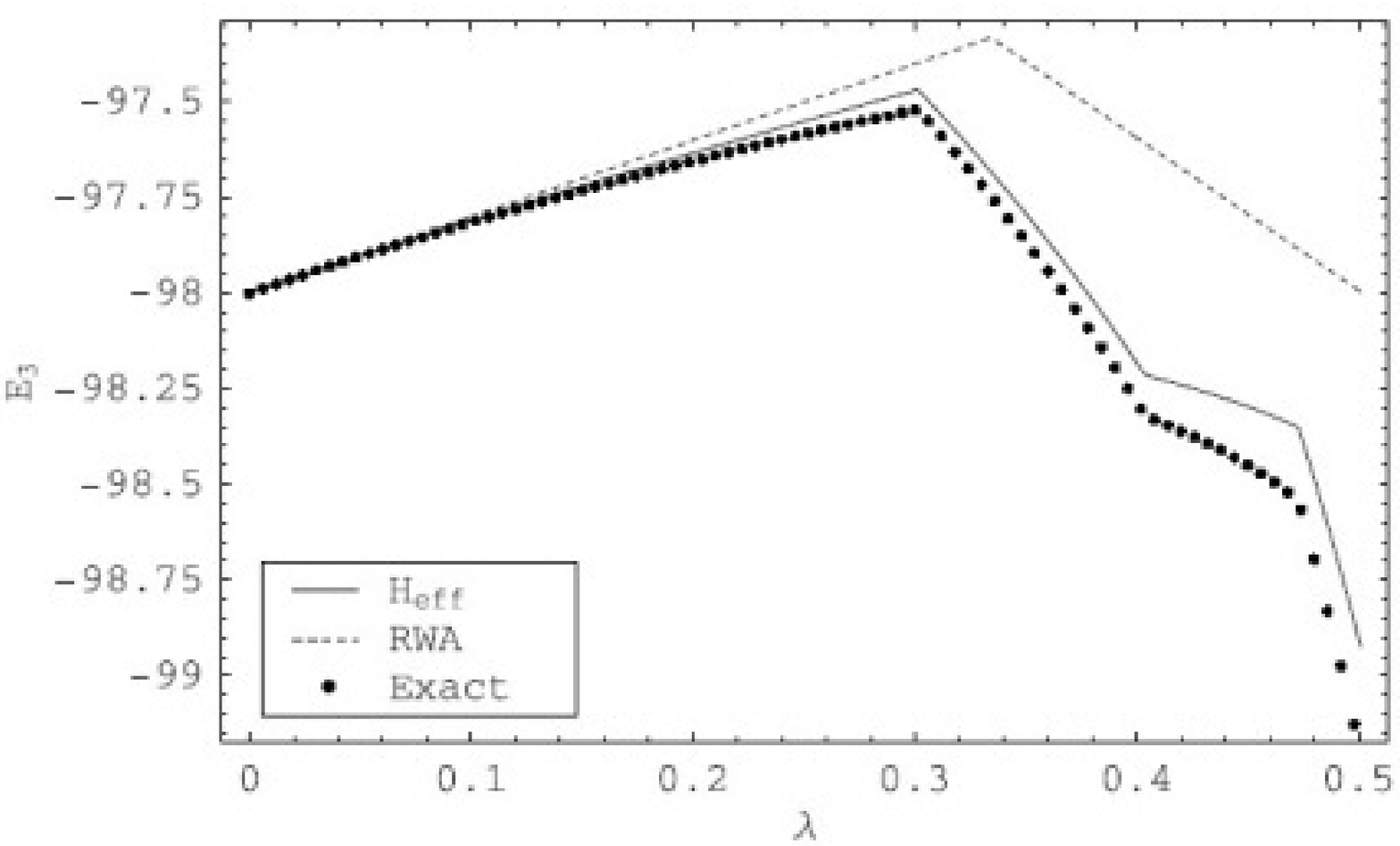,height=5cm,clip=,angle=0} &
	\epsfig{file=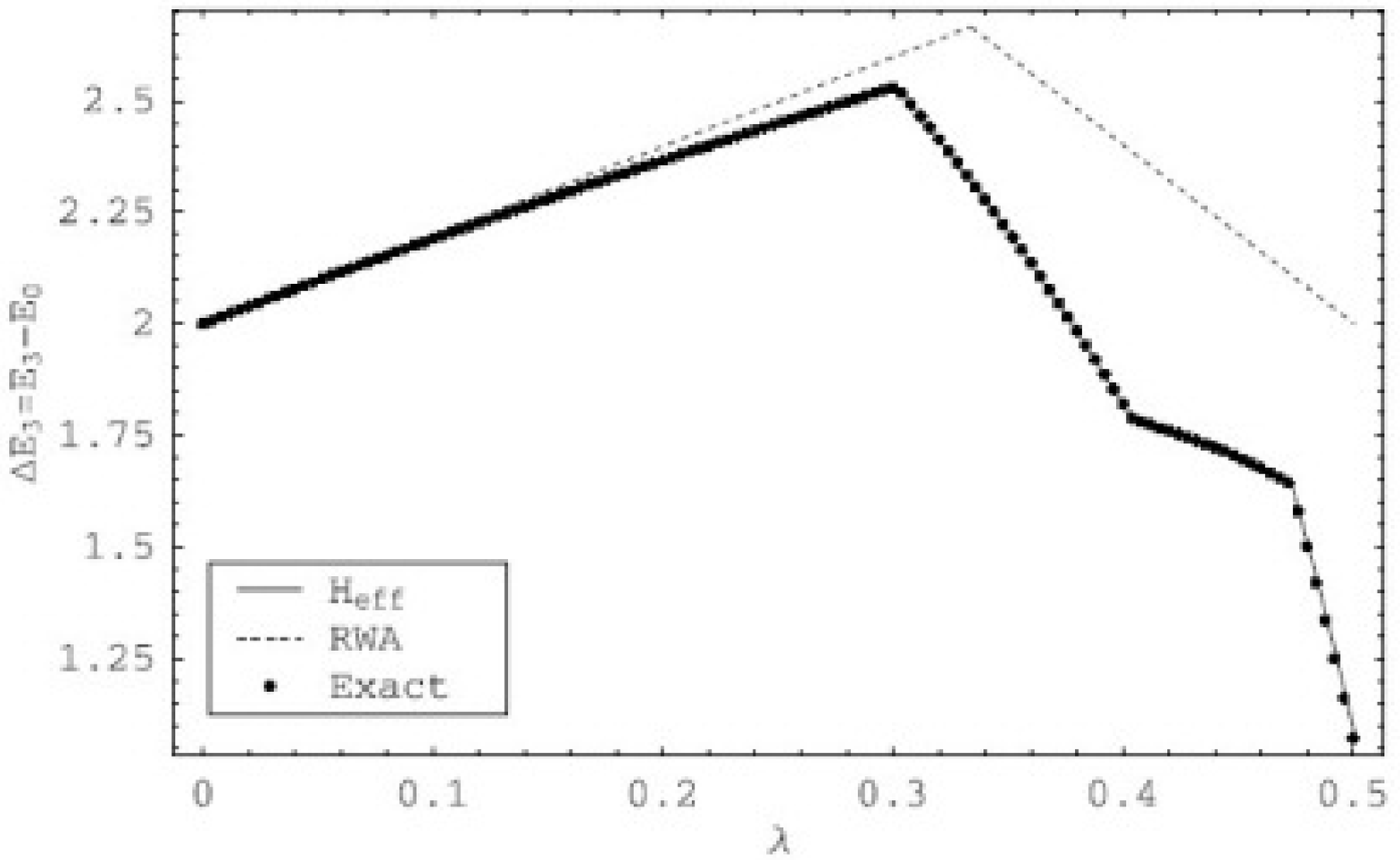,height=5cm,clip=,angle=0} \\
	(a) & (b)
\end{tabular}
\end{narrow}
	\caption[The eigenvalue $E_3$ and excitation energy $\Delta E_3$ as functions of $\lambda$.]{The (a) third excited state and (b) the gap $\Delta E_3$ as functions of $\lambda$. $H_{eff}$ was constructed for $j=100$ up to perturbation order $p_{max}=15$. }
	\label{DickeFigSet1}
\end{figure}

\begin{figure}[ht]
\begin{narrow}{-0.25in}{0in}
\begin{tabular}{cc}
  	\epsfig{file=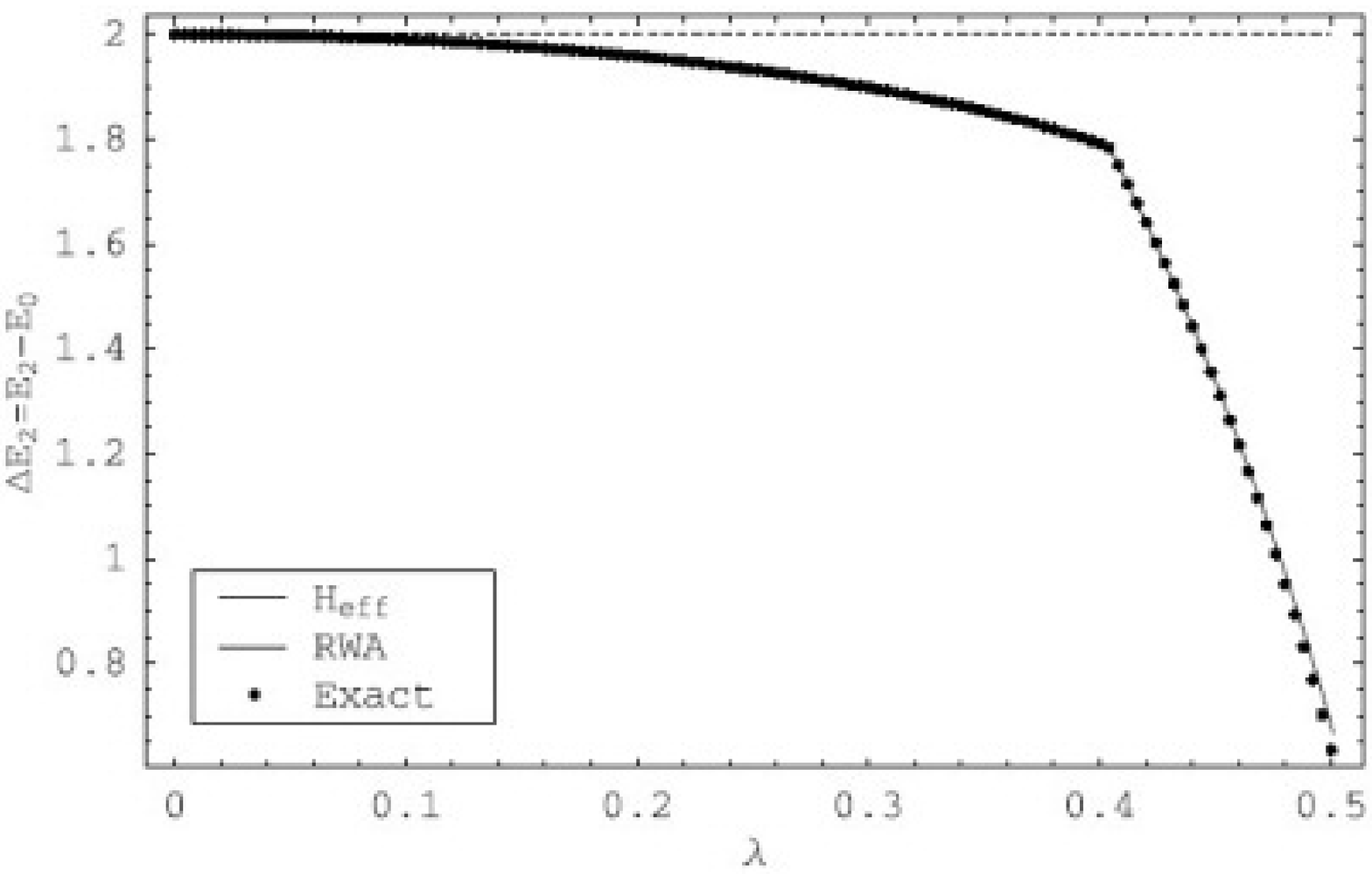,height=5cm,clip=,angle=0} &
	\epsfig{file=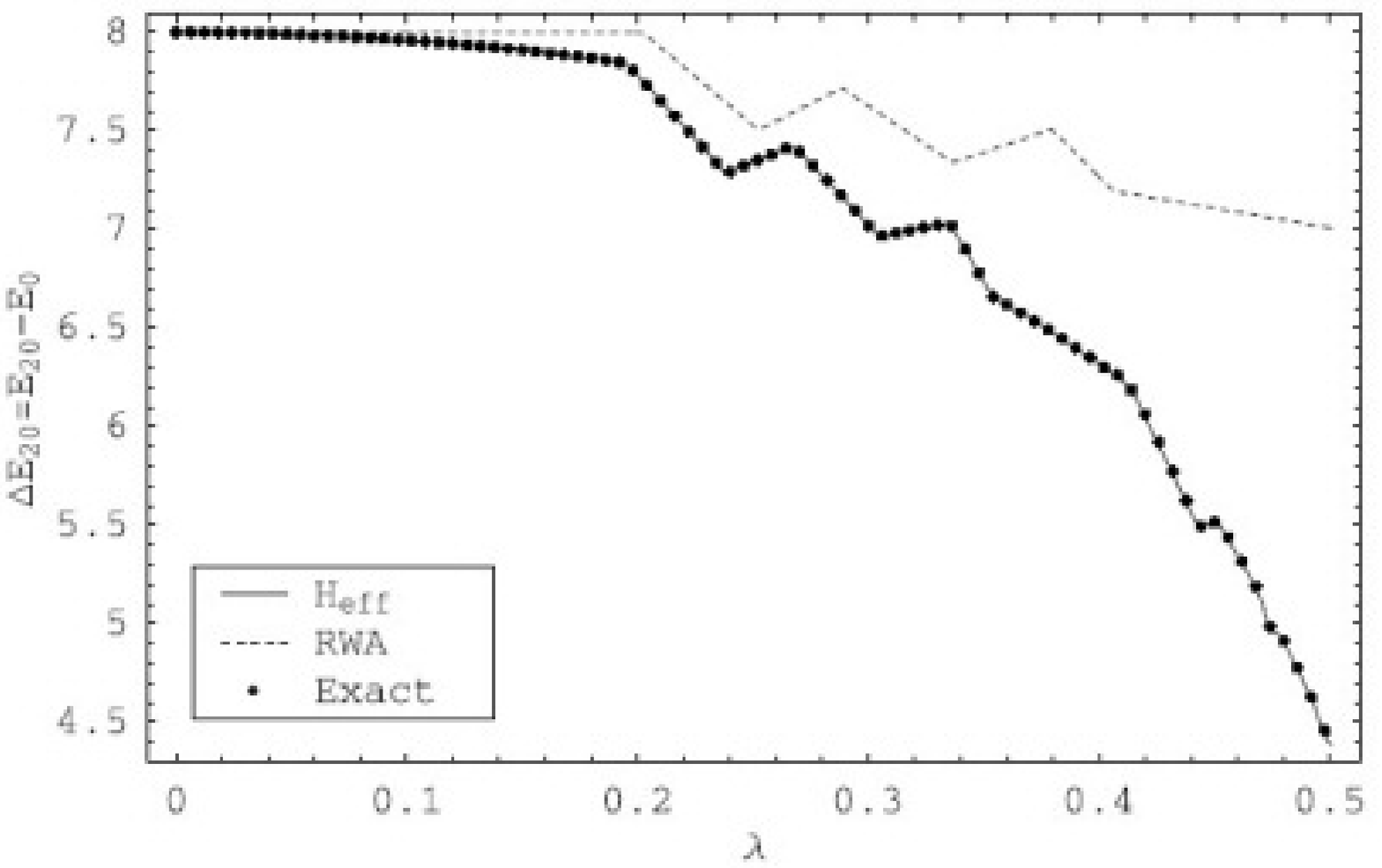,height=5cm,clip=,angle=0} \\
	(a) & (b)\\
	\epsfig{file=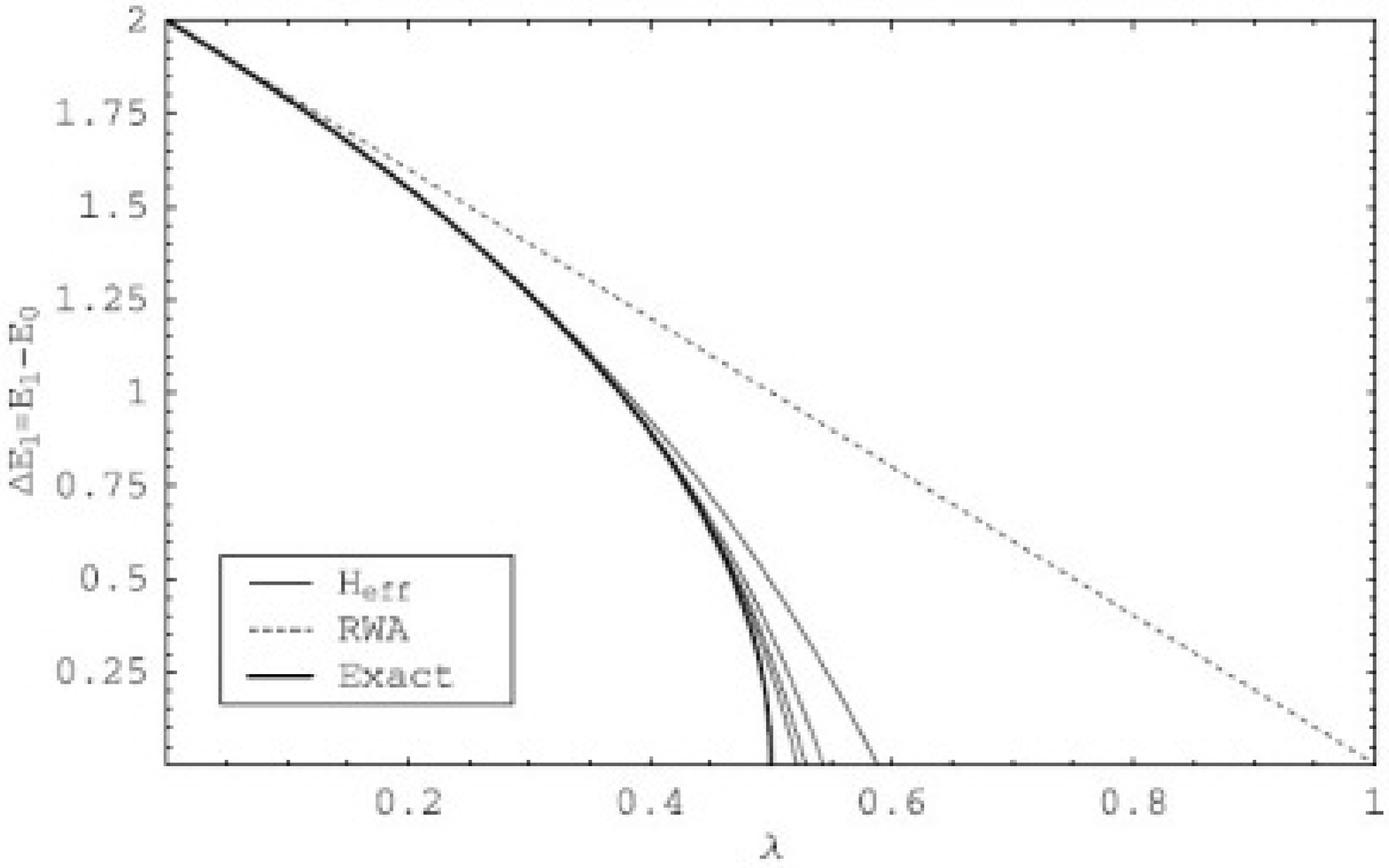,height=5cm,clip=,angle=0} &
	\epsfig{file=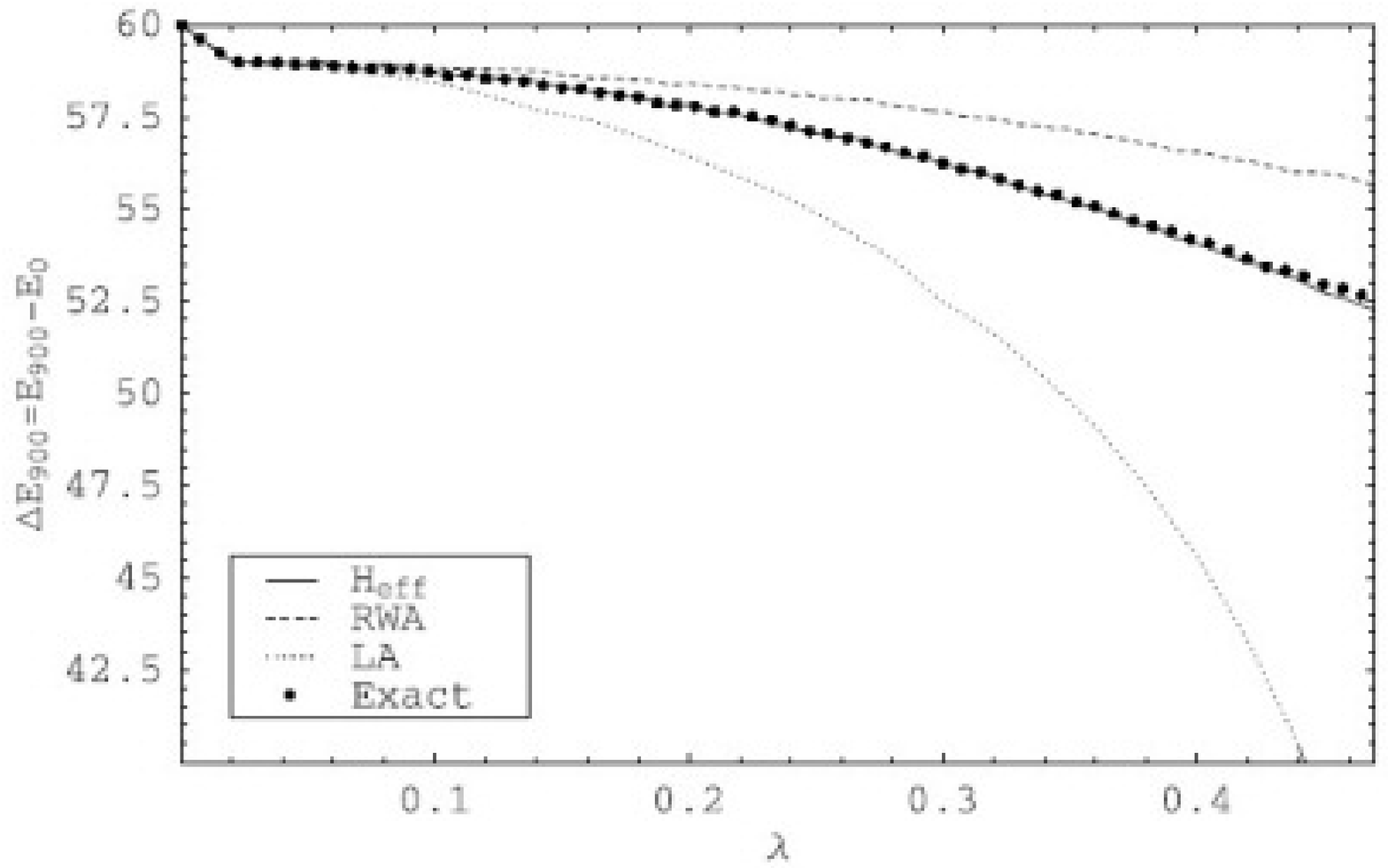,height=5cm,clip=,angle=0} \\
	(c) & (d)
	\end{tabular}
\end{narrow}
	\caption[Various excitation energies as functions of $\lambda$.]{Various excitation energies. In (a), (b) and (c) $j$ equals $100$, while in (a), (b) and (d) $p_{max}=15$. The four solid lines in (c) correspond to $p_{max}=5,10,15,20$ as they appear from right to left. The bold line corresponds to the exact solution for $j=\infty$ or, equivalently, the result of the local approximation.  In (d) $j$ equals $40$. }
	\label{DickeFigSet2}
\end{figure}
Throughout this section only the subspace corresponding to even values of $Q$, as discussed in Section \ref{DickeModelOverview}, will be considered. The first results we present are those obtained by diagonalizing $H_{eff}$ at finite $j$. The flow equation treatment of the second step will be dealt with later on. Figure \ref{DickeFigSet1} (a) shows the energy of the third excited state $\ket{E_3}$ as a function of $\lambda$ at $j=100$. Also shown are the exact values obtained using direct diagonalization, and the RWA result. Although the result of $H_{eff}$ is a marked improvement over that of the RWA, we still see ${\cal O}(1/j)$ errors present at larger values of $\lambda$. This simply the reflects the order to which the Moyal bracket was expanded in our derivation of the flow equation. However, it quickly becomes apparent that this is a systematic error and largely independent of the particular state. We attribute this to the scalar part of the higher order $1/j$-corrections that were neglected in our derivation. This only serves to shift the entire spectrum by some fixed amount relative to the exact results. For a more meaningful comparison we eliminate this shift by considering the excited energies relative to the ground state, i.e. we consider the excitation energy (or gap) $\Delta E_n=E_n-E_0$ rather than $E_n$ itself. As seen in Figures \ref{DickeFigSet1} (b) and \ref{DickeFigSet2} (a), (b) there is indeed a very good correspondence between the exact results for $\Delta E_n$ and those obtained from $H_{eff}$. From previous studies \cite{Emary} it is known that the critical point $\lambda_c=0.5$ is characterized by the vanishing of the gap between the ground state and first excited state. This agrees with the result of the local approximation that $\Delta E_1\sim\sqrt{1-2\lambda}$. Figure \ref{DickeFigSet2} (c) shows this gap together with those obtained using $H_{eff}$.

Whereas the local approximation provides reasonable results for the low-lying states, it fails to do so for the highly excited states, as is clear from Figure \ref{DickeFigSet2} (d). For these states the assumption that $\ave{J_z}+j\sim {\cal O}(j^0)$ is no longer valid. This means that the local solution within a neighbourhood of $x_s=-1$ is insufficient since the flow relevant to these states occur at points distant from $x_s=-1$.

\begin{figure}[ht]
\begin{center}
\begin{tabular}{cc}
	\epsfig{file=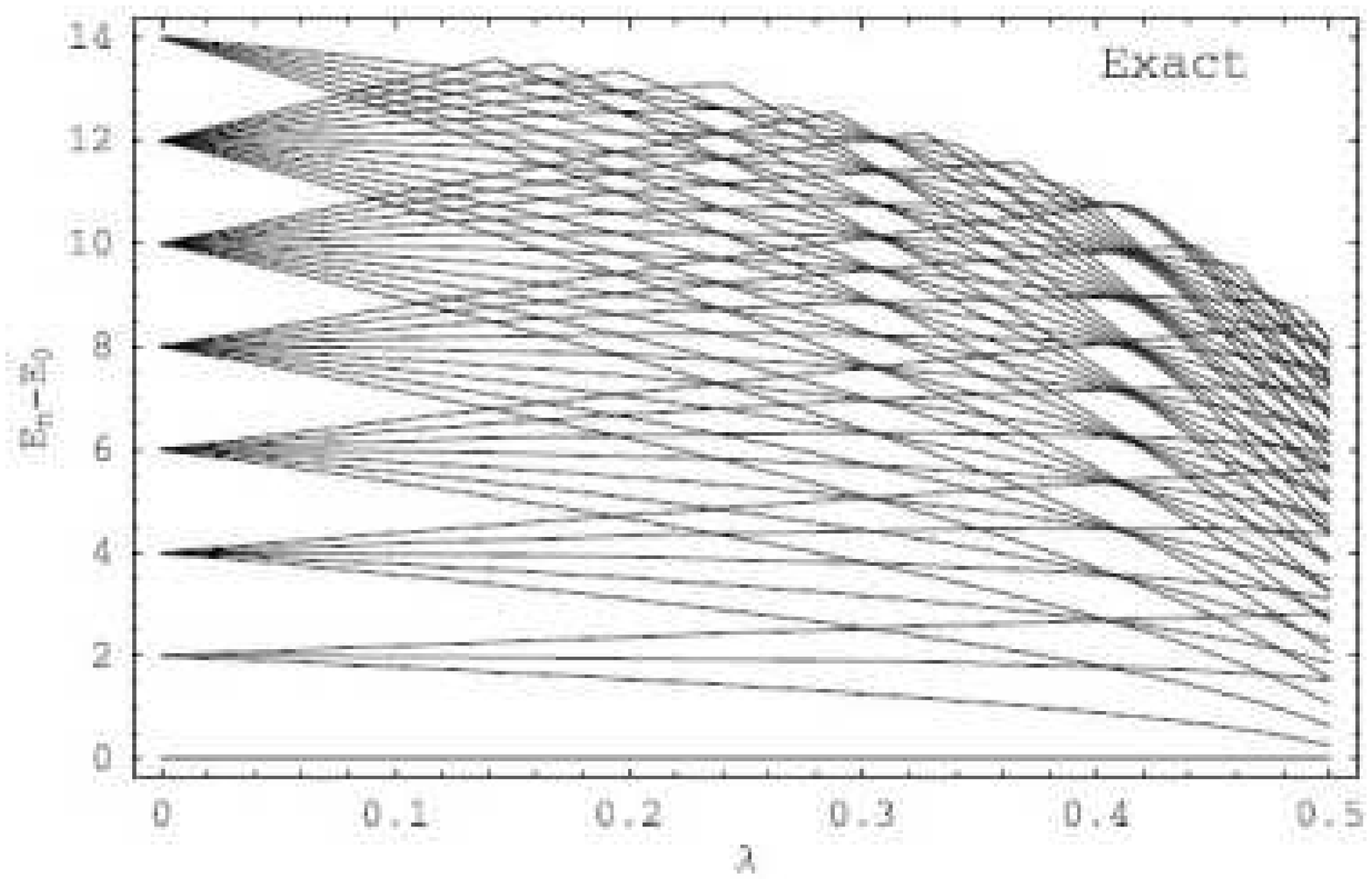,height=4.3cm,clip=,angle=0} &
	\epsfig{file=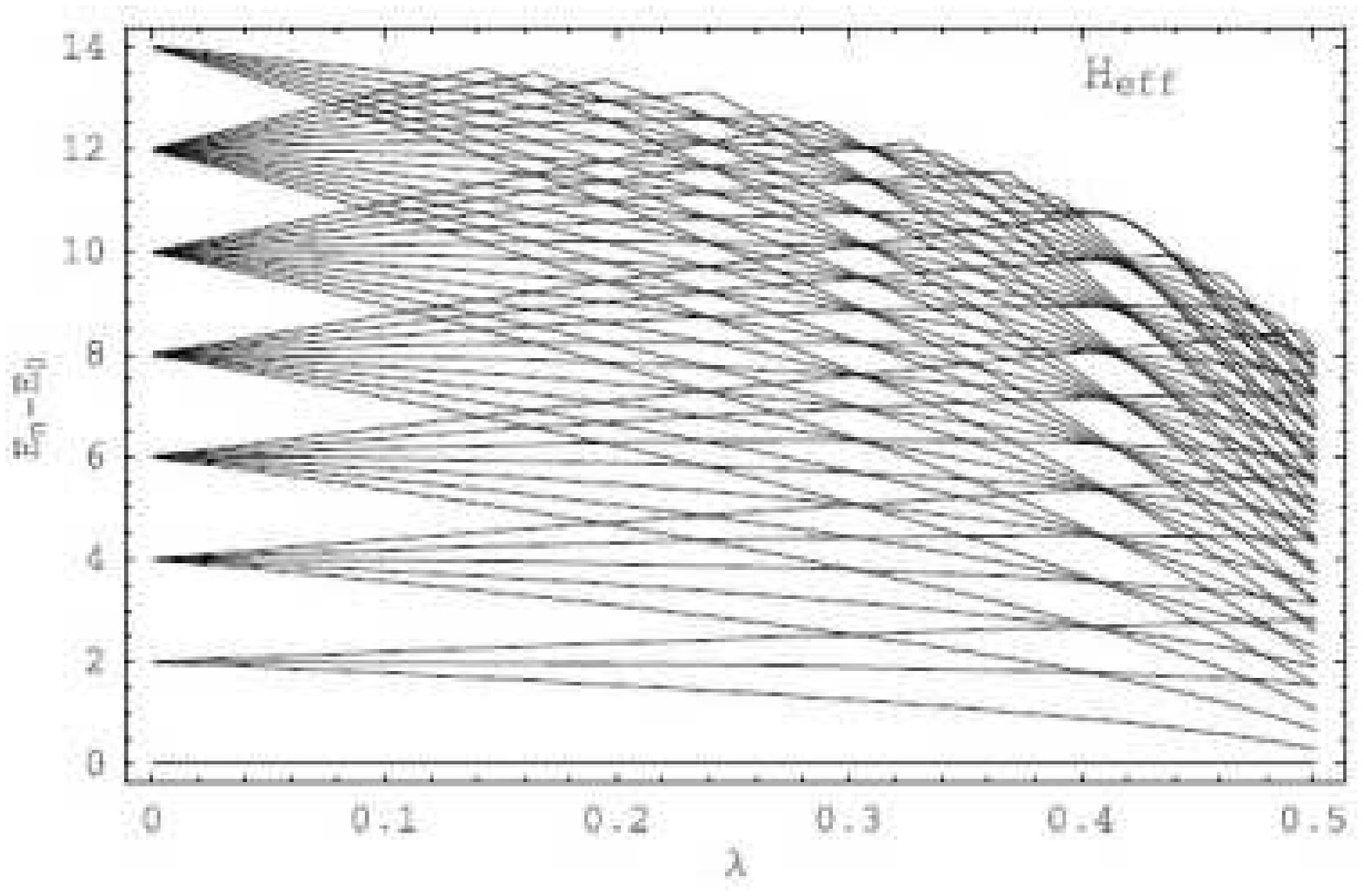,height=4.3cm,clip=,angle=0} \\
	(a) & (b)\\
	\epsfig{file=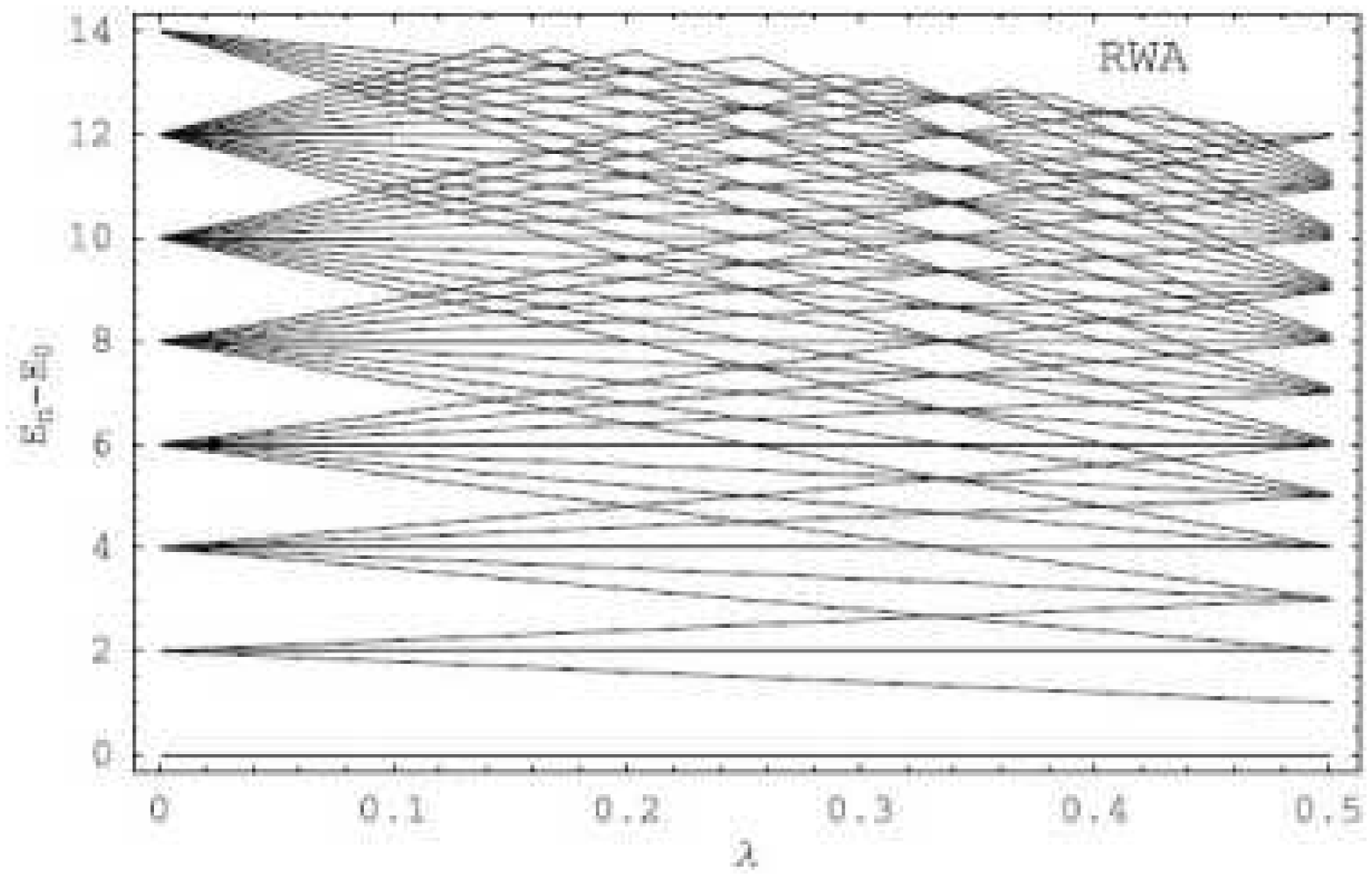,height=4.3cm,clip=,angle=0} &
	\epsfig{file=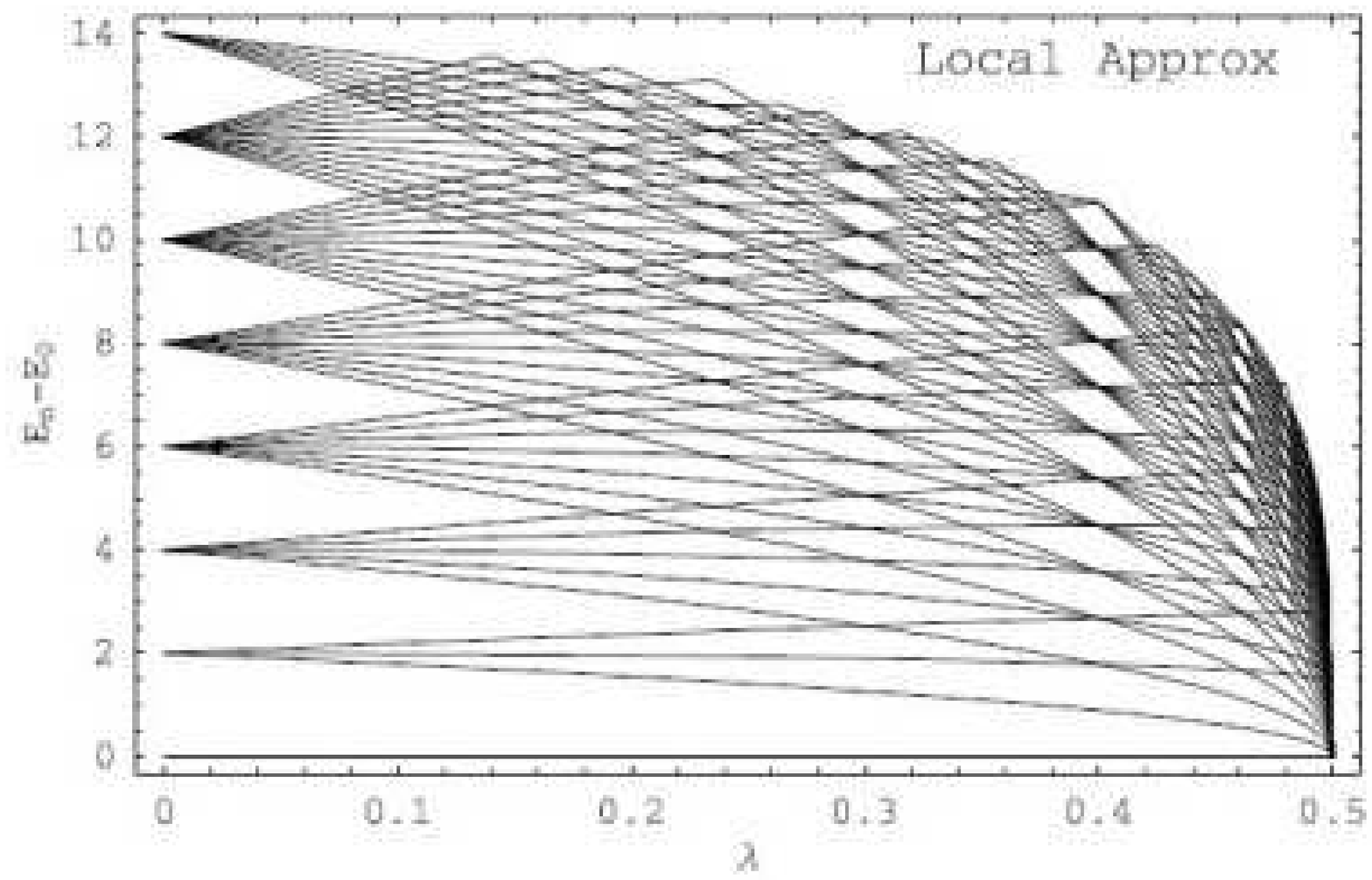,height=4.3cm,clip=,angle=0} \\
	(c) & (d)\\
	\epsfig{file=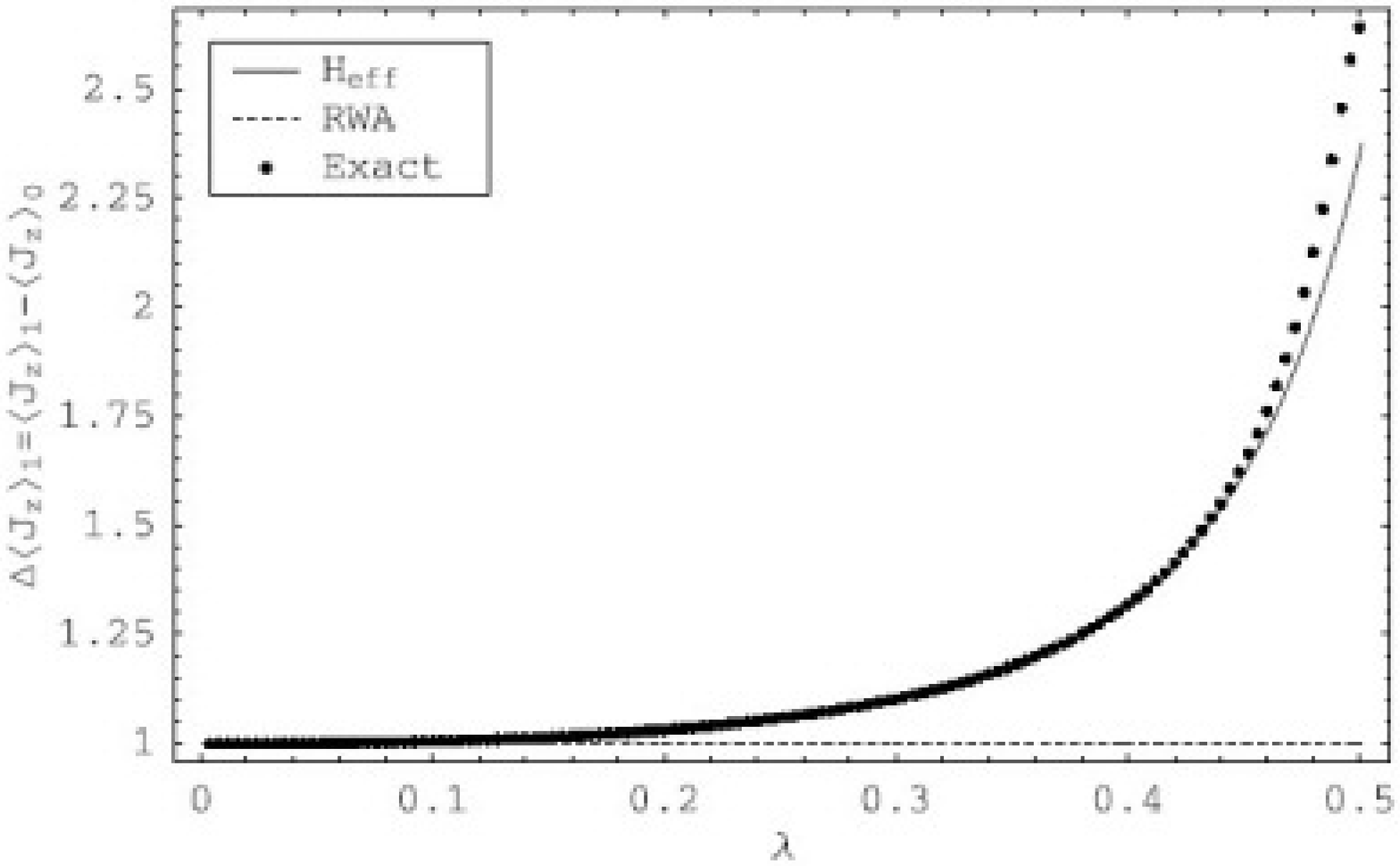,height=4.3cm,clip=,angle=0} &
	\epsfig{file=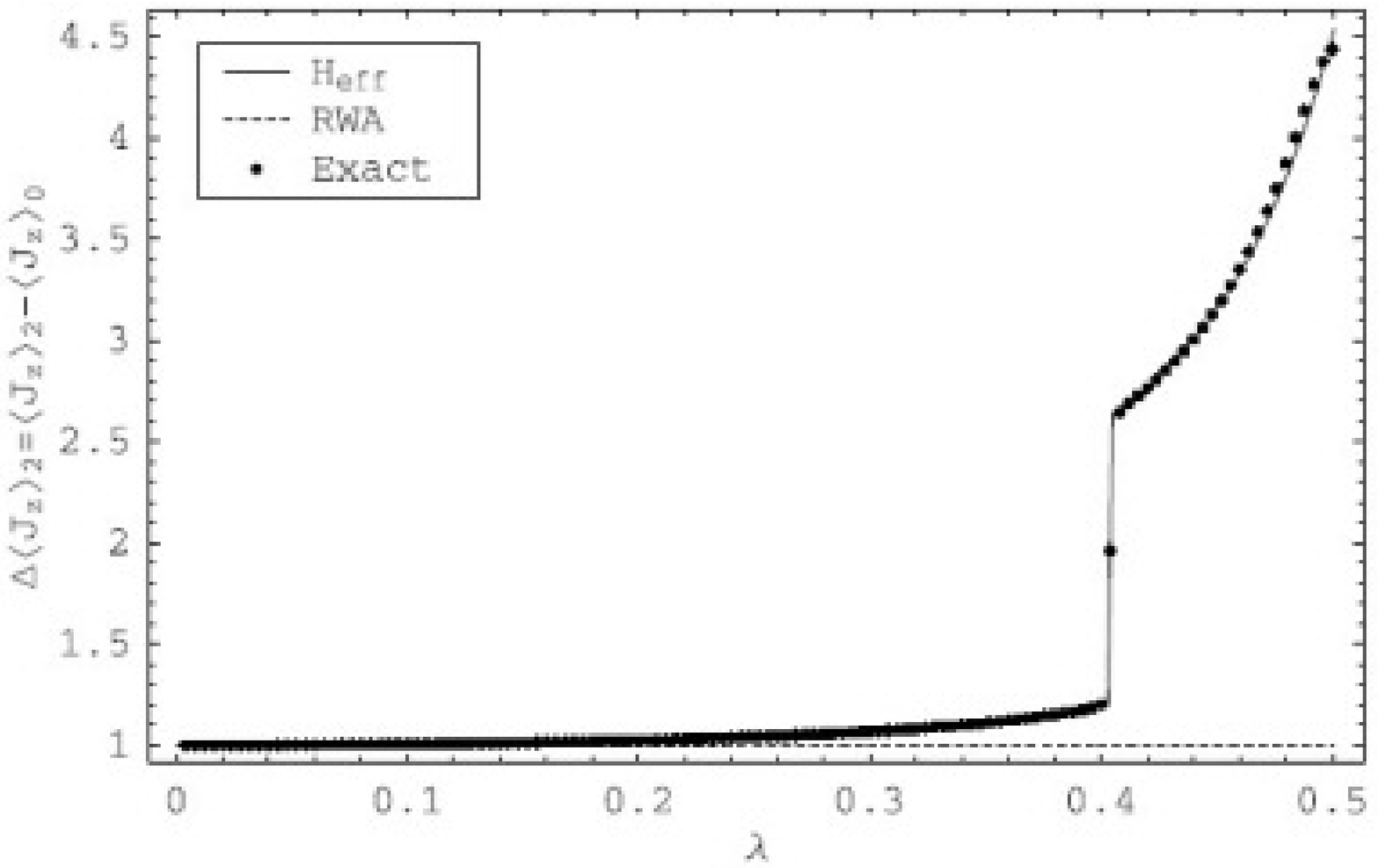,height=4.3cm,clip=,angle=0} \\
	(e) & (f)\\
	\epsfig{file=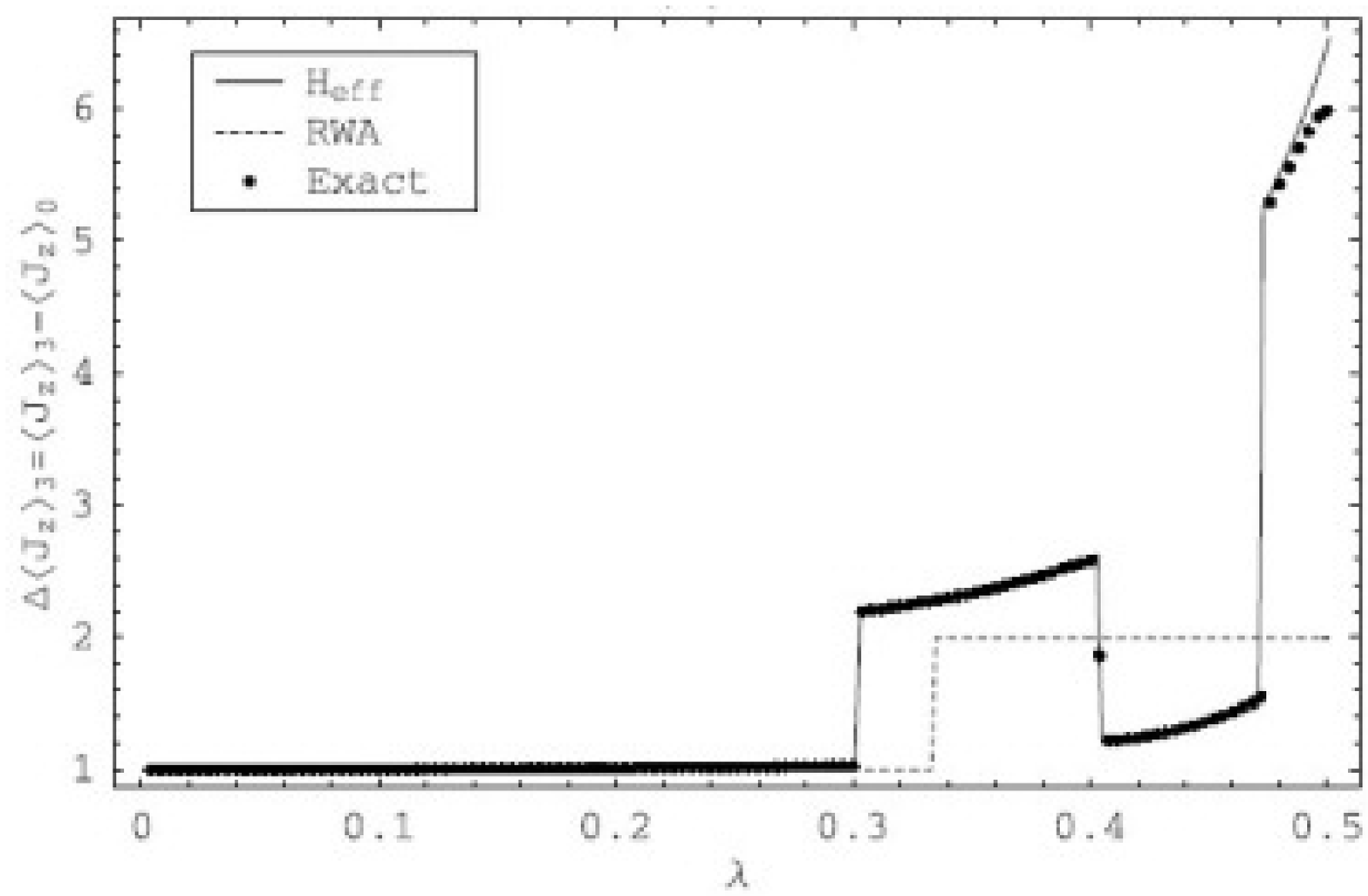,height=4.3cm,clip=,angle=0} &
  \epsfig{file=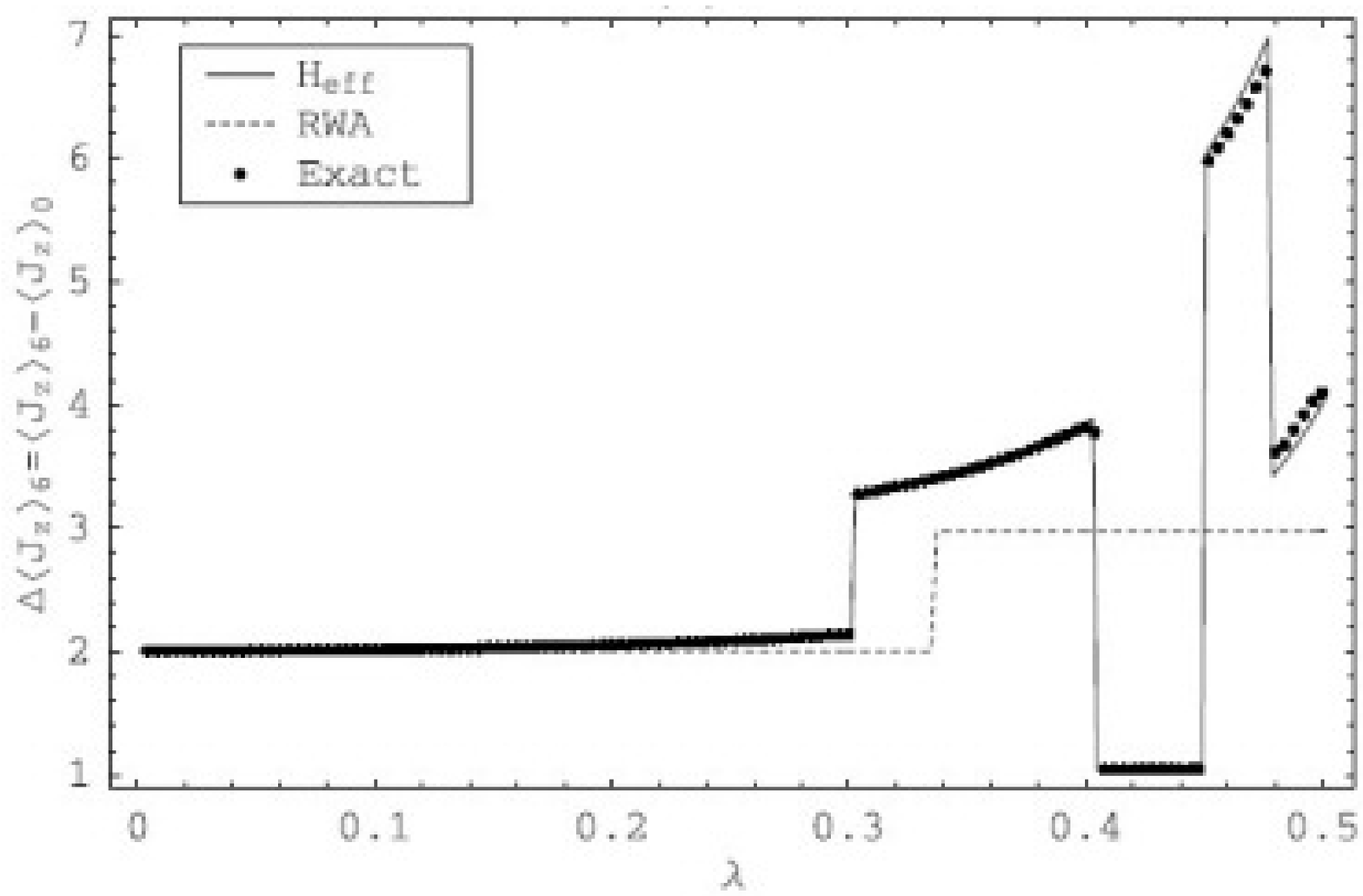,height=4.3cm,clip=,angle=0}\\
	(g) &  (h)
	\end{tabular}
	\caption[Eigenstates of the Dicke Hamiltonian obtained using different methods. Expectation values of $J_z$ as functions of $\lambda$.]{(a)-(d): The energies of the first $55$ eigenstates of the Dicke Hamiltonian obtained using different methods. For all of these $j=100$ and $p_{max}=15$. (e)-(h): Expectation values of $J_z$ relative to the ground state. $j=100$ and \mbox{$p_{max}=18$} throughout. The energies of the first $55$ eigenstates of the Dicke Hamiltonian obtained using different methods. For all of these $j=100$ and $p_{max}=15$.}
	\label{DickeFigSet3}
\end{center}
\end{figure}

Figure \ref{DickeFigSet3} (a)-(d) provides a global view the spectra obtained using the different methods. The qualitative properties of the exact result and that of $H_{eff}$ are clearly very similar. In particular  we note the level repulsion among the low-lying states, which leads to the vanishing of the gap $\Delta E_1$ at the critical point. This behaviour is clearly absent in the RWA case, where there appears to be no correlation between levels belonging to different sectors. As expected the local approximation predicts the correct behaviour for the low-lying states but fails at higher energies. In fact, the local approximation predicts a continuous spectrum at $\lambda=0.5$.

The $Q$-preserving parts of $\hat{n}_{eff}$ and $(J_z)_{eff}$ may be constructed in the same way as we did for the individual submatrices of $H_{eff}$. The calculation of the expectation values is now a straight forward procedure, the results of which appear in Figure \ref{DickeFigSet3} (e)-(h). We again find good agreement with the exact results. To eliminate any constant shift owing to higher order scalar corrections we consider the expectation values relative to the ground state, i.e. $\Delta\ave{J_z}_n=\ave{J_z}_n-\ave{J_z}_0$ where $\ave{J_z}_n=\inp{E_n}{J_z}{E_n}$.\\
\clearpage
\clearpage

\begin{figure}[th]
\begin{narrow}{-0.15in}{0in}
\begin{tabular}{cc}
	\epsfig{file=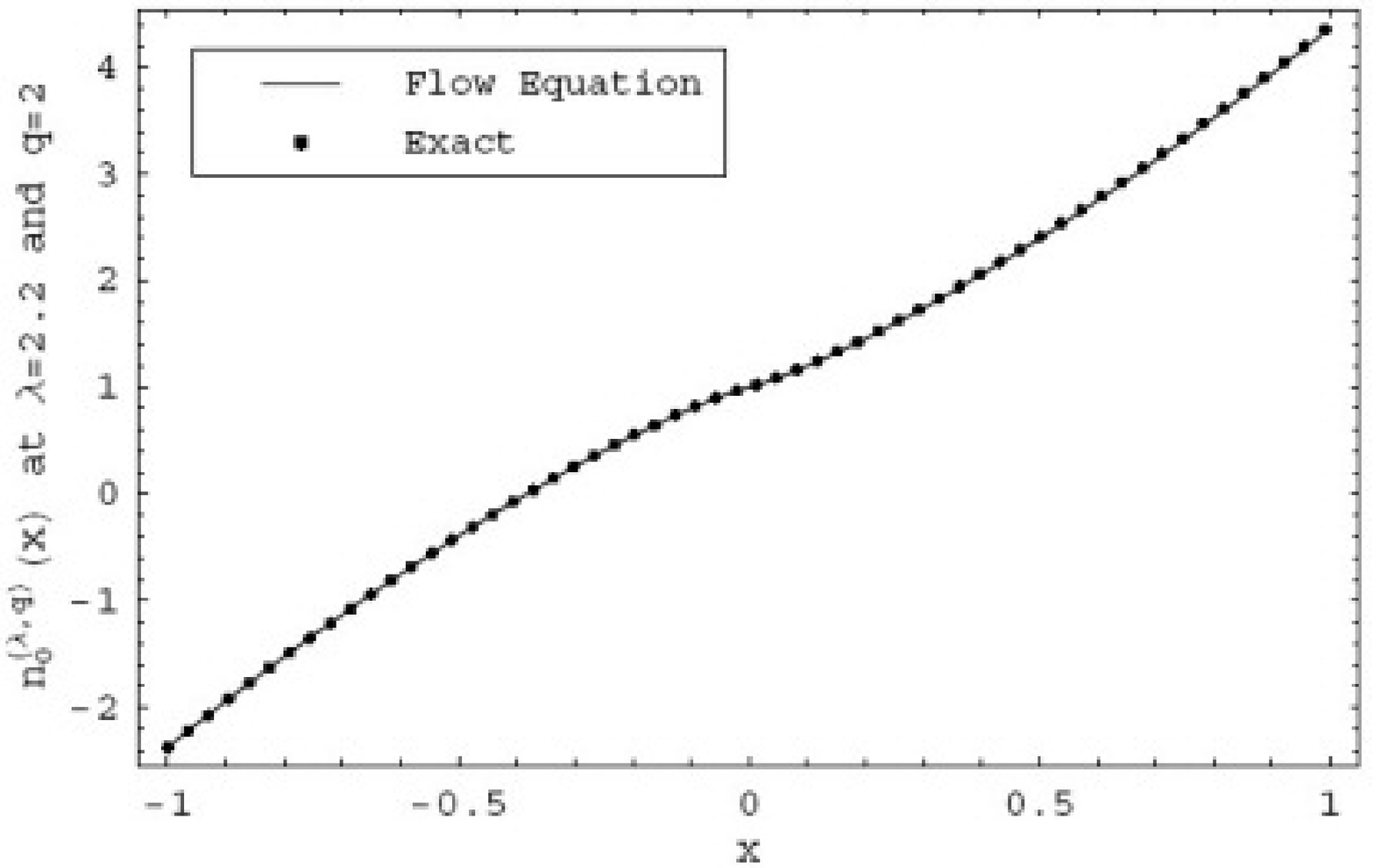,height=5cm,clip=,angle=0} &
	\epsfig{file=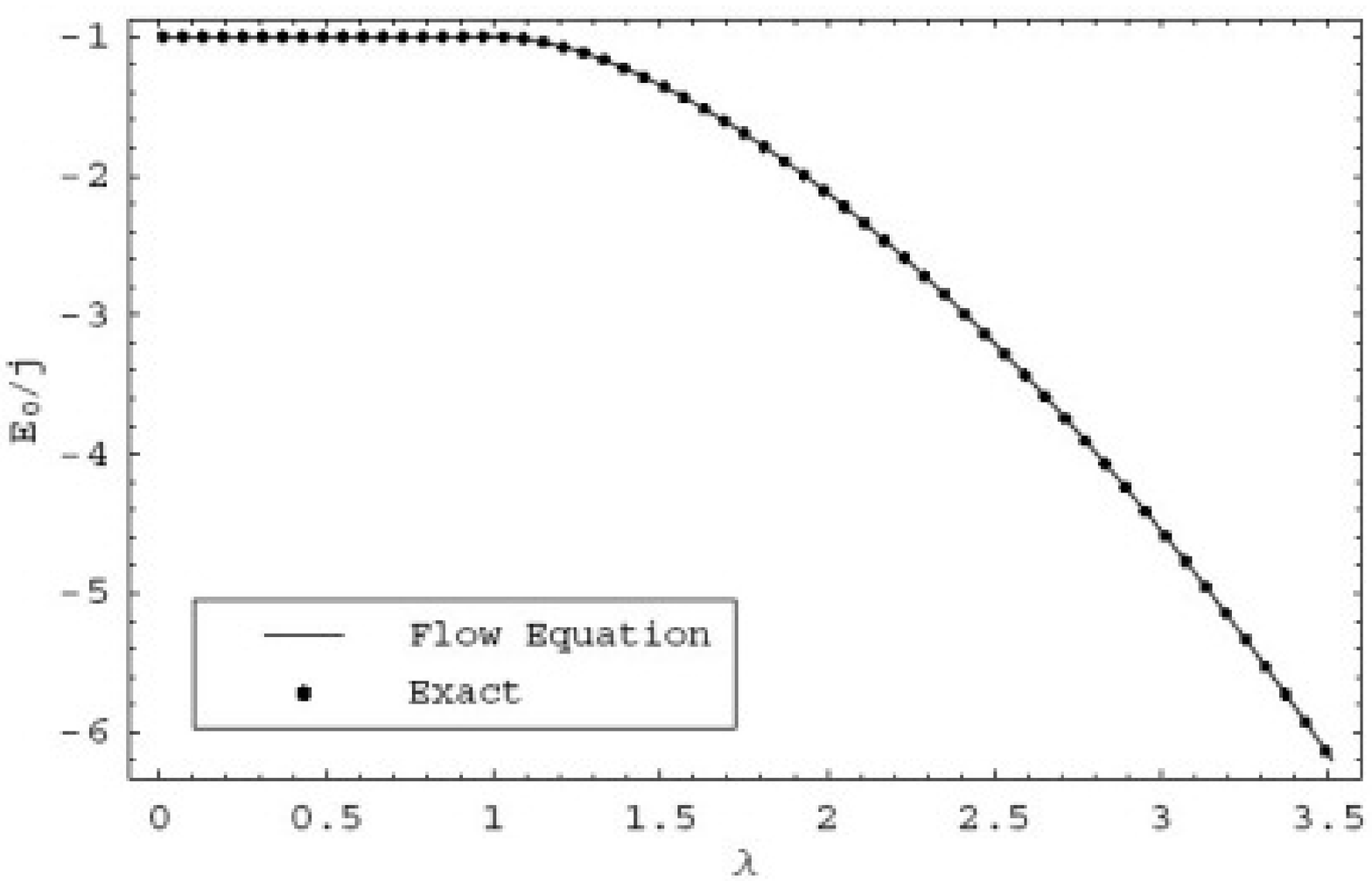,height=5cm,clip=,angle=0} \\
	(a) & (b)\\
	\epsfig{file=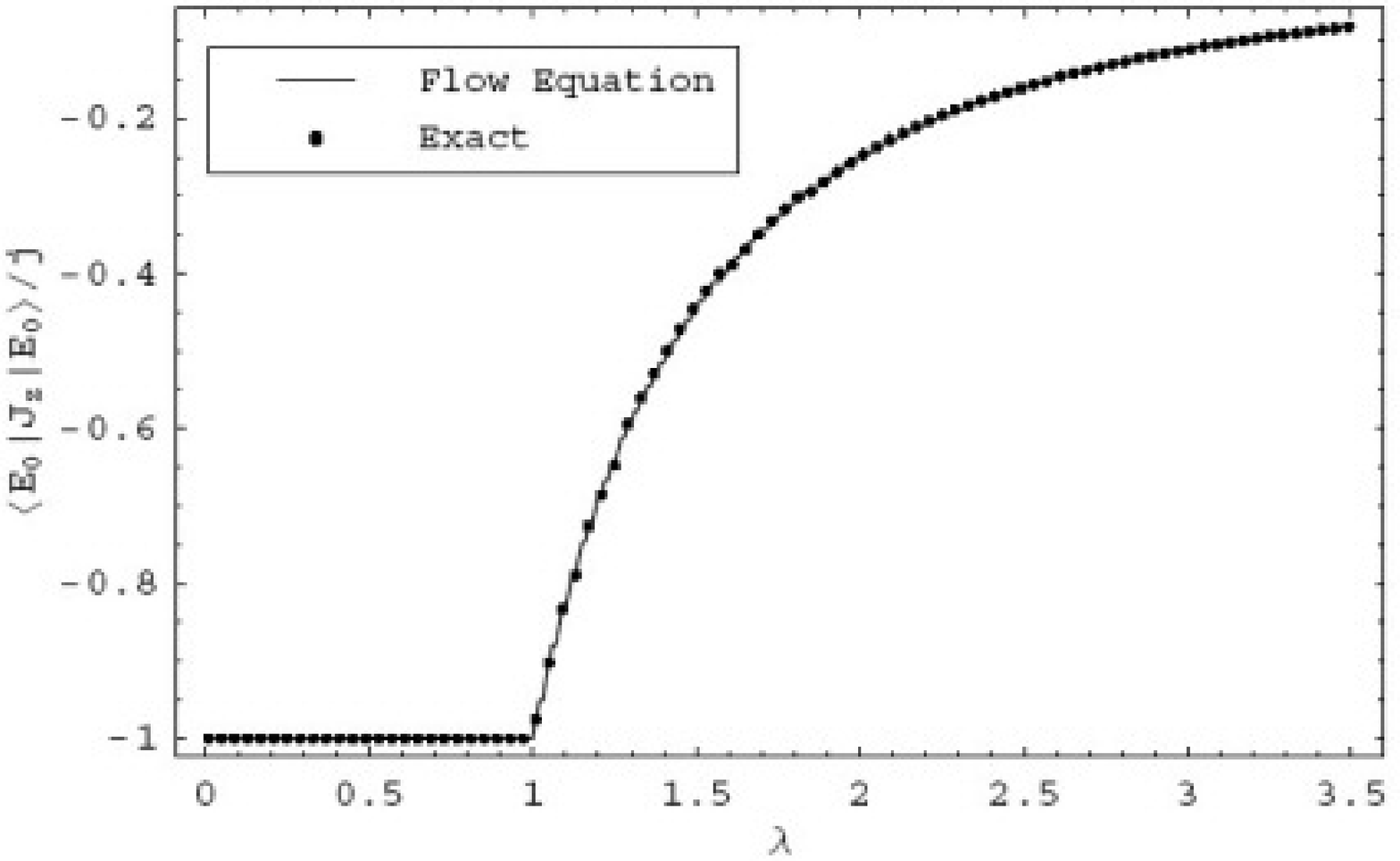,height=5cm,clip=,angle=0} &
	\epsfig{file=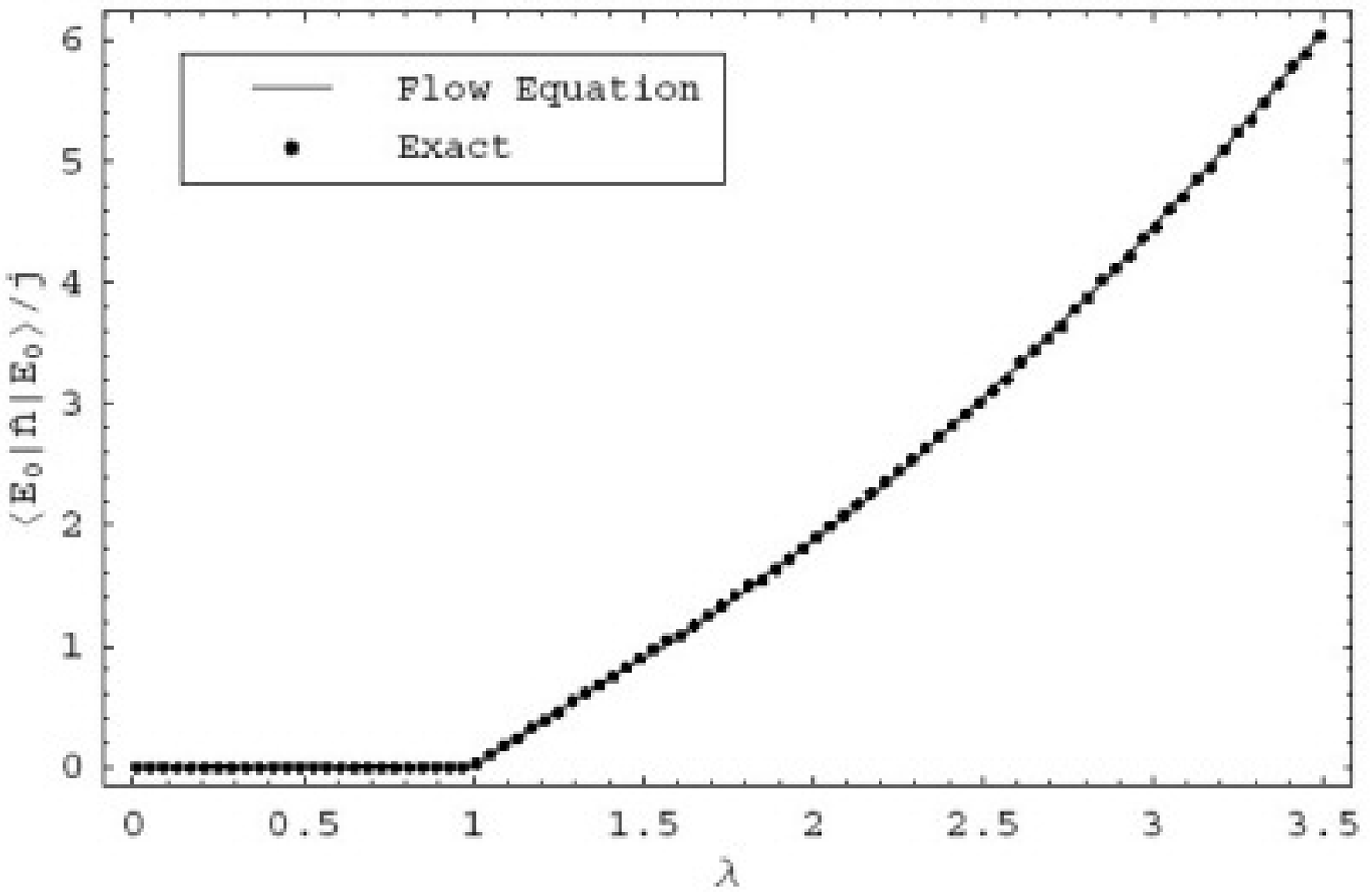,height=5cm,clip=,angle=0} \\
	(c) & (d)
	\end{tabular}
\end{narrow}
	\caption[Spectrum and expectation values of the RWA Hamiltonian]{(a) The function $n^{(\lambda,q)}_0(x)$ for $q=Q/j=2$ and $\lambda=2.2$ in the $\ell\rightarrow\infty$ limit. (b) The ground state energy together with the exact result for $j=200$. The expectation value of (c) $J_z$ and (d) $\hat{n}$ as functions of the coupling strength.}
	\label{DickeFigSet4}
\end{figure}

In the final part of this section we present the results obtained by diagonalizing the $Q$-sectors through a second application of the flow equation. All the results obtained thus far can be reproduced using this method, and we will not restate them here. Instead we focus on the RWA case, particularly with respect to the phase structure. The RWA  Hamiltonian $$H_{RWA}=J_z+\hat{n}+\frac{\lambda}{\sqrt{2j}}\left(J_+b+J_-b^\dag\right)$$
exhibits a phase transition at $\lambda=1$ \cite{Emary}, in contrast to critical value of $\lambda_c=0.5$ for the full Dicke Hamiltonian. The two phases are distinguished by the order parameter $\inp{E_0}{\hat{n}}{E_0}/j$ that becomes non-zero only in the second phase. This signals a macroscopic occupation of the bosonic mode which is responsible for the phenomenon of super-radiance. We proceed by solving the flow equation \eref{2ndStepFEH} for a range of $\lambda$ and $q$ values. In the $\ell\rightarrow\infty$ limit the off-diagonal part $n_1(x,\ell)$ vanishes and we obtain a set of functions $n^{(\lambda,q)}_0(x)\equiv n_0(x,\infty)$. Figure \ref{DickeFigSet4} (a) shows a typical example for $q=Q/j=2$ at $\lambda=2.2$, together with the exact result for $j=200$. Due to the ordering of the eigenvalues at $\ell=\infty$ we expect each $n^{(\lambda,q)}_0(x)$ to be an increasing function with $j n^{(\lambda,q)}_0(-1)$ corresponding to the lowest eigenvalue. The overall ground state energy is given by
\begin{equation}
	E_0=j \min_{q\geq0}\left\{n^{(\lambda,q)}_0(-1)\right\}.
\end{equation}
The result of this calculation appears in Figure \ref{DickeFigSet4} (b). Once it is known to which sector the ground state belongs for a certain $\lambda$ we solve the flow equation for the observables $\hat{n}$ and $J_z$ within this sector. Figures \ref{DickeFigSet4} (c) and (d) show the ground state expectation values calculated in this manner.

\specialhead{Conclusion and outlook}

We hope that our presentation has convinced the reader that flow equations obtained from continuous unitary transformations present a versatile and potentially very powerful technique for the treatment of interacting quantum systems. Our results for the Lipkin and Dicke models have clearly illustrated the myriad of information yielded by a non-perturbative solution of the flow equations. This approach is not without its difficulties however. As discussed in Section \ref{methodMoyal}, the construction of smooth initial conditions, although relatively simple for the models considered here, are generally non-trivial and in some cases possibly on par in difficulty with diagonalizing the Hamiltonian itself. Although the introduction of additional variables may aid in this construction, the resulting high dimensional PDE presents its own challenges, as was encountered in the Dicke model. Further studies into these issues are needed if the flow equations are to become a truly general and robust framework for non-perturbative calculations. It also seems inevitable that the treatment of realistic models would require further model-specific approximations which were largely absent in our approach.  One possibility is the generalization of the local approximation scheme of Sections \ref{firstPhaseSol} and \ref{dickeLocal}. The flow equation would be solved locally around an appropriate point, possibly determined by a variational method, and using a specialized generator. This may allow for a treatment of the system by considering fluctuations of the degrees of freedom around their classical values.

The flow equation is one example of a non-linear operator equation of which there are numerous others appearing in virtually all branches of physics. Although our focus has been on a small subclass of these, the solution methods developed here are much more general in nature and may in future find application in other fields as well.

\appendix
\chapter{Representations by irreducible sets}
\label{appa}
In Section \ref{fluxExpand} we made use of the fact that any flowing operator may be expressed in terms of operators coming from an irreducible set which contains the identity. Now we show why this is the case, using a result by von Neumann. First we establish some notation. Let ${\cal H}$ be a finite dimensional Hilbert space and denote by ${\cal B}({\cal H})$ the set of all matrix operators acting on it. The commutant ${\cal M}'$ of a set ${\cal M}\subseteq{\cal B}({\cal H})$ is defined as the set of operators which commute with all the elements of ${\cal M}$. The double commutant of ${\cal M}$ is defined by ${\cal M}''\equiv\left\{{\cal M}'\right\}'$. The following theorem \cite{Conway} provides the key:

\textbf{The Double Commutant Theorem:} \textit{If ${\cal A}$ is a subalgebra of ${\cal B}({\cal H})$ which contains the identity and is closed under hermitian conjugation then ${\cal A}={\cal A}''$.}

Now suppose ${\cal S}\subseteq{\cal B}({\cal H})$ is an irreducible set of hermitian operators, one of which is the identity, and ${\cal A}$ the subalgebra of ${\cal B}({\cal H})$ spanned by all the products of operators from ${\cal S}$. By Schur's lemma ${\cal S}'={\cal A}'=\left\{\lambda I\,|\,\lambda\in\mathbb{C}\right\}$ and thus ${\cal A}''={\cal B}({\cal H})$. It follows from the theorem above that ${\cal A}={\cal B}({\cal H})$, i.e. any operator acting on ${\cal B}({\cal H})$ may be written in terms of operators coming from ${\cal S}$.
\chapter{Calculating expectation values with respect to coherent states}
\label{appb}
Define $|z\rangle =(1+z z^\ast)^{-j}|z)$ where $|z)=\exp(zJ_+)\ket{-j}$ is the unnormalized coherent state. These states are known to provide a basis for the Hilbert state through the resolution of unity 
\begin{equation}
	I=\frac{2j+1}{\pi}\int\frac{{\rm d}z}{(1+z^2)^2}\ket{z}\bra{z},
\end{equation}
where the integral ranges over the entire complex plane. Furthermore, for any $z\neq0$ the state $\ket{z}$ has non-zero overlap with every state in the Hilbert space. The action of the $\texttt{su}(2)$ generators on these states may be expressed in terms of differential operators \cite{Perelomov} with respect to $z$ and $z^*$:
\begin{eqnarray}
J_z|z)=\left(-j+z\frac{\partial}{\partial z}\right)|z)=\hat{J}_z\left(z,\partial_z\right)|z)\\
J_+|z)=\frac{\partial}{\partial z}|z)=\hat{J}_+\left(z,\partial_z\right)|z)\\
J_-|z)=\left(2zj-z^2\frac{\partial}{\partial z}\right)|z)=\hat{J}_-\left(z,\partial_z\right)|z).
\end{eqnarray}
We will always consider $z$ and $z^*$ to be distinct variables, and that $(z|$ is a function of $z^*$ only. When calculating expectation values of the double commutators in Section \ref{fluxExpand}, one encounters expressions of the form 
\begin{equation}
\langle z|H_{n_1}H_{n_2}...H_{n_\alpha}|z\rangle=\frac{(z|H_{n_1}H_{n_2}\ldots H_{n_\alpha}|z)}{(z|z)}
\label{genexp}
\end{equation}
where $n_i=0  \ {\rm or} \ 1$. These can easily be computed by replacing each operator by its differential representation to obtain
\begin{equation}
	\langle z|H_{n_1}H_{n_2}...H_{n_\alpha}|z\rangle=\frac{1}{(z|z)}\hat{H}_{n_\alpha}\left(z,\partial_z\right)\ldots \hat{H}_{n_1}\left(z,\partial_z\right)(z|z),
\end{equation}
where the orders of the operators have been reversed. As a result these expectation values can be expressed as some rational function of $z=r\exp(i\theta)$ and $z^*=r\exp(-i\theta)$. Applying this method to $\ave{H_0}$ and $\ave{H_1}$, as defined in Section \ref{fluxExpand}, we obtain
\begin{eqnarray}
\langle z|H_0|z \rangle =j\left(\frac{r^2-1}{r^2 +1}\right)\ {\rm and} \ \ \langle z|H_1|z \rangle =\frac{(2j-1)r^2\cos(2\theta)}{(1+r^2)^2}.
\label{hifuncrt}
\end{eqnarray}
Using these equations to express $r$ and $\theta$ in terms of $\ave{H_i}$ we may consider expectation values of the form of equation \eref{genexp} as functions of these averages.
For example
\begin{eqnarray}
\langle z|[[\Delta H_0,\Delta H_1],(\Delta H_0)^2]|z\rangle=8jxy,
\label{cancelexample}
\end{eqnarray}
where $\langle H_0 \rangle=jx$ and $\langle H_1 \rangle =jy$.

\chapter{Scaling behaviour of fluctuations}
\label{appc}
We wish to show that 
\begin{equation}\label{aa}
\langle z|\Delta H_{n_1}\Delta H_{n_2}...\Delta H_{n_t}|z\rangle \sim j^{\lfloor \frac{t}{2} \rfloor}
 \ \ {\rm for}  \ \forall  \  t\in \mathbb{N}  \ {\rm where}  \ n_i\in \left\{0,1\right\}.
\end{equation}
From direct calculation this is found to hold for $t=2$, and we employ induction to obtain the general result. Assuming that this holds for all products of $k\leq t$ fluctuations, the induction step consists of adding either $\Delta H_0$ or $\Delta H_1$ to a general product $M=\Delta H_{n_1}\Delta H_{n_2}...\Delta H_{n_t}$ and then proving the result for $\inp{z}{M\Delta H_i}{z}$.\\

First we add an extra $\Delta H_0$. Using the results from the previous section we may write
\begin{eqnarray}
(z|M\Delta H_0|z)&=&-j(z|M|z)+(z|M z\partial_z|z)-\langle H_0 \rangle(z|M|z)\\
&=&-2j z z^\ast (1+z z^\ast)^{-1}(z|M|z)-z(z|(\partial_z M)|z)\nonumber \\
&&+z\partial_z(z|M|z)
\end{eqnarray}
where \eref{hifuncrt} and $(z|M z\partial_z|z)=z(\partial_z(z|M|z)-(z|(\partial_z M)|z))$ were used. It should be remembered that $M$ contains a $z$ dependency through the average appearing in each $\Delta{H_i}$. Dividing by $(z|z)$ leads to
\begin{eqnarray}\label{cc}
\langle z|M\Delta H_0|z\rangle=-2j z z^\ast (1+z z^\ast)^{-1}\langle z|M|z\rangle-z\langle z|(\partial_z M)|z\rangle \nonumber \\
+\frac{z}{(1+z z^\ast)^{2j}}\partial_z(z|M|z).
\end{eqnarray}
The last term can be rewritten using
\begin{equation}
\frac{1}{(1+z z^\ast)^{2j}}\partial_z(z|M|z)=\partial_z\langle z|M|z\rangle +\frac{2j z^\ast}{(1+z z^\ast)}\langle z|M|z\rangle,
\end{equation}
which is just the product rule, to obtain
\begin{equation}
\langle z|M\Delta H_0|z\rangle=z \partial_z\langle z|M|z\rangle-z\langle z|(\partial_z M)|z\rangle.
\label{finform}
\end{equation}
Note the important cancellation of terms proportional to $j\langle z|M|z\rangle$. From the induction hypothesis, and the fact that $\langle z|M|z\rangle$ is polynomial in $j$, the first term in \eref{finform} will be of order $j^{\lfloor \frac{t}{2} \rfloor}$. The second term becomes
\begin{equation}
z\sum_{i=1}^t \langle z|\Delta H_{n_1}...\Delta H_{n_{i-1}}\Delta H_{n_{i+1}}...\Delta H_{n_t}|z\rangle (\partial_z\langle H_{n_i}\rangle).
\label{sumterm}
\end{equation}
Taking into account the expressions for $\langle H_0 \rangle$ and $\langle H_1 \rangle$, we conclude that each term in equation \eref{sumterm} will have order $j^{1+\lfloor \frac{t-1}{2} \rfloor}$, which reduces to $j^{\lfloor \frac{t+1}{2} \rfloor}$ in both the cases where $t$ is odd and even. Thus, for any product $M$ of $t$ fluctuations we arrive at
\begin{equation}\label{ff}
\langle z|M\Delta H_0|z\rangle\sim j^{\lfloor \frac{t+1}{2} \rfloor},
\end{equation}
which concludes the induction step.

Exactly the same procedure is followed when adding a $\Delta H_1$, although more algebra is required as $\hat{H_1}(z,\partial_z)$ now contains second order derivatives to $z$. The final result remains unchanged:
\begin{equation}
\langle z|m\Delta H_1|z\rangle\sim j^{\lfloor \frac{t+1}{2} \rfloor}.
\end{equation}

In this case these results make exact the general notion that relative fluctuations scale like powers of one over the system size. Finally we mention that when calculating the expectation values of a double commutator there is often a cancellation of leading order terms. This can be seen in equation \eref{cancelexample}, where the sum of terms of order $j^2$ turns out to be of order $j$.
\chapter{Decomposing operators in the Dicke model}
\label{Appdecomp}
We define two classes of operators: $C_{n,m}=\left(J_+\right)^n\left(b\right)^m+\left(J_-\right)^n\left(b^\dag\right)^m$ and $T_{n,m}=\left(J_+\right)^n\left(b^\dag\right)^m+\left(J_-\right)^n\left(b\right)^m$. Note that $C_{n,m}$ changes $Q$ by $\left|n-m\right|$ while $T_{n,m}$ does so by $n+m$. We wish to find an algorithm for writing a given $C_{n,m}$ in terms of $T_q=J^+b+J_-b^\dag$ and $T_{\cdot,\cdot}$'s for which $\left|n-m\right|=n'+m'$. Central to this procedure are the relations
\begin{equation}
	C_{n,m}=\left\{\begin{tabular}{cl}
	$T_qC_{n-1,m-1}-\left(j^2-J_z^2\right)\hat{n}C_{n-2,m-2}$ & when $m,n\geq2$\\
	$T_qT_{0,m-1}-\hat{n}T_{1,m-2}$ & when $m\geq2,n=1$\\
	$T_qT_{n-1,0}-\left(j^2-J_z^2\right)T_{n-2,1}$ & when $m=1,n\geq2$\\
	$T_q$ & when $m,n=1$
	\end{tabular}\right.
\end{equation}
which follows directly from the definitions and basic properties of spin and boson operators. Given a $C_{n,m}$ we apply these relation repeatedly, until finally only $T_{\cdot,\cdot}$'s and $T_q$ appear. Note that the former can only appear linearly. Furthermore, any $C_{n',m'}$ appearing in intermittent steps has $\left|n'-m'\right|$ equal to $\left|n-m\right|$, and any $T_{n',m'}$ has $n'+m'=\left|n-m\right|$.

\chapter{Dicke model flow coefficients}
\label{Appflowcoef}
The non-zero entries of the $\Lambda^{(0)}$ matrix. For clarity the subscript indices have been raised.
\begin{eqnarray}
\Lambda^{(0)}(0,x_s)(0,x_s)&=&-32\,{x_s}\,\left( -1 + {{x_s}}^2 \right) \nonumber \\
\Lambda^{(0)}(0,x_b)(0,x_b)&=&-16\,x_b  \nonumber \\
\Lambda^{(0)}(0,x_q)(0,x_s)&=&16\,{x_s}\,{x_q} \nonumber \\
\Lambda^{(0)}(0,x_q)(0,x_b)&=&-8\,{x_q} \nonumber \\
\Lambda^{(0)}(0,x_q)(0,x_q)&=&4\,\left( -1 + 2\,x_b \,{x_s} + {{x_s}}^2 \right)  \nonumber \\
\Lambda^{(0)}(1,x_s)(0,x_s)&=&-16\,{\left( -1 + {{x_s}}^2 \right) }^2 \nonumber \\
\Lambda^{(0)}(1,x_s)(0,x_q)&=&4\,\left( -1 + {{x_s}}^2 \right) \,{x_q} \nonumber \\
\Lambda^{(0)}(1,x_b)(0,x_s)&=&-8\,\left( 2\,x_b \,\left( -1 + {{x_s}}^2 \right)  + {{x_q}}^2 \right)  \nonumber \\
\Lambda^{(0)}(1,x_b)(0,x_q)&=&-4\,x_b \,{x_q} \nonumber \\
\Lambda^{(0)}(1,x_q)(0,x_s)&=&8\,\left( -1 + {{x_s}}^2 \right) \,{x_q} \nonumber \\
\Lambda^{(0)}(1,x_q)(0,x_q)&=&8\,x_b \,\left( -1 + {{x_s}}^2 \right)  \nonumber \\
\Lambda^{(0)}(2,x_s)(0,x_b)&=&-8\,\left( 2\,x_b \,\left( -1 + {{x_s}}^2 \right)  + {{x_q}}^2 \right)  \nonumber \\
\Lambda^{(0)}(2,x_s)(0,x_q)&=&4\,\left( -1 + {{x_s}}^2 \right) \,{x_q} \nonumber \\
\Lambda^{(0)}(2,x_b)(0,x_b)&=&-16\,{x_b }^2 \nonumber \\
\Lambda^{(0)}(2,x_b)(0,x_q)&=&-4\,x_b \,{x_q} \nonumber \\
\Lambda^{(0)}(2,x_q)(0,x_b)&=&-8\,x_b \,{x_q} \nonumber \\
\Lambda^{(0)}(2,x_q)(0,x_q)&=&8\,x_b \,\left( -1 + {{x_s}}^2 \right)  \nonumber \\
\Lambda^{(0)}(3,x_s)(0,x_s)&=&-16\,{x_s}\,\left( -1 + {{x_s}}^2 \right) \,{x_q} \nonumber \\
\Lambda^{(0)}(3,x_s)(0,x_b)&=&8\,\left( -1 + {{x_s}}^2 \right) \,{x_q} \nonumber \\
\Lambda^{(0)}(3,x_s)(0,x_q)&=&8\,\left( 2\,x_b \,{x_s}\,\left( -1 + {{x_s}}^2 \right)  - {\left( -1 + {x_s}^2  \right)}^2 + {x_s}\,{{x_q}}^2 \right)  \nonumber \\
\Lambda^{(0)}(3,x_s)(1,x_b)&=&-4\,{x_q}\,\left( 4\,x_b \,\left( -1 + {{x_s}}^2 \right)  + {{x_q}}^2 \right)  \nonumber \\
\Lambda^{(0)}(3,x_s)(1,x_q)&=&4\,\left( -1 + {{x_s}}^2 \right) \,\left( 4\,x_b \,\left( -1 + {{x_s}}^2 \right)  + {{x_q}}^2 \right)  \nonumber \\
\Lambda^{(0)}(3,x_s)(2,x_b)&=&4\,{x_q}\,\left( 4\,x_b \,\left( -1 + {{x_s}}^2 \right)  + {{x_q}}^2 \right)  \nonumber
\end{eqnarray}
\begin{eqnarray}
\Lambda^{(0)}(3,x_s)(2,x_q)&=&-4\,\left( -1 + {{x_s}}^2 \right) \,\left( 4\,x_b \,\left( -1 + {{x_s}}^2 \right)  + {{x_q}}^2 \right)  \nonumber \\
\Lambda^{(0)}(3,x_b)(0,x_s)&=&16\,x_b \,{x_s}\,{x_q} \nonumber \\
\Lambda^{(0)}(3,x_b)(0,x_b)&=&-8\,x_b \,{x_q} \nonumber \\
\Lambda^{(0)}(3,x_b)(0,x_q)&=&-4\,\left(2\,x_b \,\left( -1 - 2\,x_b \,{x_s} + {{x_s}}^2 \right)  + {{x_q}}^2 \right)  \nonumber \\
\Lambda^{(0)}(3,x_b)(1,x_s)&=&4\,{x_q}\,\left( 4\,x_b \,\left( -1 + {{x_s}}^2 \right)  + {{x_q}}^2 \right)  \nonumber \\
\Lambda^{(0)}(3,x_b)(1,x_q)&=&4\,x_b \,\left( 4\,x_b \,\left( -1 + {{x_s}}^2 \right)  + {{x_q}}^2 \right)  \nonumber \\
\Lambda^{(0)}(3,x_b)(2,x_s)&=&-4\,{x_q}\,\left( 4\,x_b \,\left( -1 + {{x_s}}^2 \right)  + {{x_q}}^2 \right)  \nonumber \\
\Lambda^{(0)}(3,x_b)(2,x_q)&=&-4\,x_b \,\left( 4\,x_b \,\left( -1 + {{x_s}}^2 \right)  + {{x_q}}^2 \right)  \nonumber \\
\Lambda^{(0)}(3,x_q)(0,x_s)&=&-32\,x_b \,{x_s}\,\left( -1 + {{x_s}}^2 \right)  \nonumber \\
\Lambda^{(0)}(3,x_q)(0,x_b)&=&16\,x_b \,\left( -1 + {{x_s}}^2 \right)  \nonumber \\
\Lambda^{(0)}(3,x_q)(0,x_q)&=&4\,\left( -1 + 2\,x_b \,{x_s} + {{x_s}}^2 \right) \,{x_q} \nonumber \\
\Lambda^{(0)}(3,x_q)(1,x_s)&=&-4\,\left( -1 + {{x_s}}^2 \right) \,\left( 4\,x_b \,\left( -1 + {{x_s}}^2 \right)  + {{x_q}}^2 \right)  \nonumber \\
\Lambda^{(0)}(3,x_q)(1,x_b)&=&-4\,x_b \,\left( 4\,x_b \,\left( -1 + {{x_s}}^2 \right)  + {{x_q}}^2 \right)  \nonumber \\
\Lambda^{(0)}(3,x_q)(2,x_s)&=&4\,\left( -1 + {{x_s}}^2 \right) \,\left( 4\,x_b \,\left( -1 + {{x_s}}^2 \right)  + {{x_q}}^2 \right)  \nonumber \\
\Lambda^{(0)}(3,x_q)(2,x_b)&=&4\,x_b \,\left( 4\,x_b \,\left( -1 + {{x_s}}^2 \right)  + {{x_q}}^2 \right)   \nonumber
\end{eqnarray}
The non-zero entries of the $\Lambda^{(1)}$ matrix.
\begin{eqnarray}
\Lambda^{(1)}(1,0)(0,x_s)&=&-4\nonumber \\
\Lambda^{(1)}(3,0)(0,x_q)&=&-2\nonumber \\
\Lambda^{(1)}(3,0)(1,x_s)&=&2\, {x_q}\nonumber \\
\Lambda^{(1)}(3,0)(1,x_q)&=&2\, x_b \nonumber \\
\Lambda^{(1)}(3,0)(2,x_s)&=&-2\, {x_q}\nonumber \\
\Lambda^{(1)}(3,0)(2,x_q)&=&-2\, x_b \nonumber \\
\Lambda^{(1)}(3,x_s)(1,0)&=&-2\, {x_q}\nonumber \\
\Lambda^{(1)}(3,x_s)(2,0)&=&2\, {x_q}\nonumber \\
\Lambda^{(1)}(3,x_q)(1,0)&=&-2\, x_b \nonumber \\
\Lambda^{(1)}(3,x_q)(2,0)&=&2\, x_b \nonumber
\end{eqnarray}
The non-zero entries of the $\Lambda^{(2)}$ matrix.
\begin{eqnarray}
\Lambda^{(2)}(2,0)(0,x_b)&=&-4 \nonumber \\
\Lambda^{(2)}(3,0)(0,x_q)&=&4\, {x_s}\nonumber \\
\Lambda^{(2)}(3,0)(1,x_b)&=&-2\, {x_q}\nonumber \\
\Lambda^{(2)}(3,0)(1,x_q)&=&2\, \left( -1 + {{x_s}}^2 \right) \nonumber \\
\Lambda^{(2)}(3,0)(2,x_b)&=&2\, {x_q}\nonumber \\
\Lambda^{(2)}(3,0)(2,x_q)&=&2 - 2\, {{x_s}}^2\nonumber \\
\Lambda^{(2)}(3,x_b)(1,0)&=&2\, {x_q}\nonumber \\
\Lambda^{(2)}(3,x_b)(2,0)&=&-2\, {x_q}\nonumber \\
\Lambda^{(2)}(3,x_q)(1,0)&=&2 - 2\, {{x_s}}^2\nonumber \\
\Lambda^{(2)}(3,x_q)(2,0)&=&2\, \left( -1 + {{x_s}}^2 \right) \nonumber
\end{eqnarray}
The non-zero entries of the $\Lambda^{(3)}$ matrix.
\begin{eqnarray}
\Lambda^{(3)}(1,0)(0,x_q)&=&-2\nonumber \\
\Lambda^{(3)}(2,0)(0,x_q)&=&-2\nonumber \\
\Lambda^{(3)}(3,0)(0,x_s)&=&8\, {x_s}\nonumber \\
\Lambda^{(3)}(3,0)(0,x_b)&=&-4\nonumber \\
\Lambda^{(3)}(3,0)(1,x_s)&=&4\, \left( -1 + {{x_s}}^2 \right) \nonumber \\
\Lambda^{(3)}(3,0)(1,x_b)&=&4\, x_b \nonumber \\
\Lambda^{(3)}(3,0)(2,x_s)&=&4 - 4\, {{x_s}}^2\nonumber \\
\Lambda^{(3)}(3,0)(2,x_b)&=&-4\, x_b \nonumber \\
\Lambda^{(3)}(3,x_s)(1,0)&=&4 - 4\, {{x_s}}^2\nonumber \\
\Lambda^{(3)}(3,x_s)(2,0)&=&4\, \left( -1 + {{x_s}}^2 \right) \nonumber \\
\Lambda^{(3)}(3,x_b)(1,0)&=&-4\, x_b \nonumber \\
\Lambda^{(3)}(3,x_b)(2,0)&=&4\, x_b \nonumber 
\end{eqnarray}

\specialhead{BIBLIOGRAPHY}

\end{document}